\documentclass[fleqn,usenatbib,onecolumn,referee]{mnras}

\usepackage[T1]{fontenc}

\usepackage{ae,aecompl}

\usepackage{graphicx}	% Including figure files
\usepackage{amsmath}	% Advanced maths commands
\usepackage{amssymb}
\usepackage{times}	% Extra maths symbols
\usepackage[utf8]{inputenc}
\usepackage[english]{babel}
\usepackage{amsfonts}
\usepackage{float}
\usepackage{caption}
\usepackage{subcaption}
\usepackage{multirow}
\usepackage[dvipsnames]{xcolor}
\usepackage{mathrsfs}
\usepackage{scalerel}

\usepackage{hyperref}
\hypersetup{
  colorlinks=true,        % false: boxed links; true: colored links
  linkcolor=magenta,      % color of internal links
  citecolor=blue,         % color of links to bibliography
  urlcolor=cyan           % color of external links
}

\newcommand*{\ns}{_{\mathrm{NS}}}

\newcommand*{\lns}{_\mathrm{LNS}}
\newcommand*{\aparn}{\left(\partial_{t}-\mathscr{L}_{\boldsymbol{\beta}}\right)}

\newcommand{\eg}{e.g.,~}
\newcommand{\ie}{i.e.,~}
\newcommand{\cf}{cf.,~}

\newcommand{\spatial}[1]{{\color{PineGreen} #1}}
\newcommand{\rest}[1]{{\color{RedOrange} #1}}
\newcommand{\extrinsic}[1]{{\color{ProcessBlue} #1}}

\title[GRHD of non-perfect fluids]{General-relativistic hydrodynamics of
  non-perfect fluids: 3+1 conservative formulation and application to
  viscous black-hole accretion}

\author[M.\ Chabanov, L.\ Rezzolla and D.H.\ Rischke]{
Michail Chabanov$^{1}$, Luciano Rezzolla$^{1,2,3}$ and Dirk H.\ Rischke$^{1}$
\\
$^{1}$Institut f\"ur Theoretische Physik, Goethe-Universit\"at,
Max-von-Laue-Str.\ 1, 60438 Frankfurt am Main, Germany\\
$^{2}$Frankfurt Institute for Advanced Studies, Ruth-Moufang-Str. 1, 60438
Frankfurt am Main, Germany\\
$^{3}$School of Mathematics, Trinity College, Dublin 2, Ireland
}

\date{Accepted XXX. Received YYY; in original form ZZZ}

\pubyear{2021}

\begin{document}
\label{firstpage}
\pagerange{\pageref{firstpage}--\pageref{lastpage}}
\maketitle

% Abstract of the paper
\begin{abstract}
We consider the relativistic hydrodynamics of non-perfect fluids with the
goal of determining a formulation that is suited for numerical
integration in special-relativistic and general-relativistic
scenarios. To this end, we review the various formulations of
relativistic second-order dissipative hydrodynamics proposed so far and
present in detail a particular formulation that is fully general, causal,
and can be cast into a 3+1 flux-conservative form, as the one employed in
modern numerical-relativity codes. As an example, we employ a variant of
this formulation restricted to a relaxation-type equation for the bulk
viscosity in the general-relativistic magnetohydrodynamics code
\texttt{BHAC}. After adopting the formulation for a series of standard
and non-standard tests in 1+1-dimensional special-relativistic
hydrodynamics, we consider a novel general-relativistic scenario, namely,
the stationary, spherically symmetric, viscous accretion onto a black
hole. The newly developed solution -- which can exhibit even considerable
deviations from the inviscid counterpart -- can be used as a testbed for
numerical codes simulating non-perfect fluids on curved backgrounds.
\end{abstract}

% Select between one and six entries from the list of approved keywords.
% Don't make up new ones.

%%%%%%%%%%%%%%%%%%%%%%%%%%%%%%%%%%%%%%%%%%%%%%%%%%

%%%%%%%%%%%%%%%%% BODY OF PAPER %%%%%%%%%%%%%%%%%%

%------------------------------------------------------------------------
%------------------------------------------------------------------------
\section{Introduction}
\label{sec:intro}
%------------------------------------------------------------------------
%------------------------------------------------------------------------

The detection of the first binary neutron-star (BNS) merger event,
GW170817 \citep{Abbott2017_etal} has provided a new valuable tool to
study matter and gravity under extreme conditions. Especially the
detection of electromagnetic counterparts in the form of a short
gamma-ray burst \citep{Abbott2017dddd} and of a kilonova
\citep{Drout2017, Cowperthwaite2017} accompanying GW170817, has made this
event an incredibly rich laboratory for physics, providing a number of
constraints on the equation of state (EOS) of nuclear matter \citep[see,
  \eg][]{Margalit2017, Bauswein2017b, Rezzolla2017, Ruiz2017, Annala2017,
  Radice2017b, Most2018, De2018, Abbott2018b, Montana2018, Raithel2018,
  Tews2018, Malik2018, Koeppel2019, Shibata2019}.

BNS mergers are highly dynamical and nonlinear phenomena especially
during the first few milliseconds after merger, \cite[see, \eg][for some
  reviews]{Baiotti2016, Paschalidis2016, Burns2020}. As it has been shown
recently \citep{Alford2018, Alford2020, Alford2020b}, the highly
nonlinear oscillations that are present in this violent phase could be
damped significantly due to bulk-viscosity dissipation coming from
modified Urca processes. Thus, bulk viscosity might lead to
modifications, possibly large, to the post-merger gravitational-wave
signal. Furthermore, it was shown in high-resolution general-relativistic
magnetohydrodynamic (GRMHD) simulations that the matter after merger is
subject to MHD instabilities, \eg the Kelvin-Helmholtz instability
\citep{Baiotti08, Radice2012a} and the magnetorotational instability
\citep{Siegel2013, Kiuchi2017}. Due to these instabilities, turbulence
can develop and be maintained, which will ultimately influence BNS-merger
observables, such as the gravitational-wave signal and the ejected matter.

The emergence of turbulence clearly represents a challenge for
numerical-relativity simulations, which can only be performed with
limited resolutions and are normally carried out with resolutions that
are well above the scale at which turbulence is physically quenched
off. As a result, a number of studies have employed effective
shear-viscous models in order to capture the effects of the
magneto-turbulent motion in the remnant of BNS mergers. An approach to
handle this problem has been suggested via the use of
general-relativistic large-eddy simulations (LES) in pure
hydrodynamics. This method maps effects from numerical calculations with
high resolution to simulations with lower resolution, which would
otherwise disappear in an implicit filtering procedure \citep{Radice2017,
  Radice2018, Radice2018a, Radice2020}. The same method has been
systematically extended to the full set GRMHD equations to study the
amplification of the magnetic field shortly after merger
\citep{Vigano2020,Aguilera-Miret2020}. In this way, it was shown that by
considering the subgrid-scale model for the induction equation only, and
suitably increasing well above the expected value of order one the
phenomenological coefficients associated with the LES terms,
magnetic-field strengths can be achieved that are comparable to those
measured with a high-resolution direct simulation.

While the robustness and universality -- \ie their effectiveness to
capture multi-scale turbulence without expensive fine tuning -- remains
to be assessed \citep[see][for a careful comparison of momentum transport
  models for numerical relativity]{Duez2020}, the inclusion of genuinely
dissipative effects already in the general-relativistic hydrodynamics or
MHD equations offers the possibility to study a variety of physical
phenomena on more robust grounds, where the much larger computational
costs can be compensated by the use of high-order methods [see, \eg
  \citealt{Radice2012a, Most2019b} for standard finite-volume/difference
  methods, and \citealt{Fambri2018, Hebert2018} for more advanced
  approaches]. A first attempt in the direct solution of the dissipative
hydrodynamics equations has been made by \citet{Duez2004b}, who has
performed simulations in full general relativity of a differentially
rotating star, although employing an acausal formulation of the equations
of dissipative hydrodynamics. More recently, \citet{Shibata2017a} and
\citet{Fujibayashi2018} have employed an incomplete but causal viscous
model motivated by the work of \citet{Israel79_new} to assess the effects
of turbulence in long-term evolutions of BNS-merger remnants. In this
way, it was found that viscous effects can significantly change the
amount and composition of the matter outflow from BNS mergers, altering
the electromagnetic signal and nucleosynthetic yields \citep{Baiotti2016,
  ShibataRev19, Burns2020}. These studies, together with the prospects
suggested by the microphysical investigations of \citet{Alford2018},
clearly motivate a more comprehensive and mathematically complete use of
general-relativistic hydrodynamics of non-perfect fluids to study BNS
mergers. In addition, there is a number of phenomena that emerge
when considering the dynamics beyond the ideal MHD limit
\citep{Palenzuela:2008sf} and that can be studied in the framework of
resistive MHD \citep{Palenzuela:2008sf, Dionysopoulou:2012pp,
Ripperda2019, Wright2020}. For instance, these phenomena include
interacting magnetospheres in inspiraling BNS systems
\citep{Palenzuela2013a, Nathanail_2020, Most2020b}, or the dynamics and
magnetic-field amplification in the remnant of a BNS merger
\citep{Palenzuela2013, Dionysopoulou2015, Shibata_2021}, to name only a
few. Hereafter, however, our attention will be restricted to the study
of unmagnetized plasmas, so that we consider dissipative effects
arising from a non-perfect description of the fluid alone, setting all
electromagnetic fields to zero.

We should remark that although the use of dissipative
relativistic-hydrodynamics effects in the modelling of BNS mergers has
only just started, this is not the case when describing the
special-relativistic evolution of hot and dense strongly interacting
matter created in heavy-ion collisions. In this case, in fact, a vast
literature has been developed on the optimal way to model dissipative
effects in such collisions \citep[see, \eg][for some recent
  reviews]{romatschke_romatschke_2019,Busza2018} and extract from them
the imprint of the shear and bulk viscosity via comparison with the
experimental data \citep{Bernhard2019}.

We here make use of the experience developed when modelling numerically
dissipative effects in heavy-ion collisions (see \citealt{McNelis2021}
for some very recent overview) to propose a comprehensive description of
general-relativistic dissipative hydrodynamics (GRDHD) to be used for
modelling dissipative effects in BNS mergers. In particular, we propose a
complete formulation of the equations of GRDHD based on a second-order
description of relativistic dissipative effects and cast them into a 3+1
split of spacetime, in which they can then be coupled to the solution of
the Einstein equations. More specifically, we choose the equations first
derived by \citet{Israel76}, in the notation of \citet{Hiscock1983} (from
now on denoted as \textbf{HL83}), to serve as our reference second-order
theory. We discuss the properties of this formulation and provide a
comprehensive comparison with other and equivalent formulations of
relativistic dissipative hydrodynamics that have been proposed and
employed in the literature. In addition, we discuss the implementation of
this formulation in the general-relativistic magnetohydrodynamics code
\texttt{BHAC}, where it is subject to a number of tests in special and
general relativity.

The structure of the paper is as follows. In Section
\ref{sec:brief_overview} we first provide a brief review of relativistic
dissipative hydrodynamics, mostly to recall definitions and conventions,
while Section \ref{sec:phenomeno_approach} is dedicated to the
presentation of the set of the \textbf{HL83} equations in a general
four-dimensional manifold. Extending the work of \citet{Peitz1997} and
\citet{Peitz1999}, in Section \ref{sec:3+1} we derive a 3+1 decomposed
version of \textbf{HL83}, including all source terms and gradients of
fluid variables, which are separately listed in Section
\ref{sec:3+1sources}. In Section \ref{sec:methods} we use this
formulation to implement a simplified version for bulk viscosity in the
GRMHD code \texttt{BHAC} \citep{Porth2017} and perform tests in special
relativity in Section \ref{sec:testing}. As a first but rigorous test in
general relativity, we consider in Section \ref{sec:accretion} the
problem of stationary, spherically symmetric viscous accretion onto a
black hole. We conclude this work in Section \ref{sec:conclusions}, where
we also comment on future developments. A number of Appendices is also
used to provide additional information on the currently available
second-order descriptions of relativistic dissipative effects and on the
differences among them. In addition, details are provided on the
numerical solution of the viscous accretion problem, which is less
trivial than may appear at first sight.

Unless stated otherwise, we use geometrised units where the speed of
light $c=1$ and the gravitational constant $G=1$. Greek letters denote
spacetime indices, \ie $\mu = 0, 1, 2, 3$, while Roman letters cover
spatial indices only, \ie $i = 1, 2, 3$. Also, we make use of Einstein's
summation convention and choose the metric signature to be $
(-,+,+,+)$. Bold symbols, such as $\boldsymbol{g}$, refer to tensors of
generic rank, while the symbol $\boldsymbol{\nabla}$ denotes the
covariant derivative with respect to $\boldsymbol{g}$; we use the
following definitions for the components of a symmetric and antisymmetric
rank-2 tensor: $T^{(\mu \nu )} := \frac{1}{2}\left( T^{\mu \nu} + T^{\nu
  \mu}\right)$, $T^{[\mu \nu]} := \frac{1}{2}\left( T^{\mu \nu} - T^{\nu
  \mu}\right)$.

%------------------------------------------------------------------------
%------------------------------------------------------------------------
\section{Relativistic dissipative hydrodynamics: a brief overview}
\label{sec:brief_overview}
%------------------------------------------------------------------------
%------------------------------------------------------------------------
%% 

As anticipated, a large portion of the theory of relativistic
dissipative hydrodynamics dates back to the 70’s and is both vast and
somewhat disorienting, with multiple formulations having been
produced. Even though more than 40 years have passed since then, the
theory is still very much under development, especially when assessing
the robustness or the nonlinear stability of the various formulations
proposed. Furthermore, because the application of the various
formulations to realistic physical and astrophysical scenarios is just
starting, many of these issues remain unexplored. For this reason, we
use this section to provide a very brief overview of the formulations
proposed and of the terminology that has been introduced with them
\citep[see also][for a more systematic introduction to the relativistic
  hydrodynamics of non-perfect fluids]{Rezzolla_book:2013}.

We start by recalling that, in general, one distinguishes between
first-order and second-order theories of relativistic dissipative
hydrodynamics. \textit{``First-order''} theories are relativistic
extensions of the Navier-Stokes (NS) equations, in which the dissipative
quantities are directly related to first-order gradients of the primary
fluid variables appearing in perfect-fluid hydrodynamics, such as \eg
pressure, rest-mass density and fluid velocity\footnote{The large
majority of these works concentrate on single-fluid scenarios, but
valuable work has been done also when considering dissipative
hydrodynamics of multi-fluids [see, \eg \citealt{Carter1989,
    Andersson2015}, but also \citealt{Andersson2020} for a recent
  review].}. However, already early on \citep{Hiscock1985, Hiscock1987},
it was shown that a large class of these first-order theories suffers from
instabilities and acausal behaviour because of their partially parabolic
nature \citep{Hiscock1985, Hiscock1987, Denicol_2008}.

In view of this drawback, a class of \textit{``second-order''} theories
was developed, which take their denomination from the fact that they
extend first-order theories by including terms of second order in
quantities which describe deviations from perfect-fluid
hydrodynamics. These additional quantities can be either gradients of the
primary fluid variables or dissipative currents such as the
bulk-viscosity pressure, the heat current (also referred to as heat flux)
or the shear-stress tensor. Because the first formulations of these
second-order theories were made in the 60's and 70's, starting from the
work of \citet{Mueller1967}, \citet{Israel76} and \citet{Israel79_new},
these theories are sometimes also referred to as Mueller-Israel-Stewart
(MIS) theories. Hereafter, however, we will use the more general
denomination of second-order theories.

While there are multiple ways of constructing second-order theories, here
we focus on three conceptually rather different approaches. A first
approach to second-order theories, originally suggested by
\citet{Israel76}, extends the definition of the entropy current by
introducing all terms of second order in the dissipative currents that
are allowed by symmetry. The second-law of thermodynamics then leads to
relaxation-type equations for the dissipative currents \citep[see
  also][]{Hiscock1983, Muronga2004, Jaiswal2013}. A second approach,
instead, adds to the definitions of the dissipative currents all terms of
second order in gradients of the primary fluid variables that are allowed
by symmetry. This approach follows in spirit that of a systematic
gradient expansion and was first suggested by
\citet{Baier2008}. Unfortunately, second-order (spatial) gradients
usually destroy the hyperbolic nature of a partial differential
equation. To counter this problem and obtain a hyperbolic formulation --
that can therefore be solved numerically -- a modification of the
gradient expansion was proposed by \citet{Baier2008}, where gradients of
primary fluid variables were replaced by dissipative currents, using the
first-order Navier-Stokes relation. This effectively leads to a
resummation of gradients and, ultimately, to relaxation-type equations
for the dissipative currents.

Finally, the third approach considers hydrodynamics as an effective
theory in the long-wavelength, low-frequency limit of an underlying
microscopic theory. Taking kinetic theory for the latter, it is possible
to derive hydrodynamics from the relativistic Boltzmann equation and the
method of moments \citep{Israel79_new, Betz2009, Denicol_2012,
  Denicol2012b, Jaiswal2013b}. This approach leads to an infinite system
of coupled equations of motion for the moments of the deviation of the
single-particle distribution function away from local equilibrium. In
particular, \citet{Denicol2012b} have proposed a systematic
power-counting scheme in Knudsen and inverse Reynolds numbers that allows
for a truncation of the infinite set of moment equations at an arbitrary
order in these quantities \citep{Denicol:2012vq}.

The stability and causality of second-order theories has been studied in
a series of works. In particular, stability and causality were analysed
in a linear regime using the rest frame of the fluid \citep{Hiscock1983},
with the extension to a moving frame in the case of either bulk viscosity
\citep{Denicol_2008}, or of shear viscosity \citep{Pu:2009fj}. In this
way, it was found that stability implies causality and vice-versa, but
only if the relaxation times are larger than a certain timescale. These
studies were recently extended to include heat flow \citep{Brito:2020nou}
and a magnetic field \citep{Biswas:2020rps}. To the best of our
knowledge, the nonlinear stability of second-order theories remains to
be assessed.

While most of these works have considered either a flat (Minkowski)
background or a fixed curved background, recent works have studied the
coupling of the Einstein equations to second-order dissipative
hydrodynamics. This has been done by \citet{Bemfica2019} when considering
only the effects of bulk viscosity and by \citet{Bemfica2020} when
including also shear viscosity. These second-order formulations -- that
were shown to be causal and to admit unique solutions under rather
reasonable conditions -- effectively pave the way for applications in
numerical relativity and hence for the modelling of BNS mergers.

We conclude this overview section by remarking that it has become clear
recently that the unstable and acausal behaviour of first-order theories
discussed above is actually related to the particular choice of rest
frame of the fluid, for instance the particle frame proposed by
\citet{Eckart40_new} or the energy-density frame proposed by
\citet{Landau-Lifshitz6}. As a result, new causal and stable formulations
of first-order theories have been found by choosing different rest frames
\citep{Van:2011yn, Disconzi2017, Bemfica2019b, Kovtun2019, Hoult2020,
  Taghinavaz2020}. The prospects of using these theories, that are in
principle easier to solve numerically, are very good but no concrete
application to heavy-ion collisions or BNS mergers has been presented
yet.

%------------------------------------------------------------------------
%------------------------------------------------------------------------
\section{A comprehensive formulation of GRDHD via the entropy current}
\label{sec:phenomeno_approach}
%------------------------------------------------------------------------
%------------------------------------------------------------------------

In this section, we briefly review the phenomenological approach to
second-order GRDHD leading to the \textbf{HL83} formulation in a generic
four-dimensional manifold. The first assumption made is that the
rest-mass\footnote{Hereafter we will always refer to rest-mass as this is
a quantity normally conserved in simulations of neutron stars. However,
the rest-mass here can be replaced by any other conserved charge, \eg
baryon number.} current \(J^{\mu}\) and the energy-momentum tensor
\(T^{\mu\nu}\) continue to provide a valid description for fluids which
are out of thermodynamical equilibrium. Fluids in such an off-equilibrium
state show dissipative effects, which characterize them as non-perfect
fluids. Hereafter, we will distinguish perfect fluids from generic
(non-perfect) fluids by using the lower index ``PF''.

The conservation of rest-mass (or some associated conserved number
density), energy, and momentum are expressed by the five conservation
equations
\begin{align}
\nabla_{\mu}{J}^{\mu}&=0\,,\label{eq:continuity}\\ 
\nabla_{\mu}{T}^{\mu \nu} &= 0\,,
\label{eq:em_conservation}
\end{align}
where \(J^{\mu}=J_{_{\textrm{PF}}}^{\mu}\) and \(T^{\mu \nu} =
T_{_{\textrm{PF}}}^{\mu \nu}\). These equations can be complemented by
one equation of state (EOS) of the form
\begin{equation}
  \label{eq:EOS}
  e_{_{\textrm{PF}}} = e_{_{\textrm{PF}}} \left( \rho_{_{\textrm{PF}}}, p_{_{\textrm{PF}}} \right)\,,
\end{equation}
so that perfect fluids need only five independent components to be
described. However, under more general conditions, \ie for non-perfect
fluids, \(J^{\mu}\) and \(T^{\mu \nu}\) contain 14 independent
variables. The physical meaning of the nine additional degrees of freedom
for non-perfect fluids can be made transparent in a tensor decomposition
with respect to the fluid velocity $u^\mu$. Because in astrophysical
scenarios it is more intuitive to relate the fluid four-velocity to the
motion of particles, we hereafter adopt the ``Eckart frame'' where
$u^\mu$ is the velocity of the flow of rest-mass \citep{Eckart40_new,
  Rezzolla_book:2013} and the energy-momentum tensor reads
\begin{align}
T^{\mu \nu} = e u^{\mu}u^{\nu} + (p+\Pi)h^{\mu \nu} + 2 q^{(\mu}u^{\nu)}
+ \pi^{\mu \nu}\,,\label{eq:energy_momentum}
\end{align}
while \(J^{\mu} = \rho u^{\mu}\) maintains the form of the rest-mass
current of a perfect fluid, where $\rho$ is the rest-mass
density.

In Eq. (\ref{eq:energy_momentum}), $e$ is the energy density, $p$ the
isotropic pressure, $\Pi$ the bulk-viscosity pressure, and $h^{\mu \nu}
:= g^{\mu \nu} + u^\mu u^\nu$ the projector orthogonal to
\({u}^{\mu}\). Furthermore, \(q^{\mu}\) is the heat current, which is
orthogonal to the fluid four-velocity, \ie $q^{\mu}u_{\mu}=0$. Finally,
$\pi^{\mu \nu}$ is the shear-stress tensor and has the following
properties: it is symmetric $\pi^{\mu \nu} = \pi^{\nu \mu}$, purely
spatial $\pi^{\mu \nu}u_{\mu} = 0$, and trace-free ${\pi^{\mu}}_{\mu}
=0$, so that it contains five independent components. In addition,
it is useful to define $b^{\langle \mu \rangle} :=
h^{\mu}_{\phantom{\mu} \nu} b^\nu$ as the projection of the
contravariant components of an arbitrary vector $b^{\nu}$ in the
direction orthogonal to $\boldsymbol{u}$, and $b^{\langle \mu \nu
\rangle} := ( {h_\alpha}^{(\mu} {h^{\nu)}}_\beta - \frac{1}{3} h^{\mu
\nu} h_{\alpha \beta} ) b^{\alpha \beta}$ as the symmetric and
trace-free projection of an arbitrary rank-2 tensor $b^{\mu \nu}$ in
the direction orthogonal to $\boldsymbol{u}$.

Although we will not employ it here, we should recall that in the context
of heavy-ion collisions, and because the baryon chemical potentials are
typically small in the center of the collision zone, a different frame is
normally adopted, namely, the ``Landau frame'' \citep{Landau-Lifshitz6,
  Rezzolla_book:2013}. In such a frame, the fluid four-velocity is the
timelike eigenvector of the energy-momentum tensor, the heat current is
absent in the tensor decomposition (\ref{eq:energy_momentum}), but its
three independent degrees of freedom reappear in the form of a diffusion
current as part of the non-perfect rest-mass current.

To close the system of
Eqs. (\ref{eq:continuity})--(\ref{eq:em_conservation}), we need therefore
nine additional equations which determine the evolution of the
dissipative currents \(\Pi\), \(q^{\mu}\) and \(\pi^{\mu \nu}\).
Following \citet{Israel76} and \citet{Hiscock1983}, such relations are
obtained by ensuring the positivity of entropy production. The entropy
current $\boldsymbol{\mathcal{S}}$ depends quadratically on the
dissipative currents and reads
\begin{equation}
\mathcal{S}^{\mu}=s u^{\mu} + \frac{q^{\mu}}{T}
 -\left(\beta_0\Pi ^2+\beta_1 q_{\alpha}q^{\alpha}+\beta_2 \pi_{\alpha \beta}
   \pi^{\alpha \beta}\right)\frac{u^{\mu}}{2T}
 +\alpha_0\frac{\Pi q^{\mu}}{T} 
 + \alpha_1\frac{q_{\alpha}{\pi^{\alpha \mu}}}{T}\,,\label{eq:2o_entropycurrent}
\end{equation}
where the physical meaning of the coefficients \(\alpha_0, \alpha_1,
\beta_0, \beta_1, \beta_2\) will become clear below and $T$ is the
temperature. The latter and the baryon chemical potential $\mu$ can be
defined by matching the energy density $e$ and the number density $n$ to
the corresponding values of a fictitious equilibrium state, \ie $e =
e_{_{\textrm{PF}}}$, $n = n_{_{\textrm{PF}}}$, such that $p =
p_{_{\textrm{PF}}}$, and then from well-known thermodynamical relations
also $T$ and $\mu$, can be determined from the EOS (\ref{eq:EOS}):
\begin{align}
\frac{1}{T} = \left(\frac{\partial s}{\partial e} \right)_n\,, \qquad \qquad 
  \mu = -T\left(\frac{\partial s}{\partial n} \right)_e\,, \qquad \qquad 
s = \frac{e+p-\mu n}{T}\,,
\end{align}
where $s$ denotes the entropy density.

From the second law of thermodynamics
\begin{equation}
\nabla_{\mu}\mathcal{S}^{\mu}\geq 0\,,
\end{equation}
the following set of constitutive equations is obtained 
\begin{align}
\tau_{_{\Pi}} \dot{\Pi} &= \Pi_{\ns} - \Pi - \frac{1}{2} \zeta \Pi T
  \nabla_{\mu}\left(\frac{\tau_{_{\Pi}} u^{\mu}}{\zeta T}\right) + \alpha_0 \zeta
  \nabla_{\mu} q^{\mu} + \gamma_0 \zeta  T q^{\mu}\nabla_{\mu}\left(
  \frac{\alpha_0}{T}\right)\,,\label{eq:constitutive_full1}\\
  \tau_{\mathrm{q}} \dot{q}^{\langle \mu \rangle}&=
  q^{\phantom{\ns}\mu}_{\ns} - q^{\mu} -\frac{1}{2}\kappa T^2 q^{\mu}
  \nabla_{\nu}\left(\frac{\tau_{\mathrm{q}} u^{\nu}}{\kappa
    T^2}\right)%\notag\\
  %
  %&\phantom{=}
  +\kappa T \bigg[ \alpha_0 \nabla^{\langle \mu \rangle}\Pi +
    \alpha_1 \nabla_{\nu}\pi^{\nu \langle \mu \rangle} \notag \\
  &\phantom{=}+(1-\gamma_0)\Pi T
    \nabla^{\langle \mu \rangle}\left(\frac{\alpha_0}{T}\right) +
    (1-\gamma_1)T \pi^{\mu
      \nu}\nabla_{\nu}\left(\frac{\alpha_1}{T}\right)\bigg]\,,
 \label{eq:constitutive_full2}\\ 
 \tau_{\pi} \dot{\pi}^{\langle \mu \nu \rangle} &=
 \pi_{\ns}^{\phantom{\ns}\mu \nu} - \pi^{\mu \nu} - \frac{1}{2} \eta T
 \pi^{\mu \nu} \nabla_{\lambda}\left( \frac{\tau_{\pi} u^{\lambda}}{\eta
   T}\right) + 2\alpha_1 \eta \nabla^{\langle \mu}q^{\nu \rangle} +
 2\gamma_1\eta T q^{\langle \mu}\nabla^{\nu
   \rangle}\left(\frac{\alpha_1}{T}\right)\,,\label{eq:constitutive_full3}
\end{align}
where we have introduced two new coefficients, $\gamma_0$ and $\gamma_1$,
whose existence is related to the ambiguity when factoring out the terms
which involve the products $\Pi q^{\mu}$ and $q_{\alpha}\pi^{\alpha
  \mu}$. Also, note in the expressions above the introduction of an
operator we will often employ, \ie the \textit{``comoving derivative''}
\(\boldsymbol{\dot{A}}:= (\boldsymbol{u} \cdot \boldsymbol{\nabla})
\boldsymbol{A} = u^{\mu} \nabla_{\mu} \boldsymbol{A}\), where
\(\boldsymbol{A}\) can be an arbitrary tensor field. The comoving
derivative is naturally accompanied by \textit{``relaxation times''}
(that are zero for an inviscid fluid) for the bulk-viscosity pressure,
the heat current, and shear-stress tensor, that are respectively defined
as
\begin{equation}
  \label{eq:tau_Pi}
  \tau_{_{\Pi}} := \beta_0 \zeta \,, \qquad \qquad  \tau_{\mathrm{q}} := \beta_1
\kappa T\,, \qquad \qquad  \tau_{\pi} := 2 \beta_2 \eta \,.
\end{equation}
where $\zeta$ is the bulk viscosity, $\kappa$ the heat conductivity,
$\eta$ the shear viscosity. All of the coefficients introduced in
Eq. \eqref{eq:tau_Pi} are by definition non-negative and set the
timescales over which non-equilibrium effects push the equations of GRDHD
towards the NS solution.

A physically intuitive interpretation of the bulk-viscosity pressure, of
the heat current, and of the shear-stress tensor comes from the naive
extensions of the non-relativistic NS equations \citep{Landau-Lifshitz6},
which leads to the NS expressions of these quantities as\footnote{Note
that after using the conservation equations (\ref{eq:em_conservation})
for perfect fluids, which read $m n ha^{\mu}=-\nabla^{\left\langle \mu
  \right\rangle}p$ -- where $h$ is the specific enthalpy and $m$ the
rest-mass of the particles constituting the fluid -- and using the
Gibbs-Duhem relation, \ie $dp=sdT+nd\mu$, the NS value of the heat
current \eqref{eq:1oheat} can also be approximated by the expression
\(q^{\phantom{\ns}\mu}_{\ns}=(\kappa T^2/mh)\nabla^{\langle \mu
  \rangle}(\mu/T)\). This expression is an identity at first order in
gradients of primary fluid variables and in dissipative currents in the
sense that if higher-order terms would be used in the form of the
acceleration, then terms of order $\mathcal{O}_{_{2\mathrm{K}}}$ or
$\mathcal{O}_{_{\mathrm{2R}}}$ or $\mathcal{O}_{_{\mathrm{RK}}}$ would
appear [see Eq. (\ref{eq:classification1}) in Appendix \ref{sec:review}
  for the definitions of the symbols $\mathcal{O}_{_{2{\mathrm{K}}}}$
  , $\mathcal{O}_{_{\mathrm{2R}}}$ and
  $\mathcal{O}_{_{\mathrm{RK}}}$].}
\begin{align}
\Pi_{\ns} &= - \zeta \Theta\,,\label{eq:1obulk}\\
q^{\phantom{\ns}\mu}_{\ns}&= -\kappa T \left(\nabla^{\left\langle \mu
  \right\rangle}\ln T+a^{\mu}\right)\,,\label{eq:1oheat}\\
\pi^{\phantom{\ns}\mu \nu}_{\ns}&=-2\eta\sigma^{\mu \nu}\,.\label{eq:1oshear}
\end{align}

It is then easy to show that when the entropy current contains only
first-order dissipative currents, \ie $\beta_0 = \beta_1 = \beta_2 =
\alpha_0 = \alpha_1 = 0$, then \citep{Rezzolla_book:2013}:
\begin{align}
\Pi &= \Pi_{\ns}\,,\label{eq:fo-eckart1}\\
q^{\mu} &= q^{\phantom{\ns}\mu}_{\ns}\,,\\
\pi^{\mu \nu} &= \pi^{\phantom{\ns}\mu \nu}_{\ns}\,.\label{eq:fo-eckart2} 
\end{align}
As anticipated in Sec. \ref{sec:brief_overview}, Eqs.
(\ref{eq:fo-eckart1})--(\ref{eq:fo-eckart2}) are acausal and fall into
the class of equations investigated by \citet{Hiscock1985} that were
shown to be unstable around hydrostatic equilibrium states.

%------------------------------------------------------------------------
%------------------------------------------------------------------------
\section{General-relativistic dissipative hydrodynamics: 3+1 conservative
  formulation}
\label{sec:3+1}
%------------------------------------------------------------------------
%------------------------------------------------------------------------

We next derive a general-relativistic 3+1 flux-conservative (or simply
``conservative'') formulation of the dissipative-hydrodynamics
Eqs. (\ref{eq:constitutive_full1}) -- (\ref{eq:constitutive_full3}) so as
to provide a comprehensive and complete way of including causal
dissipative effects in general-relativistic simulations.

%------------------------------------------------------------------------
\subsection{3+1 decomposition of spacetime}
%------------------------------------------------------------------------

As customary in the so-called 3+1 decomposition of spacetime, we
decompose the four-dimensional manifold such that notions of time and
space reappear. This can be achieved by foliating spacetime in terms of a
set of non-intersecting spacelike hypersurfaces. On each hypersurface a
unit timelike four-vector field \(\boldsymbol{n}\) can be defined such
that \(\boldsymbol{n}\) is normal to its corresponding hypersurface. As
we have \(n_{\mu}n^{\mu}=-1\), its trajectory in spacetime can be
interpreted as the worldline of a ``normal'' or ``Eulerian'' observer. In
this way, it is possible to represent physical laws in terms of
projections either parallel or orthogonal to \(\boldsymbol{n}\) (see
\citet{Alcubierre:2008}, \citet{Gourgoulhon2012}, and
\citet{Rezzolla_book:2013} for details).

The generic line-element in a 3+1 decomposition can be written as 
\begin{equation}
 d s^2=-(\alpha^2-\beta_{i}\beta^{i}) d t^2+2\beta_{i} d x^{i} d
t+\gamma_{ij} d x^{i} d x^{j}\,,
\end{equation}
where \( \alpha \) is the so-called lapse function, the purely spatial
vector \( \boldsymbol{ \beta } \), \ie \( \beta ^{\mu} =
(0,\beta^{i})^{T} \), is the shift vector, and \( \gamma_{ij} \) denotes
the components of the purely spatial metric \( \boldsymbol{\gamma}\),
defined as
\begin{equation}
\gamma_{\mu \nu} := g_{\mu \nu}+ n_{\mu}n_{\nu}\label{eq:gamma}\,,
\end{equation}
which acts as the projection operator onto spatial hypersurfaces. Note
that the following identities then follow: $\beta_i := \gamma_{ij}
\beta^j$, $\sqrt{-g}=\alpha\sqrt{\gamma}$, where
\(g:=\mathrm{det}\left(g_{\mu \nu}\right)\) and
\(\gamma:=\mathrm{det}\left(\gamma_{ij}\right)\). The components of
\(\boldsymbol{n}\) are given by
\begin{equation}
n^{\mu}=\frac{1}{\alpha}(1,-\beta^{i})^T\,, \quad \quad \quad \quad n_{\mu}= (-\alpha, 0, 0, 0)\,.
\end{equation}
Having a timelike unit four-vector $\boldsymbol{n}$, we can use it to
decompose any tensor into a part that is parallel ($\parallel$) and
perpendicular ($\perp$) to $\boldsymbol{n}$. We start from the
four-velocity \(u^{\mu}\), which can then be written as
\begin{equation}
u^{\mu}=u_{\parallel}^{\phantom{\parallel}\mu} +
u_{\perp}^{\phantom{\perp}\mu}: =
(-n_{\nu}u^{\nu})n^{\mu}+{\gamma^{\mu}}_{\nu}u^{\nu} =
W(n^{\mu}+v^{\mu})\,,
\end{equation}
where \(W := (1- v_i v^i)^{-1/2}\) is the Lorentz factor and $v^\mu :=
(0, v^i)^T$ the fluid velocity measured by the normal
observer. Analogously, we can decompose the heat current \(q^{\mu}\) and
the shear-stress tensor \(\pi^{\mu \nu}\) as
\begin{align}
q^{\mu}&=q_{\parallel}^{\phantom{\parallel}\mu}+q_{\perp}^{\phantom{\perp}\mu}
: =(-n_{\nu}q^{\nu})n^{\mu}+{\gamma^{\mu}}_{\nu}q^{\nu}\,,
\label{eq:qsplit}\\
\pi^{\mu \nu}&=\pi_{\parallel}^{\phantom{\parallel}\mu
  \nu}+{\pi_{\times}}^{\mu \nu}+\pi_{\times}^{\phantom{\times}\nu \mu}
+\pi_{\perp}^{\phantom{\perp}\mu \nu} \notag
\\ &:=\left(n_{\alpha}n_{\beta}\pi^{\alpha \beta}\right)n^{\mu}n^{\nu} +
\left(-n_{\alpha}{\gamma^{\nu}}_{\beta}\pi^{\alpha \beta}\right)n^{\mu} +
\left(-n_{\beta}{\gamma^{\mu}}_{\alpha}\pi^{\alpha \beta}\right)n^{\nu} +
     {\gamma^{\mu}}_{\alpha}{\gamma^{\nu}}_{\beta}\pi^{\alpha
       \beta}\,. \label{eq:pisplit}
\end{align}
Additionally, since the time components of the parallel projections are
effectively scalar functions, we treat them as such after the following
definitions 
\begin{align}
\mathring{q}:&= -n_{\mu}q^{\mu}\, ,\\
\mathring{\pi}:&= n_{\mu}n_{\nu}\pi^{\mu \nu}\, ,\\
\mathring{\pi}^{\lambda}:&=-n_{\mu}{\gamma^{\lambda}}_{\nu}\pi^{\mu \nu}\,.
\end{align}
Note that the parallel and perpendicular components are not independent
and the following identities can be obtained after a bit of algebra and
will be used hereafter
\begin{align}
  \label{eq:restric1}
  \mathring{q}=v_{i} q^{\phantom{\perp}i}_{\perp}\,, \qquad \qquad 
  {\mathring{\pi}}^{i}=v_{j} \pi^{\phantom{\perp}ij}_{\perp}\,, \qquad \qquad 
\mathring{\pi} =v_{i}{\mathring{\pi}}^{i}=\pi_{\perp\phantom{i}
  i}^{\phantom{\perp}i}\,.
\end{align}
It is useful to remark that expressions \eqref{eq:restric1} imply that
the knowledge of \(q^{\phantom{\perp}i}_{\perp}\) and
\(\pi^{\phantom{\perp}ij}_{\perp}\) suffices to fully reconstruct
\(q^{\mu}\) and \(\pi^{\mu \nu}\) if the fluid three-velocity \(v^{i}\)
and the full metric \(g_{\mu \nu}\) are known. In particular, the
projections (\ref{eq:restric1}) can be employed in the decomposition of
the generic energy-momentum tensor (\ref{eq:energy_momentum})
\begin{equation}
T^{\mu \nu} = E n^{\mu}n^{\nu} + S^{\mu}n^{\nu} + S^{\nu}n^{\mu} + S^{\mu
  \nu}\label{eq:em_projection}\,,
\end{equation}
where the total energy density, the momentum density, and the purely spatial part
of the energy-momentum tensor are defined respectively as
\begin{align}
  E &:= n_{\mu}n_{\nu}T^{\mu \nu} = \left( e+p+\Pi \right) W^2 +
  2\mathring{q} W - \left( p + \Pi - \mathring{\pi}
  \right)\,,\label{eq:em_projection1}\\ S^{\mu} &:=
  -n_{\nu}{\gamma^{\mu}}_{\alpha}T^{\alpha \nu} = \left( e+p+\Pi
  \right)W^2 v^{\mu} + W \left( \mathring{q} v^{\mu} +
  q_{\perp}^{\phantom{\perp}\mu} \right) +
  \mathring{\pi}^{\mu}\,,\label{eq:em_projection2}\\ S^{\mu \nu} &:=
           {\gamma^{\mu}}_{\alpha}{\gamma^{\nu}}_{\beta}T^{\alpha \beta}
           = \left( e+p+\Pi \right)W^2 v^{\mu}v^{\nu} + W\left(
           q_{\perp}^{\phantom{\perp}\mu}v^{\nu} +
           q_{\perp}^{\phantom{\perp}\nu}v^{\mu} \right) + \left( p+\Pi
           \right)\gamma^{\mu \nu} + \pi_{\perp}^{\phantom{\perp}\mu
             \nu}\,.
  \label{eq:em_projection3}
\end{align}
Additionally, we will make use of the acceleration of the normal observer
and of the extrinsic curvature given respectively by
\begin{align}
\hat{a}^{\mu}& :=  n^{\nu}\nabla_{\nu}n^{\mu}=\gamma^{\mu \nu}\partial_{\nu}\ln \alpha \,, \\
K_{\mu\nu}& :=  -\frac{1}{2}\mathscr{L}_{\boldsymbol{n}}\gamma_{\mu \nu}
=-{\gamma_{\mu}}^{\lambda}\nabla_{\lambda}n_{\nu}=-\nabla_{\mu}n_{\nu}-n_{\mu}\hat{a}_{\nu}\,,
\label{eq:ext_curv}
\end{align}
where \(\mathscr{L}_{\boldsymbol{n}}\) denotes the Lie derivative along
\(\boldsymbol{n}\). Finally, we recall two important four-dimensional
tensor identities that will be useful later on to obtain a
flux-conservative formulation for the conservation of rest-mass
(\ref{eq:continuity}) and energy-momentum (\ref{eq:em_conservation}),
namely,
\begin{align}
  \nabla_{\mu}J^{\mu} &= \frac{\partial_{\mu}\left(\sqrt{-g}J^{\mu}\right)}{\sqrt{-g}}\,,
  \label{eq:vector_divergence}\\
  \nabla_{\mu}T^{\mu \nu} &= g^{\nu \lambda} \bigg[ \frac{\partial_{\mu}
      \left(\sqrt{-g}{T^{\mu}}_{\lambda}\right)}{\sqrt{-g}} - \frac{1}{2}T^{\alpha \beta}\partial_{\lambda}
    g_{\alpha \beta}\bigg]\,.
  \label{eq:tensor_divergence}
\end{align}

%------------------------------------------------------------------------
\subsection{3+1 flux-conservative formulation of HL83: general expressions}
%------------------------------------------------------------------------

We next rewrite the full set of general-relativistic dissipative
hydrodynamics, \ie Eqs. (\ref{eq:continuity}),
(\ref{eq:em_conservation}), and (\ref{eq:constitutive_full1}) --
(\ref{eq:constitutive_full3}), and which essentially represent the
equations of the \textbf{HL83} formulation, in a 3+1 flux-conservative
form. We start by recalling that a system of partial differential
equations is said to be flux-conservative if it can be written as
\begin{equation}
\partial_{t}\boldsymbol{U}+
\partial_{i}\boldsymbol{F}^{i}(\boldsymbol{U})=\boldsymbol{S}\,,
\label{eq:pde_conservative}
\end{equation}
where \(\boldsymbol{U}\) is the \textit{``state vector''},
\(\boldsymbol{F^{i}}\) are the \textit{``flux vectors''} and
\(\boldsymbol{S}\) is the \textit{``source vector''}. Employing now
Eq. (\ref{eq:vector_divergence}), it is possible to rewrite
Eq. (\ref{eq:continuity}) as
\begin{align}
  \partial_t \left(\sqrt{\gamma}D \right) + \partial_{i} \left[
    \sqrt{\gamma}D \left( \alpha v^{i} - \beta^{i} \right) \right] =
  0\,,
  \label{eq:continuity3+1}
\end{align}
where \(D:=\rho \alpha u^{t}=\rho W\) is the conserved
rest-mass. Similarly, we use Eqs. (\ref{eq:em_projection}) and
(\ref{eq:tensor_divergence}) to obtain a flux-conservative form of the
equations for the conservation of energy and momentum
(\ref{eq:em_conservation})
\begin{align}
  &\partial_t \left( \sqrt{\gamma} S_{j}\right) + \partial_{i}
  \left[\sqrt{\gamma}\left(\alpha {S^{i}}_{j} -
    \beta^{i}S_{j}\right)\right] = \sqrt{\gamma}\left( \frac{1}{2}\alpha
  S^{ik}\partial_{j}\gamma_{ik}
  +S_{i}\partial_{j}\beta^{i}-E\partial_{j}\alpha
  \right)\,,\label{eq:momentum3+1}\\
  &\partial_t \left[ \sqrt{\gamma}
    (E-D) \right] + \partial_{i} \left\lbrace \sqrt{\gamma} \left[ \alpha
    \left( S^{i}-v^{i}D \right) - \beta^{i} \left(E-D \right) \right]
  \right\rbrace = \sqrt{\gamma} \left( \alpha S^{ij} K_{ij} -
  S^{j}\partial_{j}\alpha\right)\,.\label{eq:energy3+1}
\end{align}

Note that we use $E-D$ rather than simply $E$ in Eq. (\ref{eq:energy3+1})
because the numerical conservation of $E-D$ is more accurate than that of
$E$ only. We will report the explicit expressions of \(\boldsymbol{U}\),
\(\boldsymbol{F^{i}}\), and \(\boldsymbol{S}\) in the next section, after
we have reformulated the constitutive equations
(\ref{eq:constitutive_full1})--(\ref{eq:constitutive_full3}) in a
conservative form. In order to do this, we consider the evolution
equation for the heat current (\ref{eq:constitutive_full2}) as an example
and extend the treatment to the other equations afterwards. We start with
the term
\begin{equation}
\dot{q}^{\left\langle \mu \right\rangle} := {h^{\mu}}_{\nu}u^{\lambda}\nabla_{\lambda}q^{\nu}
=u^{\lambda}\nabla_{\lambda}q^{\mu}-a_{\nu}q^{\nu}u^{\mu}\,,\label{eq:unprojecq}
\end{equation}
where we used the fact that $\boldsymbol{q}\cdot\boldsymbol{u}=0$ and
have introduced the kinematic acceleration $\boldsymbol{a}$ -- not to be
confused with the acceleration of normal observers $\boldsymbol{\hat{a}}$
-- with components $a^\mu := u^\lambda \nabla_\lambda u^\mu$. Our goal is
to obtain evolution equations in a conservative form for state variables
that are orthogonal to \(\boldsymbol{n}\). Thus, we project
Eq. (\ref{eq:unprojecq}) by multiplying it with \({\gamma^{i}}_{\mu}\)
\begin{align}
{\gamma^{i}}_{\mu}\dot{q}^{\left\langle \mu
  \right\rangle}&=u^{\lambda}\nabla_{\lambda}q^{\phantom{\perp}i}_{\perp}
-q^{\mu}u^{\nu}\nabla_{\nu}{\gamma^{i}}_{\mu}-a_{\nu}q^{\nu}{\gamma^{i}}_{\mu}u^{\mu}
\nonumber \\
& = u^{\lambda}\partial_{\lambda}q^{\phantom{\perp}i}_{\perp} -
\mathcal{G}^{\phantom{\mathrm{q}}i}_{\mathrm{q}} -
\mathcal{H}_{\mathrm{q}}^{\phantom{\mathrm{q}}i}\,,
\label{eq:projecq2}
\end{align}
where, upon using Eq. (\ref{eq:ext_curv}) and the definition of the
covariant derivative, we have introduced the new quantities
\begin{align}
  \label{eq:Gi_q}
  \mathcal{G}^{\phantom{\mathrm{q}}i}_{\mathrm{q}}&:=
  \frac{W}{\alpha}\left(K_{kj}v^{k}-\hat{a}_{j}\right)q^{\phantom{\perp}j}_{\perp}\beta^{i}
  + \mathring{q} W v_j {K^{ij}}- \mathring{q} W
  \hat{a}^{i}-\frac{W}{\alpha}{\Gamma^{i}}_{0
    j}q_{\perp}^{\phantom{\perp}j} -\frac{W}{\alpha}{\Gamma^{i}}_{j
    k}q_{\perp}^{\phantom{\perp}j}V^{k}\,,\\
  \label{eq:Hi_q}
  \mathcal{H}_{\mathrm{q}}^{\phantom{\mathrm{q}}i}&:=
  \left({a_{\perp}}_{j}q^{\phantom{\perp}j}_{\perp}-\mathring{a} \mathring{q} \right)
  Wv^{i}\,.
\end{align}
The right-hand sides of Eqs. \eqref{eq:Gi_q} and \eqref{eq:Hi_q} also
contain other new quantities, namely, the \textit{``coordinate
  velocity''} \(V^{j}:=u^{j}/u^{t}=\alpha v^{j}-\beta^{j}\) and the 3+1
split of the kinematic acceleration \(a^{\mu} :=
a_{\parallel}^{\phantom{\parallel}\mu} + a^{\phantom{\perp}\mu}_{\perp} =
\mathring{a} n^{\mu} + a_{\perp}^{\phantom{\perp}\mu}\), whose explicit
components are given by
\begin{align}
a^{\phantom{\perp}i}_{\perp}&=A^{i}+W\Lambda^{i}-2W^2v_{j}{K^{ij}}\,,\label{eq:acc_space}\\ \mathring{a}
&= v_{i} a_{\perp}^{\phantom{\perp}i} = v_{i}A^{i} +
Wv_{i}\Lambda^{i}-2W^2v_{i}v_{j}K^{ij}\,,\label{eq:acc_time}
\end{align}
with
\begin{align}
  A^{i}&:=Wv^{j}D_{j}\left(Wv^{i}\right)\,,\\
  \Lambda^{i}&:=
\frac{1}{\alpha}\left(\partial_{t}-\mathscr{L}_{\boldsymbol{\beta}}\right)Wv^{i}+W\hat{a}^{i}\,,
\end{align}
and $D_j$ being the fully spatial part of the covariant derivative
$\nabla$, \ie $D_{j}v^{i} := \partial_{j}v^{i} +
{}^{3}{\Gamma^{i}}_{jk}v^{k}$, where \({}^{3}{\Gamma^{i}}_{jk}\) represent
the Christoffel symbols related to the three-metric
\(\boldsymbol{\gamma}\) \citep[see][for details]{Gourgoulhon2012,
  Rezzolla_book:2013}
\begin{equation}
{}^{3}{\Gamma^{i}}_{j
  k}:=\tfrac{1}{2}{\gamma}^{i l}\left(\partial_{j}{\gamma}_{l k}
+\partial_{j}{\gamma}_{l k}- \partial_{l}{\gamma}_{j k}\right)\,.
\end{equation}

Note that the vector $\mathcal{G}^{\phantom{\mathrm{q}}i}_{\mathrm{q}}$
captures the influence of the choice of spacetime foliation as well as
the curvature of spacetime itself on the transport of heat on each
hypersurface. On the other hand, the tensor
$\mathcal{H}_{\mathrm{q}}^{\phantom{\mathrm{q}}i}$ is a correction term
that arises from the projection operator ${h^{\mu}}_{\nu}$, which ensures
that heat transport occurs always orthogonally to
$\boldsymbol{u}$. Furthermore,
$\mathcal{G}^{\phantom{\mathrm{q}}i}_{\mathrm{q}}$ and
$\mathcal{H}_{\mathrm{q}}^{\phantom{\mathrm{q}}i}$ as expressed by
Eqs. \eqref{eq:Gi_q}--\eqref{eq:Hi_q} are not fully decomposed due to the
presence of $\mathring{a}$, $\mathring{q}$ and of the four-dimensional
Christoffel symbols; their full 3+1 decomposition will be given in
Section \ref{sec:3+1sources} as both terms belong to the source terms.

We now exploit the continuity equation in the form
(\ref{eq:vector_divergence}) to modify the first term on the right-hand
side of Eq. (\ref{eq:projecq2})
\begin{align}
\alpha \sqrt{\gamma}\rho u^{\lambda}\partial_{\lambda}q_{\perp}^{\phantom{\perp}i}&=\partial_{\lambda}
\left(\alpha \sqrt{\gamma}\rho q^{\phantom{\perp}i}_{\perp}u^{\lambda}\right)= \partial_{t} 
\left( \sqrt{\gamma} D q^{\phantom{\perp}i}_{\perp}\right) + \partial_{j}\left( \sqrt{\gamma} D V^{j} q^{\phantom{\perp}i}_{\perp}\right)\,.
\end{align}
Using this equation, as well as Eq. (\ref{eq:projecq2}), and projecting
Eq. (\ref{eq:constitutive_full2}) with ${\gamma^i}_\mu$, we obtain
\begin{align}
\partial_t\left( \sqrt{\gamma} D
  q_{\perp}^{\phantom{\perp}i}\right)+\partial_{j}\left(\sqrt{\gamma} D V^{j}
  q_{\perp}^{\phantom{\perp}i} \right)&= \frac{\alpha \sqrt{\gamma} D}{\tau_{\mathrm{q}}W}
  \biggl\{ {\gamma^{i}}_{\mu}q^{\phantom{\ns}\mu}_{\ns}-
  q^{\phantom{\perp}i}_{\perp} -\frac{1}{2}\kappa T^2 q^{\phantom{\perp}i}_{\perp}
  \nabla_{\nu}\left(\frac{\tau_{\mathrm{q}} u^{\nu}}{\kappa T^2}\right) \notag\\
  &\phantom{=}+\kappa T \bigg[ \alpha_0 {\gamma^{i}}_{\mu}\nabla^{\langle \mu \rangle}\Pi 
  + \alpha_1 {\gamma^{i}}_{\mu}\nabla_{\nu}\pi^{\nu \langle \mu \rangle} \notag \\
  &\phantom{=}+(1-\gamma_0)\Pi T
  {\gamma^{i}}_{\mu}\nabla^{\langle \mu \rangle}\left(\frac{\alpha_0}{T}\right)
  + (1-\gamma_1)T
  {\gamma^{i}}_{\mu}\pi^{\mu \nu}\nabla_{\nu}\left(\frac{\alpha_1}{T}\right)\bigg] \notag \\
  &\phantom{=}+\tau_{\mathrm{q}} 
 \left[ \mathcal{G}^{\phantom{\mathrm{q}}i}_{\mathrm{q}} + \mathcal{H}_{\mathrm{q}}^{\phantom{\mathrm{q}}i} \right]  \biggr\}\,.\label{eq:pre_3+1_heat}
\end{align}

Proceeding in a similar way, we find that the projected version of
Eq. (\ref{eq:constitutive_full3}) is given by
\begin{align}
\partial_t\left(\sqrt{\gamma}D\pi_{\perp}^{\phantom{\perp}ij}\right)+\partial_{k}\left(\sqrt{\gamma}DV^{i}\pi_{\perp}^{\phantom{\perp}ij}\right)
&= \frac{\alpha \sqrt{\gamma}
    D}{\tau_{\pi} W}\biggl\{ {\gamma^{i}}_{\mu}{\gamma^{j}}_{\nu}\pi^{\phantom{\ns}\mu \nu}_{\ns}-\pi_{\perp}^{\phantom{\perp}ij}-\frac{1}{2}\eta
  T \pi_{\perp}^{\phantom{\perp}ij}\nabla_{\lambda}\left(\frac{\tau_{\pi}u^{\lambda}}{\eta T}\right) \notag \\
  &\phantom{=}+ 2 \alpha_1 \eta {\gamma^{i}}_{\mu}{\gamma^{j}}_{\nu}\nabla^{\langle \mu}q^{\nu \rangle} 
  + 2\gamma_1 \eta T {\gamma^{j}}_{\nu}{\gamma^{i}}_{\mu}q^{\langle \mu}\nabla^{\nu \rangle}
  \left(\frac{\alpha_1}{T}\right)  \notag \\
  &\phantom{=}+\tau_{\pi}\left[ \mathcal{G}^{\phantom{\pi}ij}_{\pi} + \mathcal{H}_{\pi}^{\phantom{\pi}ij}\right]
  \biggr\}\,,\label{eq:pre_3+1_shear}
\end{align}
with
\begin{align}
\mathcal{G}^{\phantom{\pi}ij}_{\pi}&:=
\frac{W}{\alpha}\left(K_{lk}v^{l}-\hat{a}_{k}\right)\left(\pi^{\phantom{\perp}ki}_{\perp}\beta^{j}+\pi^{\phantom{\perp}kj}_{\perp}\beta^{i}\right)
+Wv_{l}\left({K^{li}}\mathring{\pi}^{j}+{K^{lj}}\mathring{\pi}^{i}\right)
-W\left(\hat{a}^{i}\mathring{\pi}^{j}+\hat{a}^{j}\mathring{\pi}^{i}\right) \notag \\
&\phantom{:=}-\frac{W}{\alpha}\left({\Gamma^{i}}_{0k}\pi^{\phantom{\perp}kj}_{\perp}+{\Gamma^{j}}_{0k}\pi^{\phantom{\perp}ki}_{\perp}\right)
-\frac{W}{\alpha}\left({\Gamma^{i}}_{kl}\pi^{\phantom{\perp}kj}_{\perp}V^{l}+{\Gamma^{j}}_{kl}\pi^{\phantom{\perp}ki}_{\perp}V^{l}\right)\,,\\
\mathcal{H}_{\pi}^{\phantom{\pi}ij} &:= W{a_{\perp}}_{k}\left(\pi^{\phantom{\perp}k i}_{\perp}v^{j}+\pi^{\phantom{\perp}k j}_{\perp}v^{i}\right)
-W\mathring{a} \left(\mathring{\pi}^{i}v^{j}+\mathring{\pi}^{j}v^{i}\right)\,.
\end{align}

Finally, since the bulk-viscosity pressure \(\Pi\) is a scalar quantity,
there are no projections that have to be performed in order to write a
3+1 split version of Eq. (\ref{eq:constitutive_full1}), which reads
\begin{align}
\partial_t\left(\sqrt{\gamma}D\Pi\right)+\partial_{i}\left(\sqrt{\gamma}DV^{i}\Pi\right)&= \frac{\alpha
    \sqrt{\gamma} D}{ \tau_{_{\Pi}} W}\bigg[ \Pi_{\ns} -\Pi -
  \frac{1}{2}\zeta \Pi T
  \nabla_{\mu}\left(\frac{\tau_{_{\Pi}}u^{\mu}}{\zeta T}\right) \notag \\
  &\phantom{=}+ \alpha_0 \zeta
  \nabla_{\mu} q^{\mu} + \gamma_0 \zeta T q^{\mu}\nabla_{\mu}\left(
  \frac{\alpha_0}{T}\right) \bigg]\,.\label{eq:conservative_bulk}
\end{align}

In summary, Eqs. (\ref{eq:continuity3+1})--(\ref{eq:energy3+1}),
(\ref{eq:pre_3+1_heat}), (\ref{eq:pre_3+1_shear}) and
(\ref{eq:conservative_bulk}) can be combined into the flux-conservative
form (\ref{eq:pde_conservative}), with the following expressions for the
quantities \(\boldsymbol{U}\), \(\boldsymbol{F^{i}}\) and
\(\boldsymbol{S}\) 
\begin{align}
	&\boldsymbol{U} = \sqrt{\gamma}\begin{pmatrix}
	D \\[1em]
	S^{j} \\[1em]
	E-D \\[1em]
	D\Pi \\[1em]
	D q^{\phantom{\perp}j}_{\perp} \\[1em]
	D \pi^{\phantom{\perp}jk}_{\perp}
	\end{pmatrix} =\sqrt{\gamma} \begin{pmatrix}
	\rho W \\[1em]
	\left( e+p+\Pi \right) W^2 v^{j} + W \left(\mathring{q} v^{j} + q^{\phantom{\perp} j}_{\perp}\right) + \mathring{\pi}^{j} \\[1em]
	\left( e+p+\Pi \right) W^2 + 2 \mathring{q} W  - \left( p + \Pi - \mathring{\pi} \right)-\rho W \\[1em]
	\rho W \Pi \\[1em]
	\rho W q^{\phantom{\perp}j}_{\perp} \\[1em]
	\rho W \pi^{\phantom{\perp}jk}_{\perp}
        \label{eq:full_U}
	\end{pmatrix}\, ,
\end{align}
\begin{align}
	\boldsymbol{F}^{i} =\sqrt{\gamma} \begin{pmatrix}
     D V^{i}\\[1em]
	\alpha {S^{i}}_{j} - \beta^{i}S_{j} \\[1em]
	\alpha \left(S^{i}-v^{i}D\right) - \beta^{i} \left(E-D\right) \\[1em]
	D V^{i}\Pi \\[1em]
	D V^{i} q^{\phantom{\perp}j}_{\perp} \\[1em]
	D V^{i} \pi^{\phantom{\perp}jk}_{\perp}
        \label{eq:full_F}
	\end{pmatrix}\,,
\end{align}
\begin{align}
    \boldsymbol{S} = \sqrt{\gamma} \begin{pmatrix} 0 \\[1em]
      \frac{1}{2}\alpha S^{ik}\partial_{j} \gamma_{ik} +
      S_{i}\partial_{j}\beta^{i} - E \partial_{j}\alpha \\[1em] \alpha
      S^{ij}K_{ij} - S^{j}\partial_{j}\alpha \\[1em] (\alpha
      D/\tau_{_{\Pi}}W)\left(\Pi_{\ns}-\Pi + \Delta_{_\Pi}\right)\\[1em]
      (\alpha D/\tau_{\mathrm{q}}W)\left(
      {\gamma^{j}}_{\mu}q^{\phantom{\ns}\mu}_{\ns}-q^{\phantom{\perp}j}_{\perp}
      + \Delta_{\mathrm{q}}^{\phantom{\mathrm{q}}j}+\tau_{\mathrm{q}}
      \mathcal{G}^{\phantom{\mathrm{q}}j}_{\mathrm{q}} +
      \tau_{\mathrm{q}} \mathcal{H}_{\mathrm{q}}^{\phantom{\mathrm{q}}j}
      \right)\\[1em] (\alpha D/\tau_{\pi}W)\left(
            {\gamma^{j}}_{\mu}{\gamma^k}_{\nu}\pi^{\phantom{\ns}\mu
              \nu}_{\ns}-\pi_{\perp}^{\phantom{\perp}jk} +
            \Delta_{\pi}^{\phantom{\pi}jk} +\tau_{\pi}
            \mathcal{G}^{\phantom{\pi}jk}_{\pi} + \tau_{\pi}
            \mathcal{H}_{\pi}^{\phantom{\pi}jk} \right)
        \label{eq:full_S}
	\end{pmatrix}\,,
\end{align}
where we have introduced the following new quantities:
\begin{align}
\Delta_{_\Pi}&:= - \frac{1}{2}\zeta \Pi T
\nabla_{\mu}\left(\frac{\tau_{_{\Pi}}u^{\mu}}{\zeta T}\right) + \alpha_0
\zeta \nabla_{\mu} q^{\mu} + \gamma_0 \zeta T q^{\mu}\nabla_{\mu}\left(
\frac{\alpha_0}{T}\right)\,,\label{eq:sources1}\\ \Delta^{\phantom{\mathrm{q}}j}_{\mathrm{q}}&:=
-\frac{1}{2}\kappa T^2 q^{\phantom{\perp}j}_{\perp}
\nabla_{\nu}\left(\frac{\tau_{\mathrm{q}} u^{\nu}}{\kappa T^2}\right)
+\kappa T \bigg[ \alpha_0 {\gamma^{j}}_{\mu}\nabla^{\langle \mu
    \rangle}\Pi + \alpha_1 {\gamma^{j}}_{\mu}\nabla_{\nu}
  {\gamma^{j}}_{\mu}\nabla^{\langle \mu
    \rangle}\left(\frac{\alpha_0}{T}\right) \, \ \notag \\ &\phantom{:=}+
  (1-\gamma_1) T {\gamma^{j}}_{\mu}\pi^{\mu
    \nu}\nabla_{\nu}\left(\frac{\alpha_1}{T}\right)\bigg]\,,\label{eq:sources2}
\\ \Delta^{\phantom{\pi}jk}_{\pi}&:= -\frac{1}{2}\eta T
\pi_{\perp}^{\phantom{\perp}jk}\nabla_{\lambda}\left(\frac{\tau_{\pi}u^{\lambda}}{\eta
  T}\right)+ 2 \alpha_1 \eta
   {\gamma^{j}}_{\mu}{\gamma^{k}}_{\nu}\nabla^{\langle \mu}q^{\nu
     \rangle} + 2 \gamma_1 \eta T
   {\gamma^{k}}_{\mu}{\gamma^{j}}_{\nu}q^{\langle \mu}\nabla^{\nu
     \rangle}\left(\frac{\alpha_1}{T}\right)\,.
  \label{eq:sources3}
\end{align}
The evolution Eqs. (\ref{eq:pde_conservative}) for the fifteen
components of the state vector \eqref{eq:full_U}, with fluxes given by
the vectors \eqref{eq:full_F}, and source terms \eqref{eq:full_S}
represent the 3+1 flux-conservative formulation of the \textbf{HL83}
system.

A few considerations are worth making at this point. First, although the
whole solution of the set of partial differential equations becomes
computationally more expensive, the conversion from the conserved to the
primitive variables does not gain complexity, at least not beyond what is
already encountered in GRMHD. We recall that such a conversion, which
needs to be performed numerically at each grid cell and on each timelevel
of the solution, requires the solution of a set of nonlinear equations to
obtain the values of the primitive variables from the newly computed
conserved ones. Fortunately, the extension of the set of equations in
GRDHD does not require new root-finding steps to obtain the ten new
primitive variables, \ie $\Pi, q^{\phantom{\perp}i}_{\perp},
\pi^{\phantom{\perp}ij}_{\perp}$, as these are related algebraically with
the corresponding components of the state vector
$\boldsymbol{U}$. Second, the way in which the new conversion from the
conserved to the primitive variables differs from the perfect-fluid case
is in the appearance of the three-velocities $v^{i}$ in the definitions
of the conserved variables, \ie Eqs. (\ref{eq:em_projection1}) --
(\ref{eq:em_projection2}), as well as in the contractions with the
projected dissipative currents $q_{\perp}^{\phantom{\perp}i}$ and
$\pi_{\perp}^{\phantom{\perp}ij}$. While the latter are related
algebraically with the corresponding conserved variables, the
three-velocity $v^{i}$ still requires a numerical root-finding and hence
a continuous update in the root-finding process. Ultimately, this leads
to a numerical matrix inversion or additional root-finding steps in the
conversion when compared to the perfect-fluid case. Third, a different
choice of second-order theory will only change the explicit expressions
for $\Delta_{_\Pi}$, $\Delta^{\phantom{\mathrm{q}}j}_{\mathrm{q}}$, and
$\Delta^{\phantom{\mathrm{pi}}jk}_{\pi}$, since they contain
off-equilibrium contributions that are not captured by simple
relaxation-type equations and typically contain contributions that are of
second and even higher order. A few examples of other second-order GRDHD
equations are presented in Appendix \ref{sec:review}, where we compare
\textbf{HL83} to other dissipative-hydrodynamics formulations. Finally,
we remark that the sources for the dissipative quantities in
Eq. (\ref{eq:full_S}) are not yet fully 3+1 decomposed and their explicit
expressions will be presented in the following section.

%------------------------------------------------------------------------
\subsection{3+1 flux-conservative formulation of HL83: source terms}
\label{sec:3+1sources}
%------------------------------------------------------------------------

As mentioned above, in order to provide complete expressions for the 3+1
flux-conservative formulation of Eqs. (\ref{eq:pde_conservative}) we need
to obtain right-hand sides that only contain fully spatial quantities,
both for the ordinary variables -- \ie \(\rho\), \(p\), \(v^{i}\),
\(\alpha\), \(\beta^{i}\), \(\gamma_{ij}\), \(K_{ij}\), as well as their
spatial (partial) derivatives -- and for the dissipative ones, as well as
their temporal and spatial (partial and covariant) derivatives -- \ie
\(\Pi\), \(q^{\phantom{\perp}i}_{\perp}\) and
\(\pi^{\phantom{\perp}ij}_{\perp}\).

To accomplish this second goal, we need to project the corresponding
evolution equations -- and corresponding right-hand sides -- onto spatial
hypersurfaces via the metric \(\boldsymbol{\gamma}\). However, since
many terms in the source terms involve covariant derivatives [\cf
  Eqs. \eqref{eq:sources1}--\eqref{eq:sources3}], the resulting 3+1
decomposed expressions are inevitably very complicated and lengthy, so
that it is difficult to reconstruct the origin of the various source
terms. To avoid this, and hence make the calculations more transparent
and easy to follow, we collect the relevant source terms in the following
classes:
\begin{itemize}
\item[1.] \textit{Intrinsically spatial terms.} These are terms
  containing only quantities which are originally defined on a spatial
  hypersurface, \ie \(\rho\), \(p\), \(v^{i}\), \(\Pi\),
  \(q^{\phantom{\perp}i}_{\perp}\) and
  \(\pi^{\phantom{\perp}ij}_{\perp}\), as well as spatial partial or
  covariant derivatives of these quantities. Also part of this class are
  terms proportional to \(\gamma_{ij}\) and its spatial partial
  derivatives. We have marked these terms in
  \textcolor{PineGreen}{green}.
\item[2.] \textit{Terms not containing the extrinsic curvature
  \(K_{ij}\)}. These are terms involving temporal derivatives, spatial
  derivatives involving \(\alpha\) or \(\beta^{i}\), as well as terms
  containing projections parallel to the unit normal \(\boldsymbol{n}\),
  \eg \(\mathring{q}\). We have marked these terms in
  \textcolor{RedOrange}{orange}.
\item[3.] \textit{Terms containing the extrinsic curvature
  \(K_{ij}\)}. These are terms linear in the extrinsic curvature, its
  trace, $K := \gamma_{ij}K^{ij}$, but not containing spatial derivatives
  of $K_{ij}$. We have marked these terms in
  \textcolor{ProcessBlue}{light blue}. 
\end{itemize} 

However, before proceeding to this classification, we recall a number of
useful identities that will be exploited in the derivation of the source
terms and that are associated to gradients of the fluid four-velocity; in
doing so we are following in part the convention introduced by
\citet{Peitz1997,Peitz1999}. We first recall that the covariant
derivative of the four-velocity can be decomposed in terms of tensors
that describe its properties in terms of changes in volumes, shape and
vorticity, \ie as \citep{Rezzolla_book:2013}
\begin{align}
\nabla_{\mu }u_{\nu}=\omega_{\mu \nu} + \sigma_{\mu \nu} + \frac{1}{3}
\Theta h_{\mu \nu} - u_{\mu}a_{\nu}\,,
\label{eq:decompo}
\end{align}
where $\Theta$ is the \textit{``expansion''} and is defined as
\begin{align}
\Theta:&=\nabla_{\mu}u^{\mu} =\spatial{\vartheta} + \rest{\Lambda} -
\extrinsic{KW}\,,\label{eq:3+1expansion}
\end{align}
and where we have introduced
\begin{align}
\spatial{\vartheta} &:=
\spatial{D_{i}\left(Wv^{i}\right)}=\spatial{\frac{\partial_{i}
    \left(\sqrt{\gamma}Wv^{i}\right)}{\sqrt{\gamma}}}\,,\\ \rest{\Lambda}
&:= \rest{\frac{1}{\alpha}
  \left(\partial_{t}-\mathscr{L}_{\boldsymbol{\beta}}\right)
  W+Wv_{i}\hat{a}^{i}}\,.
\label{eq:Lambda}
\end{align}
Similarly, the \textit{``shear tensor''} can be written as
\begin{align}
\sigma^{\mu \nu}:&= \nabla^{\langle \mu}u^{\nu \rangle}
=\mathring{\sigma} n^{\mu}n^{\nu} + \mathring{\sigma}^{\mu}n^{\nu} + \mathring{\sigma}^{\nu}n^{\mu}
+\sigma_{\perp}^{\phantom{\perp}\mu \nu}\,,
\end{align}
where
\begin{align}
\sigma_{\perp}^{\phantom{\perp}ij}&={\gamma^{i}}_{\mu}{\gamma^{j}}_{\nu}\nabla^{\langle \mu } u^{\nu \rangle}
= \spatial{\Sigma^{ij}}+\rest{\Lambda^{ij}}-\extrinsic{W\mathcal{K}^{ij}}\,,
\end{align}
with
\begin{align}
\spatial{\Sigma^{ij}}&:=\spatial{\frac{1}{2}\left[D^{i}\left(Wv^{j}\right)+D^{j}(Wv^{i})\right]+\frac{1}{2}W\left(A^{i}v^{j}
+A^{j}v^{i}\right)-\frac{1}{3}\left(\gamma^{ij}+W^2v^{i}v^{j}\right)\vartheta}\,,\\
\rest{\Lambda^{ij}}&:=\rest{\frac{1}{2}W^2\left(\Lambda^{i}v^{j}+\Lambda^{j}v^{i}\right)
-\frac{1}{3}\Lambda\left(\gamma^{ij}+W^2v^{i}v^{j}\right)}\,,\\
\extrinsic{\mathcal{K}^{ij}}&:= \extrinsic{K^{ij}+W^2v_{k}\left({K^{ik}}v^{j}+{K^{jk}}v^{i}\right)-\frac{1}{3}K\left(\gamma^{ij}
+W^2v^{i}v^{j}\right)}\,, \\
\mathring{\sigma}^{i}&=v_{j}\sigma_{\perp}^{\phantom{\perp}ij}\,,\\
\mathring{\sigma}&=\gamma_{ij}\sigma_{\perp}^{\phantom{\perp}ij}={\sigma_{\perp i}}^{i}\,.
\end{align}
Finally, the \textit{``kinematic vorticity''} has the form:
\begin{align}
\omega^{\mu \nu}:&=\nabla^{[\mu}u^{\nu ]}+u^{[ \mu}a^{\nu ]}
= \mathring{\omega}^{\mu}n^{\nu}-\mathring{\omega}^{\nu}n^{\mu}+\omega^{\phantom{\perp}\mu \nu}_{\perp}\,,
\end{align}
where
\begin{align}
\omega_{\perp}^{\phantom{\perp}ij}&=\spatial{\Omega^{ij}}-\rest{\frac{1}{2}W^2\left(\Lambda^{i}v^{j}-\Lambda^{j}v^{i}\right)}
+ \extrinsic{ W^3v_k\left({K^{ik}}v^{j}-{K^{jk}}v^{i} \right)}\,,
\end{align}
with
\begin{align}
\spatial{\Omega^{ij}}&:=\spatial{D^{[i}Wv^{j]}-WA^{[i}v^{j]}}\,, \qquad \textrm{and} \qquad
\mathring{\omega}^{i}:=-n_{\nu}{\gamma^{i}}_{\lambda}\omega^{\lambda \nu}=v_{j}\omega_{\perp}^{\phantom{\perp}i j}\,.
\end{align}

\subsubsection{Sources for the  evolution of the bulk-viscosity pressure}

Listed below are all the spatial source terms [$6$-th component of the
  vector $\boldsymbol{S}$ in Eq. \eqref{eq:full_S}] appearing in the
evolution equation for the bulk-viscosity pressure, \ie $\partial_t
\left( \sqrt{\gamma} D\Pi \right)=\ldots$
\begin{align}
\Pi_{\ns}&= -\zeta \left( \spatial{\vartheta} + \rest{\Lambda} -
\extrinsic{KW} \right)\,,\\ \frac{1}{2}\zeta \Pi T
\nabla_{\mu}\left(\frac{\tau_{_{\Pi}}u^{\mu}}{\zeta
  T}\right)&=\spatial{\frac{1}{2}\tau_{_{\Pi}}\Pi \vartheta
  +\frac{1}{2}\zeta \Pi T
  Wv^{i}\partial_{i}\left(\frac{\tau_{_{\Pi}}}{\zeta T}\right)}
\notag \\ &\phantom{=}+\rest{\frac{1}{2}\tau_{_{\Pi}}\Pi \Lambda +
  \frac{\zeta \Pi T
    W}{2\alpha}\left(\partial_{t}-\mathscr{L}_{\boldsymbol{\beta}}\right)
  \left(\frac{\tau_{_{\Pi}}}{\zeta T}\right)}-\extrinsic{
  \frac{1}{2}\tau_{_{\Pi}}\Pi WK}\,,
\\ \nabla_{\mu}q^{\mu}&=\spatial{D_{i}q_{\perp}^{\phantom{\perp}i}}+\rest{\frac{1}{\alpha}\left(\partial_{t}
  -\mathscr{L}_{\boldsymbol{\beta}}\right)\mathring{q}+\hat{a}_{i}q_{\perp}^{\phantom{\perp}i}}-\extrinsic{\mathring{q}
  K}\,,\\ q^{\mu}\nabla_{\mu}\left(\frac{\alpha_0}{T}\right)&=\spatial{q^{\phantom{\perp}i}_{\perp}\partial_{i}\left(\frac{\alpha_0}{T}\right)}
+\rest{\frac{\mathring{q}}{\alpha}\aparn\left(\frac{\alpha_0}{T}\right)}\,.
\end{align}

\subsubsection{Sources for the evolution of the heat current}

Similarly, listed below are all the spatial source terms [components
  $7\textrm{-}9$ in Eq. \eqref{eq:full_S}] appearing in the evolution
equation for the components of the perpendicular heat current, \ie
$\partial_t \left( \sqrt{\gamma} D q^{\phantom{\perp}i}_{\perp}
\right)=\ldots$

\begin{align}
{\gamma^{i}}_{\mu}q^{\phantom{\ns}\mu}_{\ns}&= -\kappa T \bigg[
  \spatial{\left( \gamma^{ij} +W^2v^{i}v^{j}\right)
    \partial_{j}\ln\left(T\right) + A^{i}} \notag \\ &\phantom{=}+
  \rest{\frac{W^2}{\alpha}v^{i}\aparn \ln\left(T\right) + W \Lambda^{i}}
  - \extrinsic{2W^2{K^{i}}_{j}v^{j}}\bigg]\, ,\\ \frac{1}{2}\kappa
T^2q^{\phantom{\perp}i}_{\perp}\nabla_{\nu}\left(\frac{\tau_{\mathrm{q}}
  u^{\nu}}{\kappa T^2}\right)&= \spatial{\frac{1}{2}\tau_{\mathrm{q}}
  q^{\phantom{\perp}i}_{\perp}\vartheta + \frac{1}{2}\kappa T^2
  q^{\phantom{\perp}i}_{\perp}Wv^{j}\partial_{j}\left(\frac{\tau_{\mathrm{q}}}{\kappa
    T^2}\right)} \notag \\ &\phantom{=}+
\rest{\frac{1}{2}\tau_{\mathrm{q}} q^{\phantom{\perp}i}_{\perp}\Lambda +
  \frac{\kappa T^2 q^{\phantom{\perp}i}_{\perp}W}{2\alpha}\aparn
  \left(\frac{\tau_{\mathrm{q}}}{\kappa T^2}\right)}
-\extrinsic{\frac{1}{2}\tau_{\mathrm{q}}
  q^{\phantom{\perp}i}_{\perp}WK}\,,\\ {\gamma^{i}}_{\mu}h^{\mu
  \nu}\nabla_{\nu}\Pi &= \spatial{\left(\gamma^{ij}+W^2
  v^{i}v^{j}\right)\partial_{j}\Pi} +\rest{\frac{W^2}{\alpha}v^{i}\aparn
  \Pi} \,,\\
{\gamma^{i}}_{\mu}\nabla_{\nu}\pi^{\nu \left\langle \mu \right\rangle}&=
\spatial{D_{j}\pi_{\perp}^{\phantom{\perp}ij}
  +W^2v^{i}v_kD_{j}\pi_{\perp}^{\phantom{\perp}jk}} \nonumber
\\ &\phantom{=}-\rest{W^2v^{i}D_{j}\mathring{\pi}^{j}+
  \frac{1}{\alpha}\aparn
  \mathring{\pi}^{i}-\hat{a}_{j}\pi_{\perp}^{\phantom{\perp}ij}
  -\mathring{\pi} \hat{a}^{i}}\nonumber
\\ &\phantom{=}\rest{+Wv^{i}\left[\frac{v_k}{\alpha}\aparn
    \mathring{\pi}^{k}-\frac{1}{\alpha}\aparn \mathring{\pi}
    -3\hat{a}_{j}\mathring{\pi}^{j}-\mathring{\pi} v_k\hat{a}^{k}\right]}
\nonumber
\\ &\phantom{=}-\extrinsic{2{K^{i}}_{j}\mathring{\pi}^{j}-W^2v^{i}\left(2{K_{jk}}\mathring{\pi}^{j}v^{k}-K_{jk}\pi_{\perp}^{\phantom{\perp}jk}
  -\mathring{\pi} K\right)}\,,\\
\end{align}
\begin{align}
(1-\gamma_0)\Pi T
   {\gamma^{i}}_{\mu}\nabla^{\langle \mu
     \rangle}\left(\frac{\alpha_0}{T}\right) &=(1-\gamma_0)\Pi T
   \left[\spatial{\gamma^{ij}\partial_{j}+W^2v^{i}v^{j}\partial_{j}}+\rest{\frac{W^2}{\alpha}\aparn}\right]
   \left(\frac{\alpha_0}{T}\right)\,,\\ (1-\gamma_1)T{\gamma^{i}}_{\mu}\pi^{\mu
     \nu}\nabla_{\nu}\left(\frac{\alpha_1}{T}\right)
   &=\spatial{(1-\gamma_1)T\pi^{\phantom{\perp}ij}_{\perp}\partial_{j}\left(\frac{\alpha_1}{T}\right)}
   +\rest{\frac{(1-\gamma_1)T\mathring{\pi}^{i}}{\alpha}\aparn
     \left(\frac{\alpha_1}{T}\right)}\,, \\
   \mathcal{H}^{i}_{\mathrm{q}}&=Wv^{i}\bigg(\spatial{A_{j}q^{\phantom{\perp}j}_{\perp}}+\rest{W\Lambda_{j}q^{\phantom{\perp}j}_{\perp}-\mathring{q}
     Wv_{j}\Lambda^{j}-\mathring{q} v_{j}A^{j}}
   \notag \\ &\phantom{=Wv^{i}\bigg(}
   -\extrinsic{2W^2K_{kj}q^{\phantom{\perp}k}_{\perp}v^{j}+2\mathring{q}
     W^2K_{kj}v^{k}v^{j}}\bigg)\,,\\ \mathcal{G}^{i}_{\mathrm{q}}&=-\spatial{W{}^{3}{\Gamma^{i}}_{jk}q^{\phantom{\perp}j}_{\perp}v^{k}}-\rest{\mathring{q}
     W\hat{a}^{i}
     -\frac{W}{\alpha}q^{\phantom{\perp}j}_{\perp}\partial_{j}\beta^{i}}+\extrinsic{\mathring{q}
     Wv_j{K^{ij}}+W{K^{i}}_{j}q^{\phantom{\perp}j}_{\perp}}\,.
\end{align}

\subsubsection{Sources for the evolution of the shear-stress tensor}

Finally, listed below are all the spatial source terms [components
  $10\textrm{-}15$ in Eq. \eqref{eq:full_S}] appearing in the evolution
equation for the components of the perpendicular shear-stress tensor, \ie
$\partial_t \left( \sqrt{\gamma} D \pi^{\phantom{\perp}ij}_{\perp}
\right)=\ldots$

\begin{align}
{\gamma^{i}}_{\mu}{\gamma^{j}}_{\nu}\pi^{\phantom{\ns}\mu \nu}_{\ns}&=-2\eta\left(\spatial{\Sigma^{ij}}+\rest{\Lambda^{ij}}
-\extrinsic{W\mathcal{K}^{ij}}\right)\,,\\
\frac{1}{2}\eta T \pi^{\phantom{\perp}ij}_{\perp}\nabla_{\lambda}\left(\frac{\tau_{\pi}u^{\lambda}}{\eta T}\right)
&=\spatial{\frac{1}{2}\tau_{\pi}\pi^{\phantom{\perp}ij}_{\perp}\vartheta
+ \frac{1}{2}\eta T \pi^{\phantom{\perp}ij}_{\perp}Wv^{i}\partial_{i}\left(\frac{\tau_{\pi}}{\eta T}\right)} \notag \\
&\phantom{=}+\rest{\frac{1}{2}\tau_{\pi}\pi^{\phantom{\perp}ij}_{\perp}\Lambda
+\frac{\eta T \pi^{\phantom{\perp}ij}_{\perp}W}{2 \alpha}\aparn \left(\frac{\tau_{\pi}}{\eta T}\right)}
-\extrinsic{\frac{1}{2}\tau_{\pi}\pi^{\phantom{\perp}ij}_{\perp}WK}\,, \\
{\gamma^{i}}_{\mu}{\gamma^{j}}_{\nu}\nabla^{\langle \mu}q^{\nu \rangle}
&=
\spatial{D^{(i}q_{\perp}^{\phantom{\perp}j)}+\frac{1}{2}W^2v^{i}\left(v^{k}D_{k}q_{\perp}^{\phantom{\perp}j}+v_{k}D^{j}q_{\perp}^{\phantom{\perp}k}\right)
  +\frac{1}{2}W^2v^{j}\left(v^{k}D_{k}q_{\perp}^{\phantom{\perp}i}+v_{k}D^{i}q_{\perp}^{\phantom{\perp}k}\right)}
\nonumber \\ &\phantom{=}
\spatial{-W^2v^{i}v^{j}A_{k}q_{\perp}^{\phantom{\perp}k}-\frac{1}{3}\left(\gamma^{ij}+W^2v^{i}v^{j}\right)\left(D_{k}q_{\perp}^{\phantom{\perp}k}
  -A_k q_{\perp}^{\phantom{\perp}k}\right)} \nonumber
\\ &\phantom{=}+\rest{\frac{1}{2}W^2v^{i}\left[\frac{1}{\alpha}\aparn
    q_{\perp}^{\phantom{\perp}j}+\mathring{q}\hat{a}^{j}
    -\gamma^{jk}\partial_{k}\mathring{q}\right]} \notag
\\ &\phantom{=}\rest{+\frac{1}{2}W^2v^{j}\left[\frac{1}{\alpha}\aparn
    q_{\perp}^{\phantom{\perp}i}
    +\mathring{q}\hat{a}^{i}-\gamma^{ik}\partial_{k}\mathring{q}\right]}
\nonumber \\ &\phantom{=}
\rest{-W^2v^{i}v^{j}\left(W\Lambda_{k}q_{\perp}^{\phantom{\perp}k}-\mathring{q}
  W\Lambda_{k}v^{k}-\mathring{q} A_{k}v^{k}\right)}\nonumber
\\ &\phantom{=}\rest{-\frac{1}{3}\left(\gamma^{ij}+W^2v^{i}v^{j}\right)\left[\frac{1}{\alpha}\aparn
    \mathring{q} -
    W\Lambda_{k}q_{\perp}^{\phantom{\perp}k}+\hat{a}_{k}q_{\perp}^{\phantom{\perp}k}+\mathring{q}
    W\Lambda_{k}v^{k} +\phantom{=}\mathring{q} A_{k}v^{k}\right]}
\nonumber \\ &\phantom{=}-\extrinsic{\mathring{q} K^{ij}-\mathring{q}
  W^2v_k\left({K^{ik}}v^{j}+{K^{jk}}v^{i}\right)
  +2W^4v^{i}v^{j}K_{kl}\left(q_{\perp}^{\phantom{\perp}k}v^{l}-\mathring{q}
  v^{k}v^{l}\right)} \nonumber \\ &\phantom{=} \extrinsic{
  +\frac{1}{3}\left(\gamma^{ij}+W^2v^{i}v^{j}\right)K_{kl}\left(2\mathring{q}
  W^2v^{k}v^{l}
  -2W^2q_{\perp}^{\phantom{\perp}k}v^{l}+\mathring{q}\gamma^{kl}\right)}\,,
%\\
\end{align}
\begin{align}
{\gamma^{j}}_{\nu}{\gamma^{i}}_{\mu}q^{\langle \mu}\nabla^{\nu
  \rangle} \left(\frac{\alpha_1}{T}\right)&=\bigg\lbrace
\frac{1}{2}\bigg[q^{\phantom{\perp}i}_{\perp}\left(\spatial{\gamma^{jk}\partial_{k}
    +W^2v^{j}v^{k}\partial_{k}}+\rest{\frac{W^2v^{j}}{\alpha}\aparn}
  \right) \notag
  \\ &\phantom{=}+q^{\phantom{\perp}j}_{\perp}\left(\spatial{\gamma^{ik}\partial_{k}
    +W^2v^{i}v^{k}\partial_{k}}+\rest{\frac{W^2v^{i}}{\alpha}\aparn }
  \right)\bigg] \nonumber
\\ &\phantom{=}-\frac{1}{3}\left(\gamma^{ij}+W^2v^{i}v^{j}\right)\left(\spatial{q^{\phantom{\perp}k}_{\perp}\partial_{k}}+\rest{\frac{\mathring{q}}{\alpha}\aparn
}
\right)\bigg\rbrace\left(\frac{\alpha_1}{T}\right)\,,\\
\mathcal{H}^{ij}_{\pi}&=\spatial{WA_{k}\left(\pi^{\phantom{\perp}ik}_{\perp}v^{j}+\pi^{\phantom{\perp}jk}_{\perp}v^{i}\right)} \notag
\\ &\phantom{=}+\rest{W^2\Lambda_{k}\left(\pi^{\phantom{\perp}ik}_{\perp}v^{j}+\pi^{\phantom{\perp}jk}_{\perp}v^{i}\right)-W\left(v_{l}A^{l}
  +Wv_{l}\Lambda^{l}\right)\left(\mathring{\pi}^{i}v^{j}+\mathring{\pi}^{j}v^{i}\right)} \nonumber
\\ &\phantom{=}-\extrinsic{2W^3K_{lk}v^{l}\left(\pi^{\phantom{\perp}ik}_{\perp}v^{j}+\pi^{\phantom{\perp}jk}_{\perp}v^{i}\right)+2W^3K_{lk}v^{l}v^{k}\left(\mathring{\pi}^{i}v^{j}
  +\mathring{\pi}^{j}v^{i}\right)}\,,\\
%% \end{align}
%% %
%% \begin{align}
%% \hspace{1.50cm}
\mathcal{G}^{ij}_{\pi}& =
-\spatial{Wv^{l}\left({}^{3}{\Gamma^{i}}_{kl}\pi^{\phantom{\perp}kj}_{\perp}
  +{}^{3}{\Gamma^{j}}_{kl}\pi^{\phantom{\perp}ki}_{\perp}\right)} \nonumber
\\ &\phantom{=}-\rest{W\left(\mathring{\pi}^{i}\hat{a}^{j}+\mathring{\pi}^{j}\hat{a}^{i}\right)-\frac{W}{\alpha}
  \left(\pi^{\phantom{\perp}ik}_{\perp}\partial_{k}\beta^{j}+\pi^{\phantom{\perp}jk}_{\perp}\partial_{k}\beta^{i}\right)}\nonumber\\
&\phantom{=}+\extrinsic{Wv^{l}\left({K_l}^{i}\mathring{\pi}^{j}+{K_l}^{j}\mathring{\pi}^{i}\right)+W\left({K^i}_{l}\pi^{\phantom{\perp}lj}_{\perp}
  +{K^j}_{l}\pi^{\phantom{\perp}li}_{\perp}\right)}\,.
\end{align}

\subsubsection{Corollary: explicit expressions for the
  Christoffel symbols}

As a corollary to the lengthy expressions provided above and as a way to
help in the actual numerical implementation of
Eqs. \eqref{eq:pde_conservative}, we provide below also the explicit
expressions for the Christoffel symbols appearing in the source
terms \eqref{eq:full_S}
\begin{align}
{\Gamma^0}_{00}&=\rest{\partial_{t}\ln \alpha +\hat{a}_{i}\beta^{i}} -\extrinsic{\frac{1}{\alpha}K_{ij}\beta^{i}\beta^{j}}\,,\\
{\Gamma^0}_{0i}&=\rest{\hat{a}_{i}}-\extrinsic{\frac{1}{\alpha}K_{ij}\beta^{j}}\,,\\
{\Gamma^0}_{ij}&=-\extrinsic{\frac{1}{\alpha}K_{ij}}\,,\\ {\Gamma^{i}}_{00}&=
\rest{\partial_{t}\beta^{i}-\beta^{i}\partial_{t}\ln \alpha
  +\beta^{j}\partial_{j}\beta^{i}
  +\frac{1}{2}\gamma^{ij}\partial_{j}\alpha^2-\beta^{i}\beta^{j}\partial_{j}\ln
  \alpha + {}^{3}{\Gamma^{i}}_{jk}\beta^{j}\beta^{k}} \nonumber
\\ &\phantom{=}-\extrinsic{2\alpha{K^{i}}_{j}\beta^{j}
  +\frac{1}{\alpha}\beta^{i}K_{jk}\beta^{j}\beta^{k}}\,,\\ {\Gamma^i}_{0j}&=\rest{\partial_{j}\beta^{i}-\beta^{i}\partial_{j}\ln
  \alpha+{}^{3}{\Gamma^{i}}_{jk}\beta^{k}}
+\extrinsic{\frac{1}{\alpha}\beta^{i}K_{jk}\beta^{k} -\alpha
  {K^{i}}_{j}}\,,\\ {\Gamma^i}_{jk}&=\spatial{{}^{3}{\Gamma^{i}}_{jk}}+\extrinsic{\frac{1}{\alpha}\beta^{i}K_{jk}}\,.
\end{align}

In summary, the 3+1 flux-conservative formulation of the
general-relativistic Eqs. \eqref{eq:pde_conservative}, combined with the
explicit components Eqs. \eqref{eq:full_U}--\eqref{eq:full_S} and
Eqs. \eqref{eq:sources1}--\eqref{eq:sources3}, provides a complete and
ready-to-use set of equations for the numerical evaluation of dissipative
effects in special-relativistic simulations of colliding heavy ions as
well as in general-relativistic simulations of compact objects. In
Appendix \ref{sec:review} we provide a detailed comparison of the system
presented here with that of other formulations (see Table
\ref{tab:comparison}).

%------------------------------------------------------------------------
%------------------------------------------------------------------------
\section{General-relativistic dissipative hydrodynamics: numerical implementation}
\label{sec:methods}
%------------------------------------------------------------------------
%------------------------------------------------------------------------

We now turn to a numerical implementation and the strategy that needs to
be developed when Eqs. \eqref{eq:pde_conservative} with explicit
components (\ref{eq:full_U})--(\ref{eq:full_S}) have to be cast within an
already developed GRHD or GRMHD code. For simplicity, but also because
the issues that will be discussed below would apply also for more
complicated (and complete) forms of the equations, hereafter we
concentrate on a reduced set after neglecting the heat current and the
shear-stress tensor in Eqs. (\ref{eq:full_U})--(\ref{eq:full_S}), \ie
after setting
$q^{\phantom{\perp}i}_{\perp}=0=\pi^{\phantom{\perp}ij}_{\perp}$\footnote{Note
that setting to zero the spatial components of the heat current and of
the shear-stress tensor implies that also the time components are zero
[\cf Eq. \eqref{eq:restric1}].}. Furthermore, we will also assume that
the off-equilibrium contributions to the source terms are very small, \ie
$\Delta_\Pi \simeq 0$, or, equivalently, that $\Pi$ relaxes towards its
NS-value only, ignoring corrections coming from terms of order higher
than one in Knudsen number. As a result, the evolution equation for the
bulk-viscosity pressure \eqref{eq:conservative_bulk} reduces to
\begin{align}
  \partial_t\left(\sqrt{\gamma}D\Pi\right)+\partial_{i}\left(\sqrt{\gamma}DV^{i}\Pi\right)=
  \frac{\alpha \sqrt{\gamma} D}{ \tau_{_{\Pi}} W}\left( \Pi_{\ns} -\Pi
    \right)= -
  \frac{\alpha \sqrt{\gamma} D}{ \tau_{_{\Pi}} W}\left[ \zeta \left(
  \spatial{\vartheta} + \rest{\Lambda} - \extrinsic{KW} \right)  + \Pi
    \right]\,,
\label{eq:bulk_implement}
\end{align}
which is a relaxation-type equation, describing the evolution of the
bulk-viscosity pressure such that causality is not violated and stability
is guaranteed.

The specific implementation discussed here refers to the one made within
the Black Hole Accretion Code \texttt{BHAC} \citep{Porth2017}, which
solves the equations of GRMHD by means of a finite-volume approach and
high-resolution shock-capturing (HRSC) methods. \texttt{BHAC} assumes a
stationary but otherwise arbitrary curved background and it has been
employed in a number of studies of accretion onto supermassive black
holes \citep{Mizuno2018, Akiyama2019_L1} and compact objects
\citep{Olivares2020}. Of course, in our present implementation all of the
electromagnetic fields are set to zero so that the various terms in
Eqs. \eqref{eq:pde_conservative} reduce to
\begin{align}
&\boldsymbol{U} = \sqrt{\gamma}\begin{pmatrix}
	D \\[1em]
	S_{j} \\[1em]
	E-D \\[1em]
	D\Pi
	\end{pmatrix} =\sqrt{\gamma} \begin{pmatrix}
	\rho W \\[1em]
	\left( e+p+\Pi \right) W^2 v_{j} \\[1em]
	\left( e+p+\Pi \right) W^2 - \left( p + \Pi \right) - \rho W \\[1em]
	\rho W \Pi
	\end{pmatrix}\,,\label{eq:bulk_implement1}
\end{align}
\begin{align}
   \boldsymbol{F}^{i} =\sqrt{\gamma} \begin{pmatrix}
    V^{i} D\\[1em]
	\alpha {S^{i}}_{j} - \beta^{i}S_{j} \\[1em]
	\alpha (S^{i}-v^{i}D) - \beta^{i} (E-D) \\[1em]
	V^{i}D\Pi
	\end{pmatrix}\,,
\end{align}
\begin{align}
    \boldsymbol{S} = \sqrt{\gamma} \begin{pmatrix}
	0 \\[1em]
	\frac{1}{2}\alpha S^{ik}\partial_{j} \gamma_{ik} + S_{i}\partial_{j}\beta^{i} - E \partial_{j}\alpha \\[1em]
	\frac{1}{2}S^{ik}\beta^{j}\partial_{j}\gamma_{ik}+{S_i}^{j}\partial_{j}\beta^{i} - S^{j}\partial_{j}\alpha \\[1em]
	-(\alpha D/\tau_{_{\Pi}}W)
        \left[\zeta \left(
          \spatial{\vartheta} + \rest{\Lambda} - \extrinsic{KW} \right)  + \Pi\right]
  \end{pmatrix}\,.\label{eq:bulk_implement3}
\end{align}

Clearly, the extra terms that need to be handled in
Eq. \eqref{eq:bulk_implement} either involve divergences or partial
derivatives, which can be evaluated using the standard differential
operators available within \texttt{BHAC}.  For example, the divergence
appearing in the term $\spatial{\vartheta}$ is discretised at second
order as
\begin{align}
\spatial{\vartheta} =
\frac{\partial_{i}(\sqrt{\gamma}Wv^{i})}{\sqrt{\gamma}} &=
\frac{\int_V\partial_{i}(\sqrt{\gamma}Wv^{i})dV}{\int_V\sqrt{\gamma}dV} =
\frac{1}{\Delta V} \sum_{i=1}^3 \left[ \left(\overline{W}\overline{v}^{i}
  \Delta S^i\right)_{(x^i+\Delta x^i/2)} -\left(\overline{W}
  \overline{v}^{i} \Delta S^i\right)_{(x^i-\Delta x^i/2)} \right]\,,
\end{align}
where the cell volume and cell surfaces are defined respectively as \citep{Porth2017}
\begin{align}
\Delta V := \int_V \sqrt{\gamma}\,dV\,,\qquad \qquad
\Delta S^{i}_{(x^i+\Delta x^i/2)}:=\int_{(x^i+\Delta x^i/2)}\sqrt{\gamma}\,dS_{i}\,,
\end{align}  
Each integral is performed over one cell, where \(dV:=dx^1dx^2dx^3\) is
the coordinate volume, \(dS_{i}:=s_{i}dx^{j\neq i}dx^{k \neq i}\) the
coordinate surface, and the co-vector \(s_{i}\) is the $i$-th component
of the unit normal with respect to the boundary of the cell. We calculate
the boundary data through averages; for example, given a scalar function
$\phi$, we compute \(\overline{\phi}_{(x^i+\Delta x^i/2)}:=
(\phi(x^i+\Delta x^i)+\phi(x^i))/2\), \(\overline{\phi}_{(x^i-\Delta
  x^i/2)}:= (\phi(x^i-\Delta x^i)+\phi(x^i))/2\). Similarly, the
volume-averaged partial derivative of a scalar function $\phi$ (\eg $W$)
\begin{align}
\partial_{i}\phi & = \frac{\int_V\sqrt{\gamma}\partial_{i}\phi\,
  dV}{\int_V\sqrt{\gamma}\,dV} \nonumber \\ & =\frac{1}{\Delta V} \left[
  \left(\overline{\phi} \Delta S^i\right)_{(x^i+\Delta x^i/2)}
  -\left(\overline{\phi} \Delta S^i\right)_{(x^i-\Delta x^i/2)} -
  \hat{\phi} \left(\Delta S^i\right)_{(x^i +\Delta x^i/2)}+\hat{\phi}
  \left( \Delta S^i\right)_{(x^i-\Delta x^i/2)} \right]\,,
\end{align}
where $\hat{\phi}:=\phi(x^i)$. The time derivative $( \partial_{t}-
\beta^{i} \partial_{i} ) W$, such as the one appearing in the term
$\rest{\Lambda}$, is calculated using central second-order finite
differences and a two-step predictor-corrector time evolution, with
corrector step
\begin{align}
(\partial_tW)_c\approx \frac{W(t+\Delta t/2)-W(t-\Delta t/2)}{\Delta t}\,,
\end{align}
and a first-order backward-difference predictor step
\begin{align}
  (\partial_tW)_p \approx \frac{[W(t)-W(t-\Delta t/2)]}{\Delta t/2}\,.
\end{align}
Finally, for stationary spacetimes, as the one considered here, the trace
of the extrinsic curvature needed in Eq. \eqref{eq:bulk_implement3} can
be computed as
\begin{align}
K = \frac{1}{\alpha}\partial_{i}\beta^{i}+\frac{1}{2
  \alpha}\gamma^{ij}\beta^{k}\partial_{k}\gamma_{ij}\,.
\label{eq:K_cowling}
\end{align}
Note that the newly evolved bulk-viscosity pressure appears also in the
definitions of \(E\), \(S_j\), \(S^{ij}\), which need to be suitably
updated. Fortunately, \(\Pi\) only appears as a correction to the
equilibrium pressure \(p\) in the total pressure of the fluid
\(p_{t}=p+\Pi\). As a result, when considering a simple ideal-gas EOS,
where the pressure is given by
\begin{align}
p=(e-\rho)(\gamma-1)\,, \label{eq:ideal_eos}
\end{align}
and \(\gamma\) is the adiabatic index, the new effective EOS reads
\begin{align}
p_{t}=(e-\rho)(\gamma -1) + \Pi\,.
\end{align}
Note that characteristic wave-speeds need to be corrected for the sound
speed in the presence of bulk viscosity \citep{Bemfica2019}
\begin{align}
c^2_{s,t} := \left(\frac{\partial p}{\partial
  e}\right)_{n}+\frac{1}{mh_{t}} \left(\frac{\partial p}{\partial n
}\right)_{e} + \frac{\zeta}{\tau_{_{\Pi}}}\frac{ 1}{\rho h_{t}} >
c^2_{s} \,, \label{eq:bemfica}
\end{align}  
where $n$ is the number density, $m$ the particle rest-mass, $\rho=n m$,
and $h_{t}$ the total specific enthalpy
\begin{align}
h_{t}:= h + \frac{\Pi}{\rho} = \frac{e + p + \Pi}{\rho}\,.
\end{align}

As a result, in \texttt{BHAC} we make use only of the total pressure, of
the total specific enthalpy, and of the corresponding sound speed. In
other words, in the recovery of the primitive variables from the
conserved ones, we perform the following mapping
\begin{align}
\rho h = \rho + \frac{\gamma}{\gamma -1} p \quad &\rightarrow  \quad \rho h_{t} 
= \rho + \frac{\gamma}{\gamma -1} p_{t} - \frac{1}{\gamma -1} \Pi\,,\label{eq:mod_enthalpy}\\
p = \frac{\gamma -1 }{\gamma} \left( \rho h -\rho \right) \quad &\rightarrow  \quad p_{t} 
= \rho\left( \frac{\gamma -1 }{\gamma}\right) \left( h_{t} - 1 \right) + \frac{\Pi}{\gamma}\,,
\label{eq:mod_pressuren}\\
c_s^2 = (\gamma-1) \frac{h-1}{h} \quad &\rightarrow  \quad c_{s,t}^2 = (\gamma-1) 
\frac{ h_{t}-1}{h_{t}}+\frac{\zeta}{\tau_{_{\Pi}}}\frac{1}{\rho h_{t}}\,.
\label{eq:mod_sound}
\end{align}

%------------------------------------------------------------------------
%------------------------------------------------------------------------
\section{Numerical Tests: flat spacetime}
\label{sec:testing}
%------------------------------------------------------------------------
%------------------------------------------------------------------------

In this section we present the application of the relativistic
dissipative hydrodynamics
Eqs. \eqref{eq:bulk_implement1}--\eqref{eq:bulk_implement3} to two
1+1-dimensional tests in special relativity. The time integration is
carried out explicitly by using a two-step predictor-corrector
scheme. Furthermore, we employ the so-called Rusanov \citep{Rusanov1961a}
or ``TVDLF'' flux at the cell boundaries, with wave-speed given by the
expression for $c_{s,t}$ in Eq. (\ref{eq:mod_sound}), although the
differences when using instead the expressions for $c_{s}$ are at most of
the order of \(0.6\%\) for the cases considered here. Furthermore, we use
the ``minmod'' reconstruction scheme to compute state variables at cell
boundaries \citep[see, \eg][]{Rezzolla_book:2013} and \citet{Porth2017}
for more details on the numerical schemes employed. Also, to facilitate
the comparison with similar tests in the literature, in this section we
do not set $c=1$ and report the speed of light explicitly.

%------------------------------------------------------------------------
\subsection{Bjorken flow}
\label{sec:bjoflow}
%------------------------------------------------------------------------

We first consider the time-honoured Bjorken flow \citep{Bjorken1983} when
bulk viscosity is present, which still represents a well-known and
often-used test in relativistic dissipative hydrodynamics \citep[see,
  \eg][]{DelZanna2013,Inghirami16,Inghirami18} and ideal MHD with
transverse magnetic fields \citep{Roy2015,Pu2016b}. We recall that the
Bjorken flow is the idealised representation of a one-dimensional,
longitudinally boost-invariant motion of a fluid, such as the one
produced in an ultrarelativistic collision of two ions.

\begin{figure}
  \center
  \includegraphics[width=0.5\textwidth]{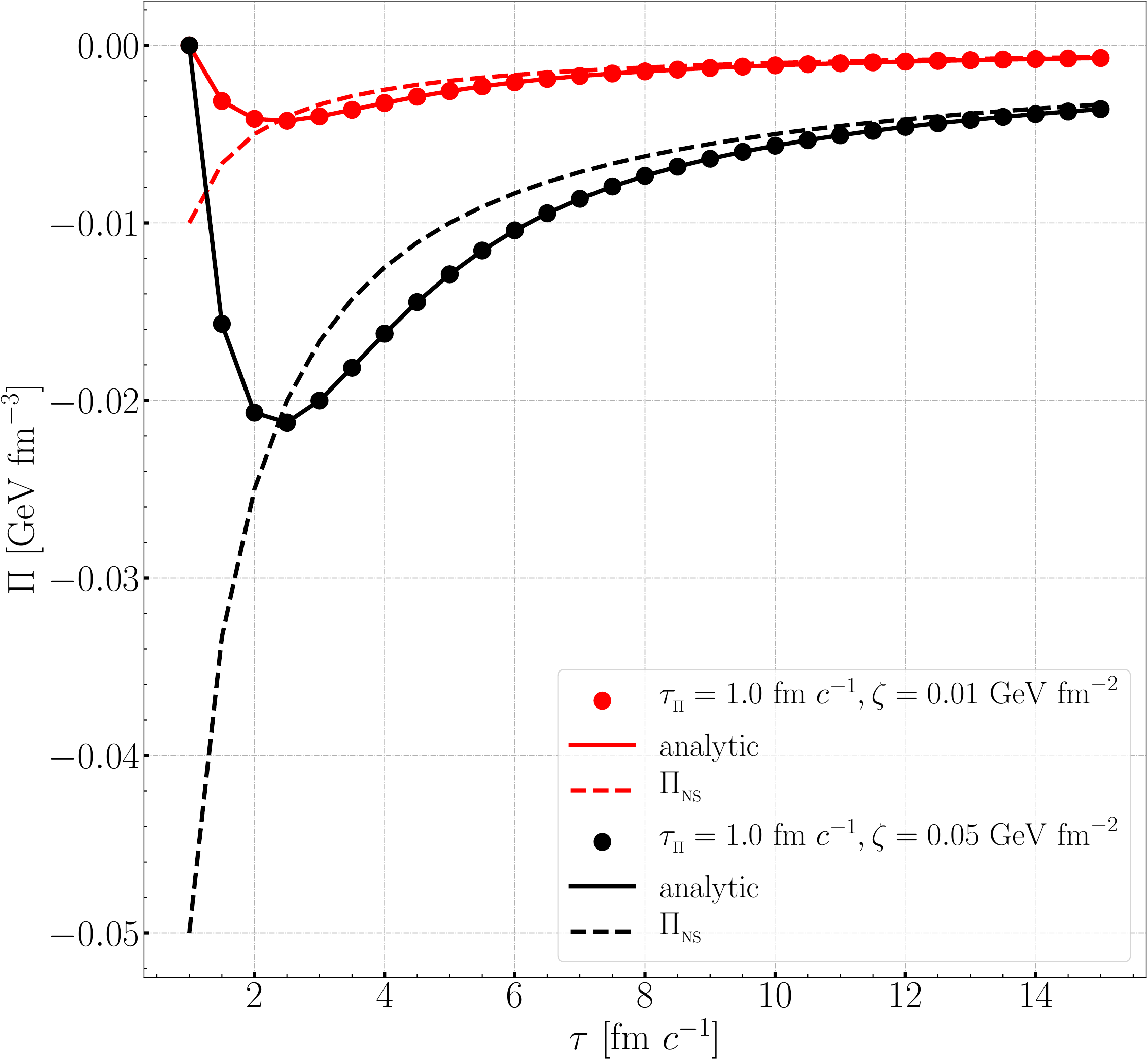}
  \caption{Evolution of the bulk-viscosity pressure in a Bjorken
    flow. Solid lines show the analytic solution, while dashed lines the
    solution in the NS approximation. Filled circles report instead the
    numerical solution from \texttt{BHAC}.}\label{fig:bjorken}
\end{figure}

As usual for the Bjorken-flow scenario, it is convenient to make use of
the so-called Milne coordinates, which are given by
\begin{align}
c\tau&:=\sqrt{(ct)^2-z^2}\, , \quad \eta:=\mathrm{artanh}\left(\frac{z}{ct}\right)=
\frac{1}{2}\mathrm{ln}\left(\frac{1+z/(ct)}{1-z/(ct)}\right)\, ,\\
t&\phantom{:}=\tau \cosh \eta \, , \quad z=c\tau \sinh \eta\, , 
\end{align}
where $\tau$ and $\eta$ are defined as the proper time and space-time
rapidity, respectively. In our setup, the evolution starts at
\(\tau=1~\mathrm{fm}~c^{-1}\) and ends at
\(\tau=15~\mathrm{fm}~c^{-1}\). The initial conditions are given by
\begin{align}
\left(~ \rho,~p,~{v},~\Pi ~\right) 
= \left(~10^{-7}~\mathrm{GeV}~c^{-2}~\mathrm{fm}^{-3}\,,~
10.0~\mathrm{GeV}~\mathrm{fm}^{-3},~0.0,~0.0~\right)\,.\label{eq:initbjo}
\end{align}
As the solution does not depend on the coordinate \(\eta\), all
solutions are taken at \(\eta = 0.0\). Finally, 
we choose constant values for $\zeta$ and $\tau_\Pi$,
as reported in the legend of Fig.\ \ref{fig:bjorken}. The analytic
solution for this problem is given by
\begin{align}
\Pi\left(\tau\right)&=\Pi\left(\tau_0\right)\exp\left[-\left(
\tau-\tau_0\right)/\tau_{_{\Pi}}\right] +\frac{\zeta}{c \tau_{_{\Pi}}}
\exp\left(-\tau/\tau_{_{\Pi}}\right) 
\left[\mathrm{Ei}\left(\tau_0/\tau_{_{\Pi}}\right)-\mathrm{Ei}\left(\tau/
\tau_{_{\Pi}}\right)\right]\, , 
\end{align}
where $\mathrm{Ei}\left(x\right)$ is the exponential integral function
\citep[see, \eg][for more details]{DelZanna2013}.

Figure \ref{fig:bjorken} shows the evolution of the bulk-viscosity
pressure \(\Pi\) for the initial data given by Eq. (\ref{eq:initbjo})
(the changes in the other hydrodynamical quantities are very small and
not particularly interesting). As it can be seen, the numerical solutions
agree very well with the analytical ones and with time \(\Pi\) converges
towards its NS value \(\Pi_{\ns}=-\zeta \Theta = -\zeta/(c\tau)\) (dashed
lines in Fig.\ \ref{fig:bjorken}). During the evolution, the relative
difference between the analytic and numerical solution is $10^{-7}$ at
most, confirming the correct implementation of the relativistic
dissipative-hydrodynamics equations for smooth flows in a flat
spacetime. Furthermore, our choices for the pair $(\zeta,\tau)$ lie well
within the range of applicability for the equations of GRDHD. This can be
seen by calculating the effective speed of sound which is given by
\begin{align}
\frac{c^2_{s,t}}{c^2} = \frac{1}{3}+\frac{\zeta}{c\tau_{_{\Pi}}}\frac{1}{4p+\Pi}\, ,
\end{align}
Requiring causality, \ie $c_{s,t}/c <1$, and expressing $p$ through the
perfect-fluid solution $p(\tau)=p(\tau_0)\left(\tau_o/\tau\right)^{4/3}$,
we obtain that
\begin{align}
  \tau <
  \left(\frac{8}{3}c\tau_{_{\Pi}}\frac{p(\tau_0)}{\zeta}\right)^{3/4}\tau_0\,,
\label{eq:estimate_acausal}
\end{align}
when $p\gg |\Pi|$. 

For our choices $(\zeta,\tau_{_{\Pi}})=(0.01\, \mathrm{GeV}\,
\mathrm{fm}^{-2}, 1.0\, \mathrm{fm}\, c^{-1})$ and
$(\zeta,\tau_{_{\Pi}})=(0.05\, \mathrm{GeV}\, \mathrm{fm}^{-2},1.0\,
\mathrm{fm}\, c^{-1})$ the dimensionless ratio $|\Pi|/p$ remains below
1\% and 5\%, respectively, up to the time $\tau = 400\, \mathrm{fm}\,
c^{-1}$. Hence, using Eq. (\ref{eq:estimate_acausal}) we find $\tau <
371\, \mathrm{fm}\, c^{-1}$ as well as $\tau < 110\, \mathrm{fm}\,
c^{-1}$, respectively; both values agree very well with the ones obtained
from the exact solution and lie clearly above the end of the simulation
at $\tau = 15\, \mathrm{fm}\, c^{-1}$.

%------------------------------------------------------------------------
\subsection{Shock-tube test}
%------------------------------------------------------------------------

We next explore the solution of a shock-tube problem for an
ultra-relativistic gas of gluons. While this is a standard
1+1-dimensional test scenario, we here use the same setup implemented by
\citet{Bouras2009} and \citet{Gabbana2020}, \ie we consider the ideal-gas
EOS relative to an ultra-relativistic fluid (\ie
\(\gamma=4/3\)). Adopting Cartesian coordinates, the spatial domain
ranges from \(x=-3.5 ~ \mathrm{fm}\) to \(x=3.5 ~ \mathrm{fm}\) and the
initial discontinuity in pressure and density is located at \(x = 0.0 ~
\mathrm{fm}\), while the velocity and bulk-viscosity pressure are assumed
to be zero initially. In other words, the initial conditions are given
by\footnote{In this section, to facilitate the comparison with codes
designed for describing heavy-ion collisions \citep{Gabbana2020} we adopt
physical units, where $k_{\mathrm{B}}$ and $\hbar$ are the Boltzmann and
reduced Planck constants, respectively.}
\begin{align}
  \left(~ T,~p,~{v},~\Pi ~\right) = \left\lbrace
  \begin{array}{ll}
     \left(~
     0.4 ~ \mathrm{GeV} ~ k_{\mathrm{B}}^{-1},~ 5.43 ~ \mathrm{GeV}~
     \mathrm{fm}^{-3},~0.0,~0.0~ \right) & x < 0.0 ~ \mathrm{fm}\,,
     \\ \left(~ 0.2 ~ \mathrm{GeV}~k_{\mathrm{B}}^{-1},~ 0.33 ~
     \mathrm{GeV}~ \mathrm{fm}^{-3},~0.0,~0.0~ \right) & x \geq 0.0 ~
     \mathrm{fm}\,.
  \end{array}
  \right.
\label{eq:initA}
\end{align}
We parametrize the bulk-viscosity coefficient $\zeta$ in terms of the
entropy density of the fluid, namely,
\begin{align}
s = \rho \frac{k_{\mathrm{B}}}{m} \left[ 4 - \ln \left(
  \frac{\pi^2\rho}{m~d_{\mathrm{F}}T^{3}}~
\frac{c^3\hbar^3}{k_{\mathrm{B}}^3} \right) \right]\,,
\end{align}
where \(d_{\mathrm{F}}\) denotes the number of degrees of freedom and is
set to \(16\) for gluons, such that
\begin{align}
  \zeta = \frac{4}{3} ~\frac{k_{\mathrm{B}}}{c\hbar} ~ \zeta_0\, s\,, 
  \label{eq:viscosity}
\end{align}
The coefficient $\zeta_0$ is a non-negative number, for which we choose
the values \(\zeta_0=\{0.002, 0.01, 0.1\}\) to obtain a direct comparison
with the data from \citet{Gabbana2020}. Note that the equations
describing the shock-tube problem with bulk viscosity in one dimension
take the same form as the corresponding equations with shear viscosity;
the latter has been investigated in the work of \citet{Bouras2010}, as
well as more recently by \citet{Gabbana2020}. This leads to the mapping
$\zeta = 4/3 ~ \eta$ between the bulk viscosity employed in this work and
the shear viscosity used in \citet{Bouras2010} and
\citet{Gabbana2020}. Furthermore, to obtain the correct rest-mass density
we assume a single-particle rest-mass \(m=0.5 ~ \mathrm{MeV}~c^{-2}\), so
that \(\rho \ll e\), as required by the ultra-relativistic
limit. Finally, the relaxation time used by \citet{Gabbana2020} is
expressed in terms of the bulk viscosity and is given by
\begin{align}
  \tau_{_{\Pi}} &= \frac{15}{16} \frac{\zeta}{pc}\,.
  \label{eq:relax_time}
\end{align}

\begin{figure}
  \center
  \includegraphics[width=0.87\textwidth]{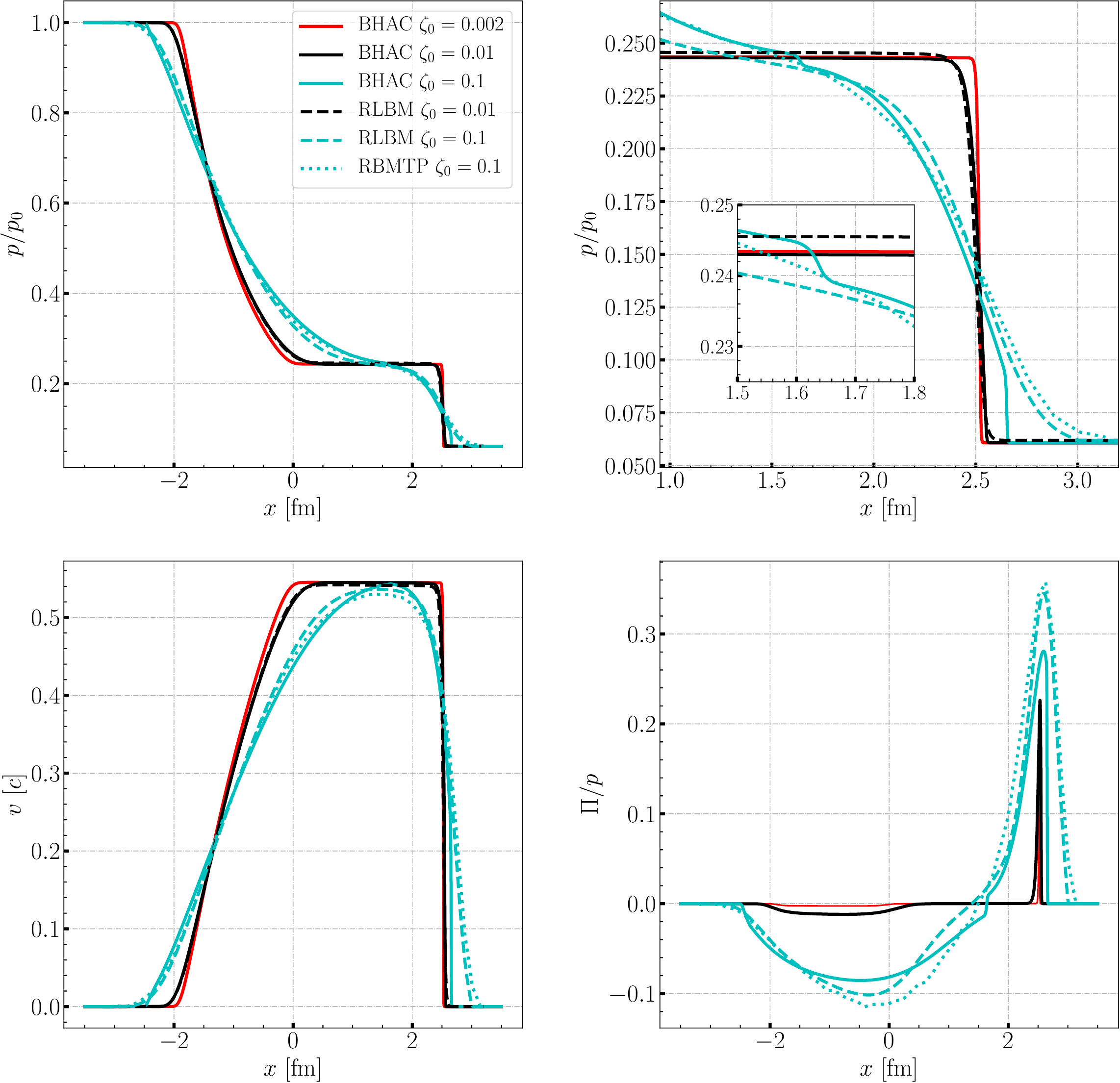}
  \caption{Solution of the shock-tube test with initial data given by
    Eq. (\ref{eq:initA}) and at time \(t=3.2~\mathrm{fm}\,c^{-1}\). 
\textit{Top left:} solution of the pressure normalized to
    \(p_0=5.43~\mathrm{GeV}~\mathrm{fm}^{-3}\); \textit{top right:} same
    as in top left but zoomed-in at the shock front; \textit{bottom
      left:} solution of the three-velocity, \textit{bottom right:}
    bulk-viscosity pressure normalised to the fluid pressure. In all
    panels, the solid lines show the results obtained with \texttt{BHAC},
    while the dashed or dotted lines present the results using the
    relativistic lattice-Boltzmann (RLBM) or the test-particle (RBMTP)
    approach, respectively.}
  \label{fig:shocktube}
\end{figure}

The numerical solution of the shock-tube problem at time
\(t=3.2~\mathrm{fm}~c^{-1}\) is shown in Fig.\ \ref{fig:shocktube}, whose
upper panels report the behaviour of the pressure normalized to $p_0 :=
5.43 ~\mathrm{GeV}~\mathrm{fm}^{-3}$ (the top right panel is a
magnification of the top left panel), while the bottom panels show the
solution of the velocity and bulk-viscosity pressure normalized to the
fluid pressure. Different lines refer either to solutions obtained with
\texttt{BHAC} for different values of $\zeta_0$, or to solutions obtained
with a relativistic lattice-Boltzmann (RLBM) approach (dashed lines) or
to solutions of the relativistic Boltzmann equation via the test-particle
(RBMTP) approach (dotted lines). Note that the case $\zeta_0=0.002$ (red
solid line) is essentially indistinguishable from an inviscid solution
with the precision shown in the figure and hence can be taken as the
perfect-fluid reference.

Figure \ref{fig:shocktube} highlights how the strong spatial gradients
present in the initial conditions tend to be washed out by the presence
of bulk viscosity and that this smearing of the discontinuities is larger
with increasing bulk viscosity. Note that the wave-pattern of
perfect-fluid hydrodynamics -- which consists of a rarefaction wave and
of a shock wave -- can still be clearly identified if the bulk viscosity is
not too large, \ie \(\zeta_0 \lesssim 0.01\) (see, \eg the pressure in
the upper left panel of Fig.\ \ref{fig:shocktube}). Furthermore, the
solution behaviour is in good agreement with the results obtained by
using the RLBM approach. However, in the case of high bulk-viscosity, \ie
\(\zeta_0 = 0.1\), the wave-pattern of perfect-fluid hydrodynamics is so
strongly smeared out that it is difficult to clearly distinguish the
rarefaction wave from the shock wave. This is not surprising, since such
large values of the bulk viscosity effectively correspond to a regime of
large Knudsen number, which is where the hydrodynamical approach -- and
hence the formation of shock waves -- is expected to fail.

Interestingly, and as pointed out by \citet{Denicol2008} and
\citet{Bouras2010}, three additional discontinuities are present in the
high-viscosity, \(\zeta_0 = 0.1\), case, two of which can be seen in the
upper right panel of Fig.\ \ref{fig:shocktube}, where one is located at
the head of the right-propagating shock, while the other near the contact
discontinuity of the corresponding inviscid (\(\zeta_0=0.002\))
case\footnote{Obviously the contact discontinuity cannot be seen in the
pressure profile, but it is apparent in the rest-mass density profile,
which is not shown in Fig.\ \ref{fig:shocktube}.}. The third additional
discontinuity is located at the head of the left-propagating rarefaction
wave which is not shown here. However, all discontinuities transition
smoothly into the wave-pattern of perfect-fluid hydrodynamics at later
times; indeed, the pressure jump located near the contact discontinuity
decays to less than 10\% of its initial size of $\simeq
2.4\,\textrm{GeV}\,\textrm{fm}^{-3}$ by $t\simeq
1.8\,\textrm{fm}\,c^{-1}$ \citep[see also][for a more detailed
  description and a possible explanation]{Bouras2010}. In addition, we
find in agreement with \citet{Bouras_2009} that at a time $t\simeq
3.2\,\textrm{fm}\,c^{-1}$ the fluid velocity downstream of the shock
front differs by less than 0.5\% from the corresponding fluid velocity of
the inviscid case. At this time, the previously mentioned pressure jump
has declined to less than 2\% of its original size.  Because the new
discontinuities are absent in the RLBM (dashed line) and RBMTP (dotted
line) solutions, their appearance may well indicate a breakdown of
dissipative hydrodynamics for such large viscosities, \ie for such large
Knudsen numbers, which is of course still correctly described by the
microscopic approaches RLBM and RBMTP.

In addition, in the high-viscosity case, the \texttt{BHAC} solution
underestimates the bulk-viscosity pressure at the head of the
right-propagating shock with respect to the values computed with the RLBM
or RBMTP approaches (see bottom right panel of
Fig.\ \ref{fig:shocktube}). While this may again be due to the breakdown
of the hydrodynamical description, part of the error may also originate
from the truncation of the relativistic evolution equation for \(\Pi\)
[we recall that we have set $\Delta_\Pi=0$ in Eq. \eqref{eq:full_S}]. As
remarked by \citet{Bouras2010}, the inclusion of additional source terms,
as well as of a coupling to a heat current, generally yields a better
description of the corresponding dissipative current (see Fig.\ 10 of
\citealt{Bouras2010}). We expect the same to be true here and hence that
a smaller deviation would be obtained with a more sophisticated source
term for Eq. \eqref{eq:bulk_implement}.

Overall, Fig.\ \ref{fig:shocktube} shows that our numerical
implementation of the relativistic dissipative-hydrody-namics equations
leads to solutions that are in very good agreement with the reference
solutions obtained by the direct solution of the relativistic Boltzmann
equation in regimes that are mildly dissipative, \ie \(\zeta_0 \lesssim
0.01\), and in regimes that are highly dissipative, \ie \(\zeta_0
\lesssim 0.1\). The relative differences remain below 8\% for $p/p_0$ in
the region $x \in [-2.5,2.6]~\mathrm{fm}$, but larger differences can be
found when considering also other regions of the computational domain
including the rightmost shock wave and the head of the left-propagating
rarefaction wave.

%------------------------------------------------------------------------
%------------------------------------------------------------------------
\section{Numerical Tests: curved spacetime}
\label{sec:accretion}
%------------------------------------------------------------------------
%------------------------------------------------------------------------

In this section we present a stationary solution of the spherically
symmetric equations of GRDHD in a Schwarzschild spacetime.  For perfect
fluids, the fully general-relativistic solution is known as the so-called
``Michel solution'' \citep{Michel72_new} and, together with the
``Bondi-Hoyle'' solution \citep{Bondi52_new}, serves as the reference
solution for models of accreting nonrotating black holes in spherical
symmetry, \citep[see \eg ][and references therein]{Nobili1991} as well as
a testbed for GRHD and GRMHD codes \citep[see,
  \eg][]{Hawley84a,Porth2017,Weih2020b}.

The effects of shear viscosity -- such as the one arising from turbulent
motion -- have first been considered by \citet{Turolla1989}, who however
adopted a description in terms of the general-relativistic NS equations.
Already in their simplified setup, \citet{Turolla1989} have pointed out
the numerous subtleties and highly nontrivial behaviour of the problem of
stationary viscous accretion onto a black hole. Hence, to the best of our
knowledge, the problem of stationary, spherically symmetric accretion of
bulk viscous fluids onto nonrotating black holes using second-order
dissipative-hydrodynamics framework has not been considered before. We
here use the solution of this problem obtained from the corresponding
system of ordinary differential equations (ODEs) to test our
implementation of bulk viscosity in \texttt{BHAC} in a curved spacetime
geometry. Details on the derivation and solution of the ODEs can be found
in Appendix \ref{sec:vis_acc_details}.

We assume the fluid to be a mixture of ionised non-relativistic hydrogen
coupled to photons by employing the following EOS
\citep[see, \eg][]{Rezzolla_book:2013}
\begin{align}
  p = p_{_{\mathrm{M}}}(1+\alpha)\,,
  \label{eq:EOSalpha}
\end{align}
where \(p_{_{\mathrm{M}}}\) denotes the pressure of the matter component,
which is assumed to be an ideal gas, while the contribution from the
radiation component is fixed using the parameter \(\alpha\). By
rearranging Eq. (\ref{eq:EOSalpha}), we find that the EOS of the total 
mixture takes the same form as the EOS for an ideal gas having an 
effective adiabatic index \(\gamma_e\):
\begin{align}
  p = (\gamma_e-1)(e-\rho)\,,
  \label{eq:EOS1}
\end{align}
where \(\gamma_e=1+2(1+\alpha)/[3(1+2\alpha)]\). Note that $e$ is the
total energy density of the mixture, while $\rho$ denotes the rest-mass
density of the hydrogen ions. The effective adiabatic index $\gamma_e$
should not be confused with the generalized adiabatic exponent
$\Gamma_1$, which instead is defined through a thermodynamic relation:
$\Gamma_1 := \left(\partial \ln p/\partial \ln \rho\right)_{\tilde{s}}=
\left(5/2+20\alpha+16\alpha^2\right)/\left[(3/2+12\alpha)(1+\alpha)\right]\,$,
where $\tilde{s}$ denotes the specific entropy \citep[see,
  \eg][]{Mihalas84,Rezzolla_book:2013}. However, the effective adiabatic
index $\gamma_e$ can be expressed through the generalized adiabatic
exponent $\Gamma_3$ as
\begin{align}
\gamma_e = \frac{20-16\Gamma_3}{11-9\Gamma_3}\,,
\end{align}
where $\Gamma_3:=1+\left(\partial \ln T/\partial \ln
\rho\right)_{\tilde{s}}=
\left(5+32\alpha\right)/\left(3+24\alpha\right)$. The temperature can be
obtained from the ideal-fluid EOS relative to the matter component
$p_{\mathrm{M}}=2 (k_{\mathrm{B}}/m_{\mathrm{p}})\rho T $ which yields:
\begin{align}
T = \frac{1}{2(1+\alpha)}\frac{m_{\mathrm{p}}}{k_{\mathrm{B}}}\frac{p}{\rho}\, ,
\label{eq:temperature_ideal_plasma}
\end{align}
where \(m_{\mathrm{p}}\) denotes the proton mass. Note the appearance of a 
factor $2$ in the denominator coming from the electrons in our charge-neutral 
plasma. Furthermore, we use a modification of the formula for radiative
bulk viscosity given by \citet{Weinberg1971} and \citet{Sawyer2006}
\begin{equation}
\zeta =
4\zeta_0\sigma_{_{\mathrm{SB}}}T^4\tau_{\mathrm{mfp}}\left(\frac{4}{3}-\gamma_e\right)^2\,,
\end{equation}
where \(\sigma_{_{\mathrm{SB}}}\) is the Stefan-Boltzmann constant and
\(\tau_{\mathrm{mfp}}:=m_{\mathrm{p}}\rho^{-1}\sigma_{\mathrm{T}}^{-1}\)
the mean-free-path for a photon in the mixture, with
\(\sigma_{\mathrm{T}}\) being the Thomson scattering cross-section. The
dimensionless constant \(\zeta_0\) is essentially arbitrary and used here
to explore the regimes of low and high bulk viscosities. For the
relaxation time \(\tau_{_{\Pi}}\), we choose the parametrization:
\begin{equation}
  \tau_{_{\Pi}}=\tau_0\frac{M}{|\dot{M}|}\left(\frac{r}{2M}\right)^3\,,
  \label{eq:tau3}
\end{equation}
where \(r\) denotes the circumference radius in Schwarzschild
coordinates, \(M\) the mass of the black hole, \(\dot{M}\) the accretion
rate, and \(\tau_0\) is a dimensionless parameter to study short and long
relaxation times. Note that in Eq. (\ref{eq:tau3}) the relaxation time
increases cubically with radius and this is necessary to prevent the
rapid growth of non-equilibrium effects near the sonic point, which can
yield unphysical solutions.

All of the models considered have the sonic point $r_s$ at $200\,M$, a
black-hole mass of $M=3M_{\odot}$, and $\alpha=1$, which then yields
$\gamma_e=1.4\overline{4}$. Furthermore, we set the constants of motion,
\ie the mass-accretion rate $\dot{M}$ and the viscous analogue of the
relativistic Bernoulli constant $\mathcal{B}$ (see Appendix
\ref{sec:vis_acc_details} for a definition), from their inviscid values
computed at $r_s=200\,M$, \ie $\dot{M}=-0.01582$ and
$\mathcal{B}=-1.00192$. Note that in the absence of dissipative losses
the accretion process of a perfect fluid is isentropic, so that it can
be described by a polytropic EOS $p=k\rho$, where the polytropic constant
of the fluid mixture is given by
$k=2(1+\alpha)T_{\infty}\rho_{\infty}^{1-\gamma_e}$ with asymptotic
values $\rho_{\infty}=2\times 10^{-9}\,\mathrm{g}~\mathrm{cm}^{-3}$ and
$T_{\infty}=1.5\times 10^5 \,\mathrm{K}$\footnote{The units employed in
the code are such that $m_{\mathrm{p}}/k_{\mathrm{B}}=1$ and $k=1$.}. In
the following, we consider five different cases with different
viscosities and relaxation times, whose parameters are given in Table
\ref{tab:models}, employing a grid of 10,000 cells ranging from \(1.5M\)
to \(1,\!000\,M\) in the radial direction and using horizon-penetrating
Kerr-Schild coordinates.

\begin{figure*}
\centering
  \includegraphics[width=0.45\textwidth]{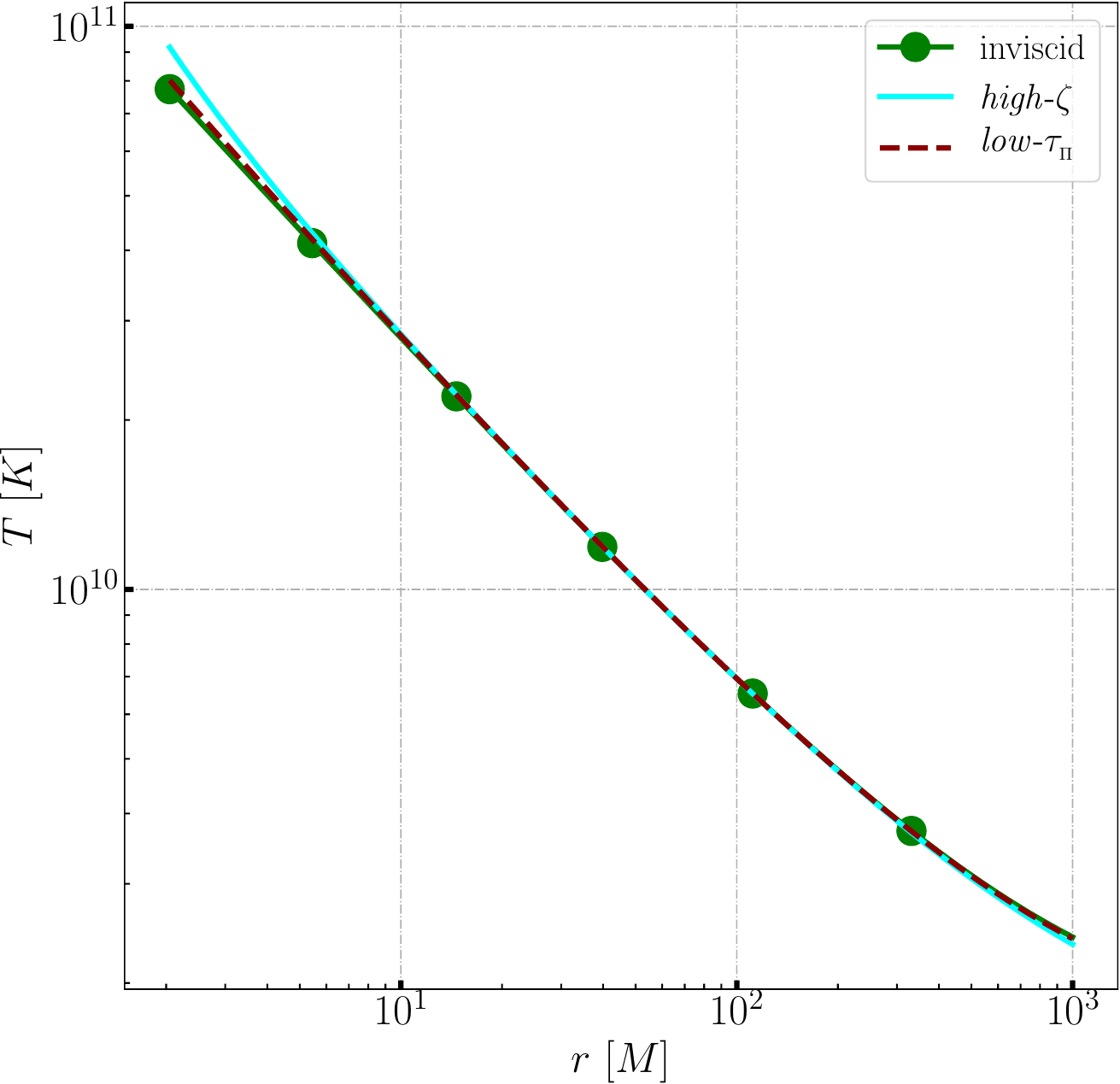}
  \hspace{1.0cm}
  \includegraphics[width=0.45\textwidth]{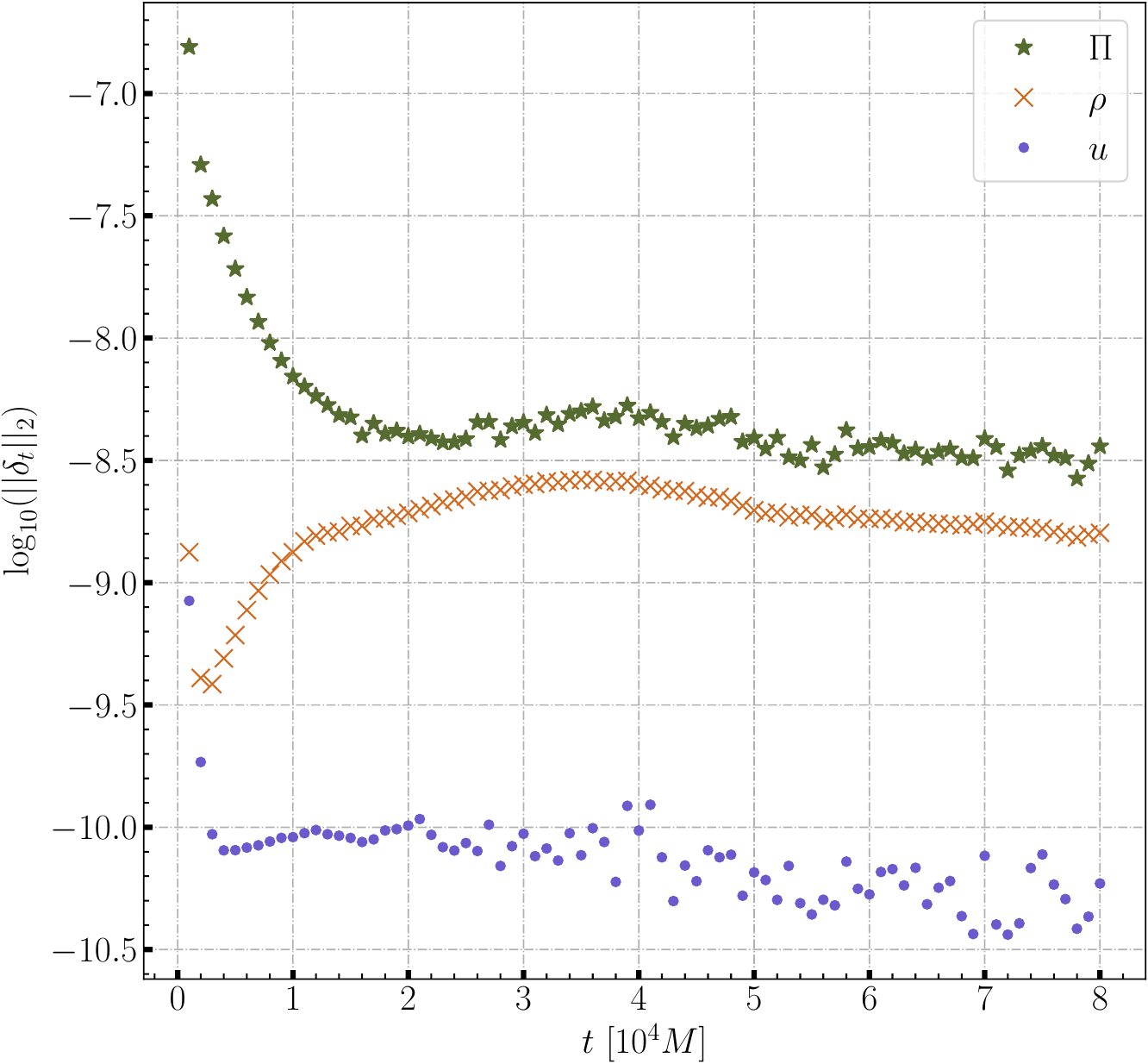}
  \caption{\textit{Left:} Temperature \(T\) as a function of the
    circumference radius \(r\) in units of \(M\) and at time
    $10,\!000~M$. Shown are the inviscid solution (filled circles) and
    the models with \textit{high-$\zeta$} (solid line) and
    \textit{low-$\tau$} (dashed line), respectively. Note that deviations
    from the inviscid solution increase towards the event horizon and
    that the sonic point is at \(r_s=200\,M\). \textit{Right:} $L_2$-norm
    of the relative time variation \(\delta_t\) for the bulk-viscosity
    pressure \(\Pi\) (stars), the rest-mass density \(\rho\) (crosses),
    and the primitive fluid velocity \(u\) (filled circles) shown as a
    function of time for the \textit{medium-$\zeta$} model set using
    \(20,\!000\) grid cells.}
  \label{fig:absoluteT}
\end{figure*}

\begin{table}
\centering
\begin{tabular}{lrr}
\textbf{Model}
& $\zeta_0$ & $\tau_0\ [10^{-12}]$ \\
\hline
\hline
\rule{0pt}{5pt}
\textit{low-$\zeta$} & $160$ & $1.0$ \\[3pt] 
\hline
\rule{0pt}{5pt}
\textit{medium-$\zeta$} & $16,\!000$ & $1.0$ \\[3pt]
\hline
\rule{0pt}{5pt}
\textit{high-$\zeta$} & $104,\!000$ & $1.0$ \\[3pt]
\hline 
\rule{0pt}{5pt}
\textit{low-$\tau_{_{\Pi}}$} & $16,\!000$ & $0.06$ \\[3pt]
\hline
\rule{0pt}{5pt}
\textit{high-$\tau_{_{\Pi}}$} & $16,\!000$ & $50$\\
\hline
\hline
\end{tabular}
\caption{Summary of the various models evolved and their corresponding
  parameters $\zeta_0$ and $\tau_0$.}
\label{tab:models}
\end{table}

The inviscid solution for the temperature $T$, together with the
\textit{high-$\zeta$} and \textit{low-$\tau$} models are shown in the
left panel of Fig.\ \ref{fig:absoluteT} at time $t=10,\!000\,M$. To
verify that our calculation reaches a stationary state, we show in the
right panel of Fig.\ \ref{fig:absoluteT} the logarithm of the $L_2$-norm
of the relative time variation \(\delta_t\) for the rest-mass density
\(\rho\), the primitive fluid velocity \(u\) and the bulk-viscosity
pressure \(\Pi\) as a function of time for the \textit{medium-$\zeta$}
model using 20,000 grid cells. For each quantity $\phi$, the relative
time variation is defined as $\delta_t\phi(t) :=
1-{\phi(t-10\,M)}/{\phi(t)}$.
In essence, the right panel of \ref{fig:absoluteT} shows that, although
the fluid was not stationary at the beginning of the evolution, it
reaches an approximately stationary state at late times. The small but
nonzero value of the relative differences at late times is due to
low-amplitude, small-scale oscillations generated at the outer boundary
of the numerical domain. We note that after performing a (global)
self-convergence test we were able to recover the correct global
convergence order of \texttt{BHAC}, \ie \(\approx 2\)
\citep[see][]{Porth2017}; on the other hand, local self-convergence tests
show that the convergence order is very close to two for small radii,
while it exhibits small oscillations around two for large radii, with
amplitudes that increase towards the outer boundary.

\begin{figure}
  \center
  \includegraphics[width=1.0\textwidth]{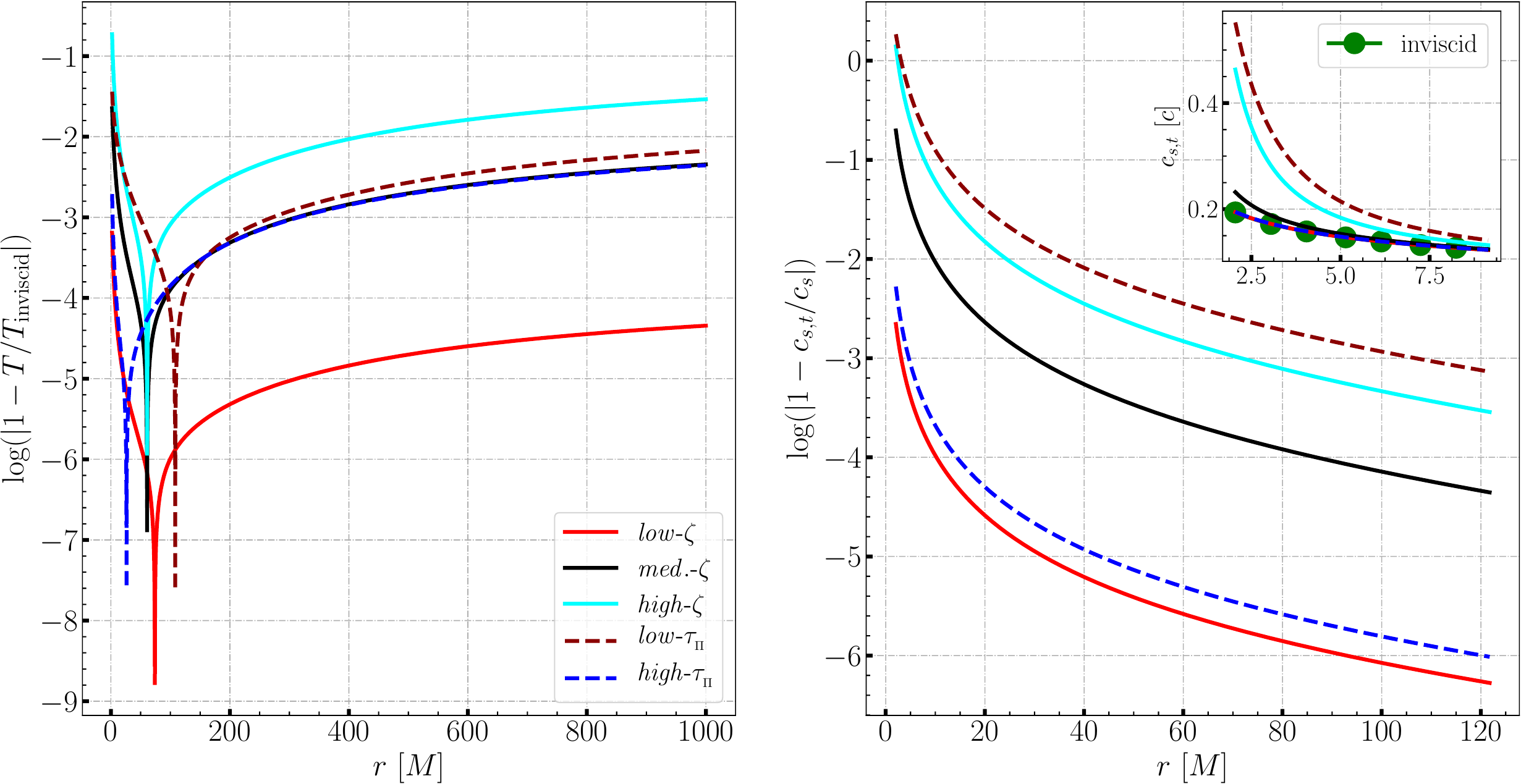}
  \caption{\textit{Left:} Relative difference between the temperature
    \(T\) of the viscous models and the corresponding inviscid
    model. Note that the differences clearly increase with the increase
    of the bulk viscosity and that the relative difference changes sign
    somewhere outside the event horizon. \textit{Right:} Same as on the
    left but for the sound speed. Reported in the inset is the actual
    value of the total sound speed near the event horizon, which shows
    considerable deviations from the inviscid
    solution.}\label{fig:tmodsound}
\end{figure}

The left panel of Fig. \ref{fig:tmodsound} displays the radial profiles
of the temperature for the various viscous models when compared to the
values obtained from the inviscid solution. As can be seen from the solid
lines -- which refer to the \textit{low-, medium-}, and
\textit{high-$\zeta$} models -- the viscous fluid is hotter near the
horizon and colder at larger radii than the corresponding inviscid
fluid. In particular, for the \textit{high-$\zeta$} model, the
temperature at the horizon can be up to \(\sim 18 \%\) larger than in the
inviscid model. A similar behaviour can be seen also for the dashed
lines, which refer to the \textit{low} and \textit{high
  \(\tau_{_{\Pi}}\)} models.
Shown instead in the right panel of Fig.\ \ref{fig:tmodsound} is the
corresponding comparison in terms of the viscous sound speed very close
to the event horizon, which can be larger by a factor $\sim 3$ for the
\textit{low-$\tau_{_{\Pi}}$} model and by a factor $\sim 2.3$ for the
\textit{high-$\zeta$} model, reaching viscous sound speeds above \(0.5\)
and \(0.4\), respectively (see inset). In general, and as can be
intuitively expected, solutions with high bulk viscosities and low
relaxation times tend to have larger temperatures and viscous sound
speeds, and hence larger deviations from the inviscid case.

\begin{figure}
  \center
  \includegraphics[width=1.0\textwidth]{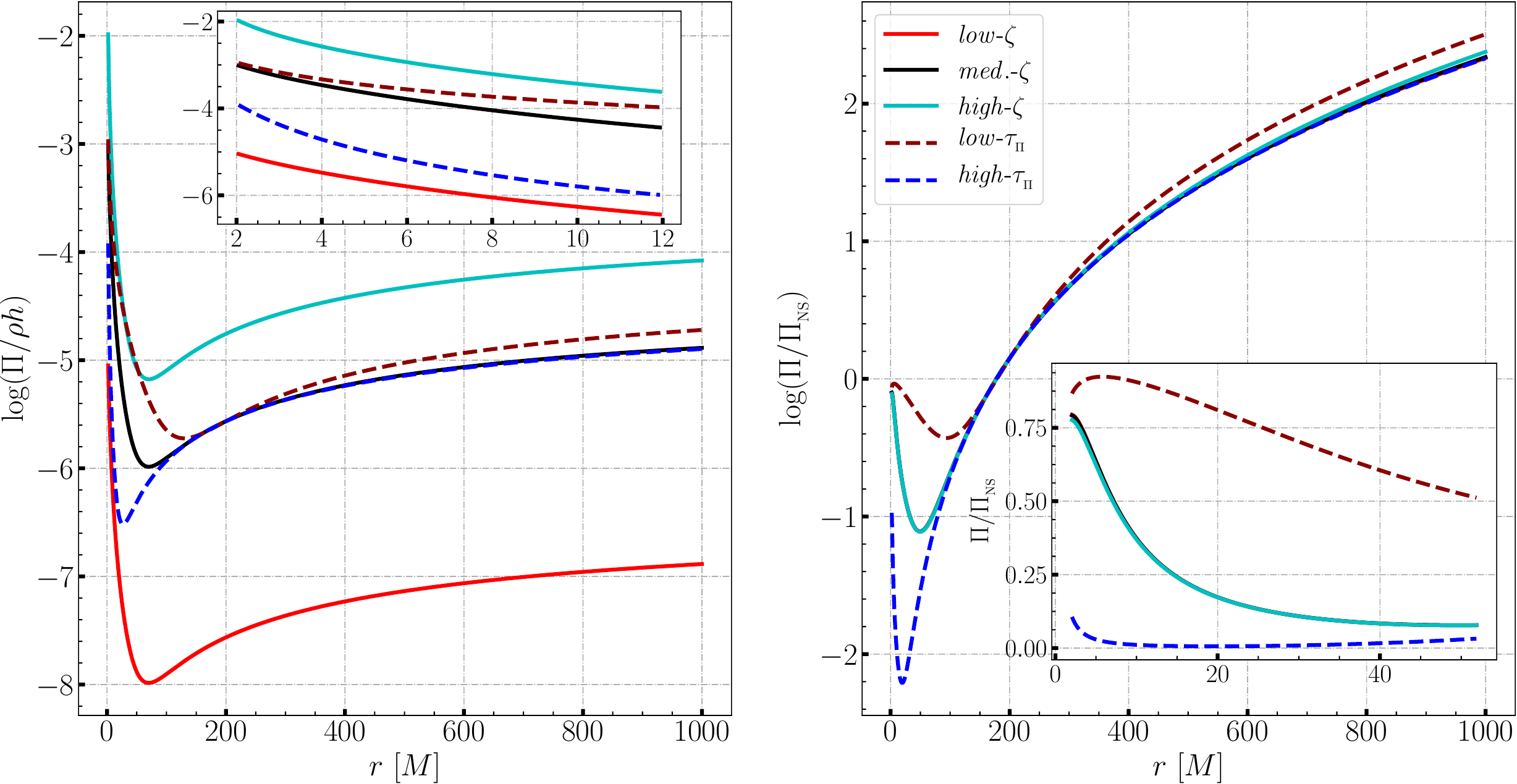}
  \caption{\textit{Left:} radial profiles of the ratio of the
    bulk-viscosity pressure over the enthalpy density (\ie inverse Reynolds
    number). Solid lines of different colours refer to models with
    \textit{low-$\zeta$}, \textit{medium-$\zeta$}, and \textit{high-$\zeta$}; 
    dashed lines of different colours refer to models with
    \textit{low-$\tau_{_{\Pi}}$} and \textit{high-$\tau_{_{\Pi}}$},
    respectively. The inset reports the same quantities but near the
    event horizon. \textit{Right:} radial profiles of the ratio of the
    bulk-viscosity pressure over the corresponding NS value. Solid and
    dashed lines follow the same convention as in the left panel, and
    inset reports the same quantities but near the event horizon.}
  \label{fig:doublepi}
\end{figure}

Figure \ref{fig:doublepi} shows radial profiles of the relativistic
``inverse Reynolds number'' \(\Pi/\rho h\) (left panel, solid and dashed
lines) and of the bulk-viscosity pressure normalized to the corresponding
NS value \(\Pi/\Pi_{_{\mathrm{NS}}}\) (right panel, solid and dashed
lines). Note that for all models, \(\Pi/\rho h\) assumes small but finite
values at large radii, it decreases when moving inwards, and then
increases again sharply close to the horizon. The overall magnitude of
\(\Pi/\rho h\) is very sensitive to the parameter \(\zeta_0\) (\cf
\textit{low}, \textit{medium}, and \textit{high-$\zeta$} models), while
\(\tau_0\) affects the location of the sharp increase (\cf cases
\textit{medium-$\zeta$}, \textit{low-$\tau_{_{\Pi}}$}, and
\textit{high-$\tau_{_{\Pi}}$}). Note that most models approach their
corresponding NS values near the horizon, while the rate at which this
happens is again controlled by \(\tau_0\). This can be seen in the right
panel of Fig. \ref{fig:doublepi}, where models with different \(\tau_0\)
show different behaviour. In particular, for the models
\textit{low-$\zeta$}, \textit{medium-$\zeta$}, and \textit{high-$\zeta$},
the bulk-viscosity pressure reaches nearly \(\sim 80 \%\) of the NS
value, while the corresponding value for the
\textit{high-$\tau_{_{\Pi}}$} model is considerably smaller and of the
order of \(\sim 10 \%\). Note also that the \textit{low-$\tau_{_{\Pi}}$}
model reaches a maximum of \(\sim 92 \%\), but not exactly at the
horizon; this is most likely a behaviour due to a cancellation error near
the horizon for the solution obtained by \texttt{BHAC}, since it is
absent in the initial solution.

As a concluding remark we note that there are analogies between the
late-time behaviour realised in longitudinally expanding fluids, such as
the Bjorken flow, and the near-horizon properties of the accretion
solution considered here. In both cases, in fact, the solution tends to
the corresponding NS value (for late times in the case of the Bjorken
flow and for $r \sim 2\,M$ in the case of accretion). This behaviour
suggests that while the parameter \(\zeta_0\) controls the magnitude of
first-order non-equilibrium effects -- which in the case of the accretion
develop mostly in strong-gravity regions -- \(\tau_0\) controls the
degree to which \(\Pi\) approaches its NS value. Of course, these
considerations are based on the examination of the simplest form of a
bulk viscosity. A more extensive investigation of possible initial
conditions, transport coefficients and additional source terms in the
bulk-viscosity pressure equation will yield a deeper insight into the
accretion process of viscous matter that, as pointed out by
\citet{Turolla1989}, is far from being trivial.

%------------------------------------------------------------------------
%------------------------------------------------------------------------
\section{Summary and Conclusion}
\label{sec:conclusions}
%------------------------------------------------------------------------
%------------------------------------------------------------------------

After having reviewed the various approaches developed over the years to
model relativistic dissipative fluids, we have derived a
general-relativistic 3+1 flux-conservative formulation of the
second-order dissipative-hydrodynamics equations first suggested by
\citet{Israel76} and \citet{Hiscock1983}, \ie the \textbf{HL83} set of
GRDHD equations. The new set of equations provides a comprehensive and
complete way of including causal dissipative effects in
general-relativistic calculations.

Although a 3+1 formulation of a reduced version of the \textbf{HL83}
equations was already proposed by \citet{Peitz1997} and
\citet{Peitz1999}, the work presented here extends the results of Peitz
and Appl in three important ways. First, our set of equations is complete
and does not neglect terms that are considered to play a less significant
role\footnote{In \citet{Peitz1997} and \citet{Peitz1999}, all terms
including products of dissipative currents, \ie $\Pi$, $\boldsymbol{q}$
and $\boldsymbol{\pi}$, and first-order gradients of the primary fluid
variables are set to zero in Eqs. (\ref{eq:constitutive_full1}) --
(\ref{eq:constitutive_full3}).} (see Appendix \ref{sec:review} for a
comparison of the system presented here with other formulations). Second,
the equations presented are cast into a flux-conservative form suitable
for numerical implementation. Finally, also the coupling terms between
the different dissipative currents are rewritten in a 3+1 form. As a result,
the full system can now be readily implemented in modern
numerical-relativity codes and evolved numerically.

As a way to test the new set of equations, we have proceeded with the
implementation in the GRMHD code \texttt{BHAC} of a reduced version of
the equations describing fluids with zero shear and heat currents, and
used them against a number of tests in flat and curved background
metrics.

In the first case, we have first considered the one-dimensional,
longitudinally boost-invariant motion of a viscous fluid, such as the one
produced in an ultrarelativistic collision of two ions. This expansion,
first proposed by Bjorken for an inviscid fluid, is a standard testbed
and can be solved analytically. Overall, we find that the numerical
solutions agree well with the analytical solutions, with a relative
difference that is $\sim10^{-7}$ for all the values of viscosity
considered. As an additional flat-spacetime test, we have explored the
solution of a shock-tube problem for an ultra-relativistic gas of gluons
for different values of the ratio $\zeta/s$, with $\zeta$ and $s$ being
bulk-viscosity coefficient and the entropy density, respectively. Also in
this case, when comparing the solutions with those obtained with methods
based on the direct solution of the relativistic Boltzmann equation, we
find a very good agreement up to a ratio of \( \zeta/s \lesssim 0.133\),
which is already in a regime where a dissipative-hydrodynamics framework
breaks down.

Finally, as a general-relativistic test we have considered for the first
time the problem of stationary, spherically symmetric accretion of bulk
viscous fluids onto nonrotating black holes within second-order
dissipative hydrodynamics. Starting from initial conditions obtained from
the solution of a non-trivial set of ODEs (see Appendix
\ref{sec:vis_acc_details} for details), we evolved the fluid with
different values of the bulk-viscosity coefficient and of the relaxation
time. Given the non-triviality of this testbed, we recommend it as a
standard benchmark for those codes wishing to include dissipative effects
in the general-relativistic modelling of compact objects. Overall, we
found that the solution obtained by \texttt{BHAC} can deviate from the
corresponding inviscid solution with differences \( \lesssim 19 \%\) for
the temperature and \(\sim 200 \%\) for the sound speed. In addition, the
bulk-viscosity pressure is highly sensitive to the bulk-viscosity
coefficient, exhibiting deviations of up to three orders of magnitude
near the event horizon depending on whether the viscosity is large or
small. We also showed that although \texttt{BHAC} is able to maintain a
quasi-stationarity in the solution with a global second-order convergence
it does not converge to the reference solution with increasing grid
resolution because of the influence of a finite-size computational
domain. Interestingly, we note analogies between the late-time behaviour
in the Bjorken and accretion flows in the sense that in both cases the
solutions tend to the corresponding NS values, with this happening at
late times in the Bjorken flow, and near the horizon in the case of
accretion onto a black hole.

As a concluding remark we note that although the 3+1 formulation
presented here, and its corresponding discretisation, offers a viable
path to the inclusion of non-equilibrium effects in general-relativistic
simulations of compact objects, short relaxation times as well as 
the required temporal and spatial discretization of the source terms 
in the full system may lead to stiff equations, whose solution will 
not be feasible with simple explicit schemes. We leave the examination 
of more sophisticated numerical techniques for future work, where 
mixed implicit-explicit (IMEX) time integrators 
\citep[see, \eg][]{Palenzuela:2008sf, Dionysopoulou:2012pp, Weih2020b} 
will be considered.

%------------------------------------------------------------------------
%------------------------------------------------------------------------
\section*{Acknowledgements}
%------------------------------------------------------------------------
%------------------------------------------------------------------------

We thank Masoud Shokri, Elias Most, Hector Olivares and Lukas Weih for
useful discussions. Support comes in part from HGS-HIRe for FAIR; the
LOEWE-Program in HIC for FAIR; ``PHAROS'', COST Action CA16214; the ERC
Synergy Grant ``BlackHoleCam: Imaging the Event Horizon of Black Holes''
(Grant No. 610058); the Deutsche Forschungsgemeinschaft (DFG, German
Research Foundation) through the CRC-TR 211 ``Strong-interaction matter
under extreme conditions'' - project number 315477589 - TRR 211.

%------------------------------------------------------------------------
%------------------------------------------------------------------------
\section*{Data availability}
%------------------------------------------------------------------------
%------------------------------------------------------------------------

The data underlying this article will be shared on request to the
corresponding author.

%%%%%%%%%%%%%%%%%%%%%%%%%%%%%%%%%%%%%%%%%%%%%%%%%%

%%%%%%%%%%%%%%%%%%%% REFERENCES %%%%%%%%%%%%%%%%%%

\bibliographystyle{mnras}
\bibliography{aeireferences}

%%%%%%%%%%%%%%%%%%%%%%%%%%%%%%%%%%%%%%%%%%%%%%%%%%

%%%%%%%%%%%%%%%%% APPENDICES %%%%%%%%%%%%%%%%%%%%%

\appendix

%------------------------------------------------------------------------
%------------------------------------------------------------------------
\section{Review of second-order relativistic dissipative hydrodynamics}
\label{sec:review}
%------------------------------------------------------------------------
%------------------------------------------------------------------------

This appendix provides a guidance in comparing our reference second-order
formulation of general-relativistic dissipative hydrodynamics originally
suggested by \citet{Hiscock1983} with other formulations that have
appeared in the literature, \ie the formulations by \citet{Israel79_new}
(hereafter \textbf{IS79}), by \citet{Denicol_2012} (hereafter
\textbf{DMNR12})\footnote{Note that there is another frequently used
second-order SRDHD formulation proposed by \citet{Denicol2012b} and often
referred to as the ``DNMR'' formulation. We will not use this formulation
for our comparison here since the formulation by \citet{Denicol2012b} is
in the Landau frame and we instead consider, if possible, the
formulations adopting the Eckart frame for our comparison.}, and by
\citet{Baier2008} (hereafter \textbf{rBRSSS08}). Since these formulations
often adopt different notations that make it hard to compare them, we
introduce a generalized notation for the various transport coefficients,
whose terminology is given in Table
\ref{tab:transport_coefficients}. Furthermore, since the equations of
\textbf{HL83} are chosen as our reference equations, we use upper-case
letters for transport coefficients appearing in \textbf{HL83}; the only
exception to this rule will be made for the relaxation times.
Furthermore, since the comparison needs to distinguish terms of first and
second order, we introduce the following dimensionless numbers: the
Knudsen number
\begin{equation}
\mathrm{Kn}:=\frac{\ell_{\mathrm{micro}}}{L_{\mathrm{macro}}}\,,
\end{equation}
where $\ell_{\mathrm{micro}}$ is the microscopic lengthscale given by the
mean-free-path of the microscopic constituents of the fluid, while
$L_{\mathrm{macro}}$ is the macroscopic lengthscale over which gradients
of the primary fluid variables appear. Furthermore, we introduce the
following inverse Reynolds numbers given by
\begin{equation}
\mathrm{R}_{\Pi}^{-1}:=\frac{|\Pi|}{p+e}, \quad \mathrm{R}_{n}^{-1}:=\frac{|n^{\mu}|}{n}, 
\quad \mathrm{R}_{\pi}^{-1}:=\frac{|\pi^{\mu \nu}|}{p+e}\,,\label{eq:DMNR_reynolds}
\end{equation}
where in the Eckart frame $\mathrm{R}^{-1}_{n}$ is replaced by
\begin{equation}
\mathrm{R}^{-1}_{q}:= \frac{|q^{\mu}|}{p+e} \sim \mathcal{O}\left(R^{-1}_{n}\right)\,,
\end{equation}
Using these dimensionless numbers, we can identify the order of the terms
given in Table \ref{tab:comparison} according to the following
classification:
\begin{align}
  \label{eq:classification1}
  \mathcal{O}\left(\mathrm{R}^{-1}_{i},\mathrm{Kn}\right) := \mathcal{O}_1\,,\quad
  \mathcal{O}\left(\mathrm{R}^{-1}_{i}\mathrm{Kn}\right) := \mathcal{O}_{_{\mathrm{RK}}}\,,\quad
  \mathcal{O}\left(\mathrm{R}^{-1}_{i}\mathrm{R}^{-1}_{j}\right) := \mathcal{O}_{_{2\mathrm{R}}}\,,\quad
  \mathcal{O}\left(\mathrm{Kn}^2\right) := \mathcal{O}_{_{2\mathrm{K}}}\,.
\end{align}
In this way, following the convention of \citet{Denicol2012b}, we refer
to terms of order \(\mathcal{O}_1\) as being of first order, while all
other terms are referred to as being of second order.

With these definitions made, we next proceed to the actual comparison
which will take place by first briefly reviewing each of the formulations
considered (Sec. \ref{app:IS79}--\ref{app:BRSSS08}) and then proceeds with the
actual comparison (Sec. \ref{app:grdhd_comparison}).

\begin{table}
\begin{center}
\begin{tabular}{|c|l|}
\hline \textbf{transport coefficients}  & \textbf{description}\\
\hline
\(\tau_{_{\Pi}},\tau_{q},\tau_{\pi}\) & relaxation times for bulk viscosity, heat conduction and shear viscosity, respectively  \\
%\hline
\(\delta^{i}_{j}\) or \(\Delta^{i}_{j}\) & transport coefficients for gradients of the fluid velocity \(\boldsymbol{u}\)\\
%\hline
\(l^{i}_{j}\) or \(L^{i}_{j}\) & transport coefficients for gradients of the dissipative currents\\
%\hline
\(\lambda^{i}_{j}\) or \(\Lambda^{i}_{j}\) & transport coefficients for the contribution of gradients of  \(\rho\) and \(p\)\\
%\hline
\(\varphi^{i}_{j}\) & transport coefficients for the contribution of contracted dissipative currents\\
%\hline
\(g^{i}_{j}\) & transport coefficients in front of geometrical quantities, \ie \(R_{\mu \nu \lambda \rho}\), \(R_{\mu \nu}\) or \(R\)\\
\hline
\end{tabular}
\caption{Generalized notation for the transport coefficients. Shown on
  the left are the newly introduced symbols, while a description of which
  fluid field it appears in combination with is shown on the right. The
  upper index indicates which dissipative current the transport
  coefficient belongs to, \eg \(\delta^{^\Pi}_{2}\) is the second
  transport coefficient expressing the coupling to a gradient of the
  fluid velocity in the constitutive equation for the bulk-viscosity
  pressure. Upper-case letters indicate that the corresponding transport
  coefficient occurs in the \textbf{HL83} formulation, while lower case
  letters indicate that it is set to zero in \textbf{HL83}.}
\label{tab:transport_coefficients}
\end{center}
\end{table}

%------------------------------------------------------------------------
\subsection{Israel and Stewart 1979 (IS79)}
\label{app:IS79}
%------------------------------------------------------------------------

The equations for the dissipative currents of \textbf{IS79} read as
follows
\begin{align}
\tau_{_{\Pi}} ~ \dot{\Pi} &= \Pi_{\lns} -\Pi
+\delta^{^\Pi}_{1} ~ q^{\mu}a_{\mu}
+L^{^\Pi}_1 ~ \nabla_{\mu}q^{\mu}\,,\label{eq:IS79-1}\\
\tau_{\mathrm{q}} ~ \dot{q}^{\left\langle \mu \right\rangle} &= q_{\ns}^{\phantom{\ns}\mu}
- q^{\mu}
+\delta^{q}_{1} ~ \Pi a^{\mu}
+\delta^{q}_{2} ~ \pi^{\mu \nu}a_{\nu}
+\delta^{q}_{3} ~ \omega^{\mu \nu}q_{\nu}
+ L^{q}_{1} ~ \nabla^{\left\langle \mu \right\rangle}\Pi
+L^{q}_{2} ~ \nabla_{\nu}\pi^{\left\langle \mu \right\rangle \nu }\,,\label{eq:IS79-2} \\
\tau_{\pi} ~ \dot{\pi}^{\left\langle \mu \nu \right\rangle}  &= \pi_{\lns}^{\phantom{\lns}\mu \nu} 
- \pi^{\mu \nu}
+\delta^{\pi}_{1} ~ q^{\left\langle\mu \right.}a^{\left. \nu \right\rangle}
+\delta^{\pi}_{2} ~ \pi^{\lambda \left\langle \mu \right.}{\omega^{\left. \nu \right\rangle}}_{\lambda}
+ L^{\pi}_1 ~ \nabla^{\left\langle \mu \right.} q^{\left. \nu \right\rangle}\,.\label{eq:IS79-3}
\end{align}
Here, \(\Pi_{\lns}\) and \(\pi_{\lns}^{\phantom{\lns}\mu \nu}\) denote
the NS values of the dissipative currents evaluated in the Landau
frame. Equations (\ref{eq:IS79-1}), (\ref{eq:IS79-2}) and
(\ref{eq:IS79-3}) correspond to Eqs. (7.1a), (7.1b) and (7.1c) of
\citet{Israel79_new}, respectively. This set of equations was derived
using the Boltzmann equation for the single-particle distribution
function $f$, parametrized in the form
\begin{align}
  y(x^{\mu},p^{i})&:=\ln\left[\frac{f\left(x^{\mu},p^{i}\right)}{\mathrm{A}_1}\right]
  \,,\\ y(x^{\mu},p^{i})&:= \frac{\mu}{T} +
  \epsilon+\left(\frac{u_{\lambda}}{T} +\frac{\epsilon_{\lambda}}{m}
  \right) p^{\lambda}+\frac{1}{m^2}\epsilon_{\lambda
    \rho}p^{\lambda}p^{\rho}\,,\\ \mathrm{A}_1&:= 1+\mathrm{A}_2
  f\left(x^{\mu},p^{i}\right), \qquad \mathrm{A}_2 \in \{-1,0,+1\}\,,
\end{align}
where $p^0= p^0 (p^i,m)$ is the on-shell energy of the particles,
$\mathrm{A_2}=-1 (+1)$ refers to fermions (bosons) so that
\(\mathrm{A}_2\rightarrow 0\) corresponds to the limit yielding the
Boltzmann distribution, and \(m\) is the rest-mass of the particles.

The functions \(\epsilon\), \(\epsilon_{\lambda}\) and
\(\epsilon_{\lambda \rho}\) are off-equilibrium corrections and we
recover the standard Fermi, Bose, and Boltzmann local-equilibrium
distributions for \(\epsilon, \epsilon_{\lambda}, \epsilon_{\lambda
  \rho}=0\). The single-particle distribution $f$ contains 14 independent
variables ($T, \mu$, the three independent components of $u^\mu$ and the
nine independent components of the functions \(\epsilon\),
\(\epsilon_{\lambda}\) and \(\epsilon_{\lambda \rho}\)). These are
one-to-one matched to the components of $J^\mu$ and $T^{\mu \nu}$. The
system of conservation equations, corresponding to the equations of
motion for the first and second moment of $f$, is closed by employing the
equation of motion for the third moment of $f$, leading to
relaxation-type equations for the dissipative currents. The fact that 14
independent variables occur in the equations of motion is also referred
to as the \textit{``14-moment approximation''}.

We should recall that in the derivation of \textbf{IS79} two
approximations were made. First, they neglected second-order terms
proportional to $\nabla^{\mu} \rho$ and $\Theta$. Following the
derivation from the second law of thermodynamics, these terms arise from
gradients of the transport coefficients and are generally of the form
(see, \eg the equation for the shear-stress tensor) \(\propto \pi^{\mu
  \nu}u_{\lambda}\overline{I}^{\lambda}\) and \(\propto q^{\langle
  \mu}\overline{I}^{\nu \rangle}\), where \(\overline{I}^{\mu}\) is a
linear combination of gradients of transport coefficients that can be
related to gradients of the chosen pair of thermodynamical variables
\(\{T, \mu \}\) or \(\{\rho, p \}\). In general, these terms can be
further decomposed by using the first-order version of the conservation
laws (\ref{eq:continuity}) and (\ref{eq:em_conservation}):
\begin{align}
\dot{\rho} &\propto \Theta \,,\label{eq:firstorder1}\\
a^ {\mu} &\propto \nabla^{\left\langle \mu \right\rangle}p\,,\label{eq:firstorder2}\\  
\dot{e} &\propto \Theta \label{eq:firstorder3}\,.
\end{align}
By choosing $p$ and $\rho$ as our thermodynamical variables and
exploiting Eqs. (\ref{eq:firstorder1}) -- (\ref{eq:firstorder3}), the
following relations are valid to second-order
\begin{align}
\pi^{\mu \nu}u_{\lambda}\overline{I}^{\lambda} &\propto \pi^{\mu \nu}\Theta \,,\\
q^{\langle \mu}\overline{I}^{\nu \rangle} &\propto  b_1 q^{\langle \mu}a^{\nu \rangle} 
+ b_2 q^{\langle \mu}\nabla^{\nu \rangle}\rho \,,
\end{align}
where \(b_1\) and \(b_2\) denote scalar functions of $p$ and
$\rho$. Hence, the missing terms in \textbf{IS79} proportional to
$\nabla^{\mu} \rho$ and $\Theta$ are of the form \(\propto \pi^{\mu
  \nu}\Theta\) and \(\propto q^{\langle \mu}\nabla^{\nu \rangle}\rho\),
while the term \(b_1 q^{\langle \mu}a^{\nu \rangle}\) could be absorbed
in the transport coefficient \(\delta_1^{\pi}\).

Second, there are terms of the type \(\propto \pi^{\lambda \langle
  \mu}{\sigma^{\nu \rangle}}_{\lambda}\) and \(\propto \Pi \sigma^{\mu
  \nu}\), appearing for instance in the equation for the shear-stress
tensor that are also missing in the \textbf{IS79} formulation. The
existence of these terms was first pointed out by \citet{Betz:2010cx},
which were subsequently included in the \textbf{DMNR12} formulation (see
below).

%------------------------------------------------------------------------
\subsection{Hiscock and Lindblom 1983 (HL83)}
\label{app:HL83}
%------------------------------------------------------------------------

When using a generalized notation, the evolution equations
(\ref{eq:constitutive_full1})--(\ref{eq:constitutive_full3}) for the
dissipative currents in the \textbf{HL83} formulation can be written as
\begin{align}
\tau_{_{\Pi}} ~ \dot{\Pi} &= \Pi_{_{\mathrm{NS}}} -\Pi
+\Delta^{^\Pi}_1 ~ \Pi \Theta
+L^{^\Pi}_1 ~ \nabla_{\mu}q^{\mu}
+\Lambda^{^\Pi}_{1} ~ \Pi u_{\mu}{I_1}^{\mu}
+\Lambda^{^\Pi}_{2} ~ q_{\mu}{I_2}^{\mu}\,, 
\label{eq:hl-1}\\ 
\tau_{\mathrm{q}} ~ \dot{q}^{\left\langle \mu \right\rangle} &= q_{_{\mathrm{NS}}}^{\mu}
-q^{\mu}
+ \Delta^{q}_1 ~ q^{\mu}\Theta
+ L^{q}_{1} ~ \nabla^{\left\langle \mu \right\rangle} \Pi
+ L^{q}_{2} ~ \nabla_{\nu}\pi^{\left\langle \mu \right\rangle \nu}
+ \Lambda^{q}_{1} ~ q^{\mu}u_{\nu}{I_3}^{\nu}  \notag \\
&\phantom{=}+ \Lambda^{q}_{2} ~ \Pi {I_2}^{\left\langle \mu \right\rangle}
+ \Lambda^{q}_{3} ~ {\pi^{\mu}}_{\nu}{I_4}^{\nu}\,, \label{eq:hl-2}\\
\tau_{\pi} ~ \dot{\pi}^{\left\langle \mu \nu \right\rangle} &= \pi_{_{\mathrm{NS}}}^{\mu \nu}
-\pi^{\mu \nu}
+\Delta^{\pi}_1 ~ \pi^{\mu \nu} \Theta
+L^{\pi}_1 ~ \nabla^{\left\langle \mu \right.}q^{\left. \nu \right\rangle}
+\Lambda^{\pi}_{1} ~ \pi^{\mu \nu}u_{\lambda}{I_5}^{\lambda}
+\Lambda^{\pi}_{2} ~ q^{\langle \mu}{I_4}^{\nu \rangle}\,.
\label{eq:hl-3}
\end{align}
where the currents \({I_{i}}^{\mu}\) are given by
\begin{align}
& {I_1}^{\mu}:=\nabla^{\mu}\left({\tau_{_{\Pi}}}/{\zeta T}\right)\,,\\
& {I_2}^{\mu}:=\nabla^{\mu}\left({\alpha_0}/{T}\right)\,,\\
& {I_3}^{\mu}:=\nabla^{\mu}\left({\tau_{\mathrm{q}}}/{\kappa T^2}\right)\,,\\
& {I_4}^{\mu}:=\nabla^{\mu}\left({\alpha_1}/{T}\right)\,,\\
& {I_5}^{\mu}:=\nabla^{\mu}\left({\tau_{\pi}}/{\eta T}\right)\,.
\end{align}

In similarity with Sec. \ref{app:IS79}, we can use the first-order
relations (\ref{eq:firstorder1})--(\ref{eq:firstorder3}) to further
decompose the terms in (\ref{eq:hl-1})--(\ref{eq:hl-3}) that include the
currents \({I_{i}}^{\mu}\) and obtain
\begin{align}
\Pi u_{\mu}{I_1}^{\mu} &\propto \Pi \Theta \,,\label{eq:redHL1}\\
q_{\mu} {I_2}^{\mu} &\propto b_3 q_{\mu}a^{\mu} + b_4 q_{\mu}\nabla^{\mu}\rho\,,\\
q^{\mu} u_{\nu}{I_3}^{\nu} &\propto q^{\mu} \Theta \,,\\
\Pi {I_2}^{\langle \mu \rangle } &\propto b_4 \Pi a^{\mu} + b_5 \Pi \nabla^{\langle \mu \rangle }\rho\,,\\
{\pi^{\mu}}_{\nu}{I_4}^{\nu} &\propto b_6 {\pi^{\mu}}_{\nu}a^{\nu} + b_7 {\pi^{\mu}}_{\nu} \nabla^{\nu}\rho\,,\\
\pi^{\mu \nu}u_{\lambda}{I_5}^{\lambda} &\propto \pi^{\mu \nu}\Theta\,,\\
q^{\langle \mu}{I_4}^{\nu \rangle} &\propto  b_8 q^{\langle \mu}a^{\nu \rangle} 
+ b_9 q^{\langle \mu}\nabla^{\nu \rangle}\rho \,,\label{eq:redHL7}
\end{align}
where, again, the quantities \(b_3\textrm{-}b_9\) are scalar functions of
\(\rho\) and \(p\).

%------------------------------------------------------------------------
\subsection{Denicol et al.\ 2012a (DMNR12)}
\label{app:DNMR12}
%------------------------------------------------------------------------

When the Eckart frame is chosen as the frame of reference, \ie when
\(V^{\mu}=0\) and \(W^{\mu}=q^{\mu}\), the equations for the dissipative
currents of the \textbf{DMNR12} formulation are
\begin{align}
\tau_{_{\Pi}} ~ \dot{\Pi} &= \Pi_{_{\mathrm{NS}}} - \Pi
+ \Delta^{^\Pi}_1 ~ \Pi \Theta
+ \delta^{^\Pi}_{1} ~ q^{\mu}a_{\mu}
+ \delta^{^\Pi}_{2} ~ \pi^{\mu \nu}\sigma_{\mu \nu}
+ L^{^\Pi}_1 ~ \partial_{\mu}q^{\mu}  
+ \Lambda^{^\Pi}_{2} ~ q_{\mu}{I_{6}}^{\mu} \,,\label{eq:DMNR1}\\
\tau_{\mathrm{q}} ~ \dot{q}^{\left\langle \mu \right\rangle} &= q_{_{\mathrm{NS}}}^{\mu}
- q^{\mu}
+ \Delta^{q}_1 ~ q^{\mu}\Theta
+ \delta^{q}_{1} ~ \Pi a^{\mu}
+ \delta^{q}_{2} ~ \pi^{\mu \nu}a_{\nu}  
+ \delta^{q}_{3} ~ \omega^{\mu \nu}q_{\nu} 
+ \delta^{q}_{4} ~ \sigma^{\mu \nu}q_{\nu} \notag \\
&\phantom{=}+ L^{q}_{1} ~ \partial^{\left\langle \mu \right\rangle} \Pi  
+ L^{q}_{2} ~ \partial_{\nu}\pi^{\left\langle \mu \right\rangle \nu } 
+ \Lambda^q_{2} ~ \Pi {I_6}^{\left\langle \mu \right\rangle}
+ \Lambda^q_{3} ~ {\pi^{\mu}}_{\nu}{I_6}^{\nu}\,,\label{eq:DMNR2}\\
\tau_{\pi} ~ {\dot{\pi}}^{\left\langle \mu \nu \right\rangle}  &= \pi_{_{\mathrm{NS}}}^{\mu \nu} - \pi^{\mu \nu}
+ \Delta^{\pi}_1 ~ \pi^{\mu \nu} \Theta 
+\delta^{\pi}_{1} ~ q^{\left\langle \right. \mu}a^{\nu \left. \right\rangle}
+ \delta^{\pi}_{2} ~ \pi^{\lambda \left\langle \right. \mu}{\omega^{\nu \left. \right\rangle}}_{\lambda} 
+\delta^{\pi}_{3} ~ \pi^{\lambda \left\langle \right. \mu}{\sigma^{\nu
    \left. \right\rangle}}_{\lambda} \notag \\
&\phantom{=}+ \delta^{\pi}_{4} ~ \Pi \sigma^{\mu \nu}+  L^{\pi}_1 ~ \partial^{\left\langle \right. \mu}q^{\nu \left. \right\rangle} 
+ \Lambda^{\pi}_{2} ~ q^{\left\langle \mu \right.}{I_6}^{\left. \nu \right\rangle}\,.\label{eq:DMNR3}
\end{align}

Equations (\ref{eq:DMNR1}), (\ref{eq:DMNR2}) and (\ref{eq:DMNR3})
correspond to Eqs. (128), (138) and (153) of \cite{Denicol_2012},
respectively. Note that the terms including the current ${I_6}^{\mu}$ can
be decomposed as
\begin{align}
q_{\mu}{I_6}^{\mu} &\propto b_{10} q_{\mu}a^{\mu} + b_{11}
q_{\mu}\nabla^{\mu} \rho\,,\label{eq:redDMNR1}\\ \Pi {I_6}^{\langle \mu
  \rangle } &\propto b_{10} \Pi a^{\mu} + b_{11} \Pi \nabla^{\langle \mu
  \rangle }\rho\,,\\ {\pi^{\mu}}_{\nu}{I_6}^{\nu} &\propto b_{10}
{\pi^{\mu}}_{\nu}a^{\nu} + b_{11} {\pi^{\mu}}_{\nu}
\nabla^{\nu}\rho\,,\\ q^{\langle \mu}{I_6}^{\nu \rangle} &\propto b_{10}
q^{\langle \mu}a^{\nu \rangle} + b_{11} q^{\langle \mu}\nabla^{\nu
  \rangle}\rho \,,\label{eq:redDMNR4}
\end{align}
where \({I_6}^{\alpha}:=\partial^{\alpha}\left({\mu}/{T}\right)\) and
\(b_{10}\), \(b_{11}\) are scalar functions of \(\rho\) and \(p\). We
should remark that Eqs. (\ref{eq:DMNR1})--(\ref{eq:DMNR3}) have been
derived for a flat spacetime with signature $(+,-,-,-)$\footnote{This
choice introduces a sign difference in some quantities, \eg
$\pi^{\phantom{\ns}\mu \nu}_{\ns}$, which are instead computed with the
signature $(-,+,+,+)$.}, so that the comoving derivative is
\(\boldsymbol{\dot{A}} = u^{\mu}\partial_{\mu}\boldsymbol{A}\). Finally,
note that in the derivation of the \textbf{DMNR12} formulation, the full
off-equilibrium distribution function is decomposed as
\begin{align}
f(x^{\mu},p^{i})=f_0(x^{\mu},p^{i})+\delta f(x^{\mu},p^{i})\,,\label{eq:DMNRf}
\end{align}
where \(f_0\) is the local-equilibrium distribution function and \(\delta
f\) denotes the deviation from it. One then defines the so-called
generalized irreducible moments of order $r$ of \(\delta f\) as 
\begin{equation}
\rho_r^{\alpha_1 \cdots \alpha_l} :=\int\frac{g  d^3 p}{(2\pi)^3 p^0}(E)^r
p^{\langle \alpha_1}\cdots p^{\alpha_l \rangle} \delta
f\,,\label{eq:irredmoments}
\end{equation}
where \(g\) is the number of internal degrees of freedom and \(E\) is
defined by \(p^{\mu} =: Eu^{\mu} + p^{\left\langle \mu
  \right\rangle}\). The full off-equilibrium distribution function can
now be expanded in momentum space in a basis of irreducible tensors
$p^{\langle \alpha_1} \cdots p^{\alpha_l \rangle}$ and orthogonal
polynomials in $E$, where the generalized irreducible moments appear as
coefficients. Some of these moments are directly connected to the
dissipative currents, \eg
\begin{align}
\Pi &= -\frac{m^2}{3}\rho_0= -\frac{m^2}{3}\int\frac{g d^3
  p}{(2\pi)^3p^0}\delta f\,,\label{eq:DMNRpi}\\ q^{\mu}&= \rho^{\mu}_1=
\int\frac{g d^3 p}{(2\pi)^3p^0}Ep^{\mu}\delta
f,\label{eq:DMNRq}\\ \pi^{\mu \nu}&=\rho^{\mu \nu}_0=\int\frac{g d^3
  p}{(2\pi)^3p^0}p^{\langle \mu}p^{\nu \rangle}\delta f\,.
\label{eq:DMNRlpi}
\end{align}

Inserting Eq. (\ref{eq:DMNRf}) into the special-relativistic version of
the Boltzmann equation with collision term $\mathcal{C}[f]$ leads to an
equation of motion for $\delta f$,
\begin{equation}
\delta\dot{f}=-\dot{f}_0-\frac{1}{E}p^{\langle \mu
  \rangle}\partial_{\mu}\left(f_0+\delta f\right) +
\frac{1}{E}\mathcal{C}[f]\,,
\label{eq:boltzmann_DMNR}
\end{equation}
which can be used to evaluate the comoving derivatives of the generalized
irreducible moments (\ref{eq:irredmoments}) and leads to an infinite set
of evolution equations for the latter. When this infinite system is
truncated at the lowest order -- corresponding to the 14-moment
approximation -- it leads to \textbf{IS79}-like equations, namely,
Eqs. (\ref{eq:DMNR1})--(\ref{eq:DMNR3}).

Note that this procedure is ambiguous because once the 14-moment
approximation is applied, it is possible to obtain an evolution equation
of the desired dissipative current from any choice of $r$ in
Eq. (\ref{eq:irredmoments}), for the irreducible moment of tensor rank
corresponding to that of the respective dissipative
current\footnote{Stated differently, it is possible to derive a
relationship between the dissipative currents $\Pi, q^\mu$, and
$\pi^{\mu\nu}$, which are essentially the irreducible moments $\rho_0,
\rho_1^\mu$, and $\rho_0^{\mu \nu}$, and all the other $\rho_r,
\rho_r^\mu$, and $\rho_r^{\mu\nu}$, for any index $r$. As a result, there
is ambiguity because it is possible to derive an equation of motion for
$\Pi$, etc. from the equation of motion for $\rho_r$, etc. for any $r$
and not only for $r=0$.}. This can be seen from the $r$-dependence of the
transport coefficients in Tables \ref{tab:transport1} and
\ref{tab:transport2}, which reflects the choice of the moment equation.
The transport coefficients of \textbf{IS79} are obtained if one sets
\(r=3\) for the scalar moment, \(r=2\) for the vector moment and \(r=1\)
for the tensor moment. The microphysical properties of the system are
encoded in the transport coefficients obtained from a moment expansion of
the collision integral of the Boltzmann equation. Instead of applying
the 14-moment approximation, in \citet{Denicol2012b} all moments of a
given tensor rank were resummed, removing the aforementioned
ambiguity. Subsequently, a power-counting scheme in Knudsen and inverse
Reynolds numbers was used to truncate the system of equations of motion
for the irredubible moments, in order to finally obtain the equations of
motion for the dissipative currents.

%------------------------------------------------------------------------
\subsection{Baier et al.\ 2008 (rBRSSS08)}
\label{app:BRSSS08}
%------------------------------------------------------------------------

The equations for the dissipative currents of the \textbf{rBRSSS08}
formulation read as follows [here we use the version presented by
  \citet{romatschke_romatschke_2019}, where they appear as Eq. 
(2.122)]\footnote{In \citet{Baier2008} the equations for the dissipative currents 
have been derived for the case of conformal fluids. Here we present their 
complete extension to the case of non-conformal fluids.}
\begin{align}
\tau_{_{\Pi}} ~ \dot{\Pi} &= \Pi_{_{\mathrm{NS}}} - \Pi
+\delta^{^\Pi}_{3} ~ \omega^{\mu \nu}\omega_{\mu \nu}
+\lambda^{^\Pi}_1 ~ {I_7}_{\left\langle \mu \right\rangle}{I_7}^{\left\langle \mu \right\rangle}
+\varphi^{^\Pi}_{1} ~ \pi^{\mu \nu}\pi_{\mu \nu} 
+ \varphi^{^\Pi}_{2} ~ \Pi^2  \notag \\ 
&\phantom{=}+g^{^\Pi}_1 ~ R
+g^{^\Pi}_2 ~ u^{\mu}u^{\nu}R_{\mu \nu}\,,\label{eq:RR1}\\
\tau_{\pi} ~ {\dot{\pi}}^{\left\langle \mu \nu \right\rangle}  &= \pi_{_{\mathrm{NS}}}^{\mu \nu} 
- \pi^{\mu \nu}
+ \Delta^{\pi}_1 ~  \pi^{\mu \nu}\Theta
+ \delta^{\pi}_{2} ~ \pi^{ \lambda \left\langle \right. \mu }{\omega^{\nu \left. \right\rangle}}_{\lambda}
+\delta^{\pi}_{5} ~  {\omega^{\left\langle \right. \mu}}_{\lambda}{\omega^{\nu \left. \right\rangle \lambda}}
+ \lambda^{\pi}_1 ~ {I_7}^{\left\langle \mu \right.}{I_7}^{\left. \nu \right\rangle} \notag \\
&\phantom{=}+ \varphi^{\pi}_1 ~ \pi^{\lambda \left\langle \right. \mu} {\pi^{\nu \left. \right\rangle}}_{\lambda}
+ g^{\pi}_1 ~  R^{\left\langle \mu \nu \right\rangle} 
+ g^{\pi}_2 ~  u_{\lambda}u_{\rho}R^{\lambda \left\langle \mu \nu \right\rangle \rho }
\,.\label{eq:RR2}
\end{align}
Note that we have defined \({I_7}^{\mu}:=\nabla^\mu \ln e\) and that the
heat current is absent because these expressions refer to the Landau
frame as the reference frame, where the heat currents are zero by
definition.

As for the previous formulations, we can use the first-order relations
(\ref{eq:firstorder1})--(\ref{eq:firstorder3}) to further decompose the
terms in Eqs. (\ref{eq:RR1}), (\ref{eq:RR2}) that include the current
\({I_{7}}^{\mu}\) as
\begin{align}
{I_7}_{\left\langle \mu \right\rangle}{I_7}^{\left\langle \mu \right\rangle} &=
{I_7}_{\mu}{I_7}^{\mu} + u_{\mu}{I_7}^{\mu}u_{\nu}{I_7}^{\nu} \notag\\
&\propto (b_{12}a_{\mu}+b_{13}\nabla_{\mu} \rho )
(b_{12}a^{\mu}+b_{13}\nabla^{\mu} \rho) + b_{14}\Theta^2 \notag \\ 
&\propto ({b_{12}})^2
a_{\mu}a^{\mu} + 2b_{12}b_{13}a^{\mu}\nabla_{\mu}\rho +
({b_{13}})^2\nabla_{\mu}\rho \nabla^{\mu}\rho  + b_{14}\Theta^2 
\,,\label{eq:redRR1}\\ {I_7}^{\left\langle \mu \right.}{I_7}^{\left. \nu
  \right\rangle} &\propto ({b_{12}})^2 {a}^{\left\langle \mu
  \right.}{a}^{\left. \nu \right\rangle} + 2b_{12}b_{13}{a}^{\left\langle
  \mu \right.}{\nabla}^{\left. \nu \right\rangle}\rho +
({b_{13}})^2{\nabla}^{\left\langle \mu \right.}\rho{\nabla}^{\left. \nu
  \right\rangle}\rho \,,\label{eq:redRR2}
\end{align}
where, again, the quantities $b_{12}, b_{13}$, and $b_{14}$ are scalar
functions of $\rho$ and $p$.

The \textbf{rBRSSS08} formulation is based on a systematic expansion in
terms of gradients of the fluid variables in equilibrium and of the
metric \(g_{\mu \nu}\), hence inheriting purely geometric terms involving
the Ricci tensor $R_{\mu \nu }$ and the Ricci scalar
$R:=R^{\mu}_{\phantom{\mu} \mu }$. The dissipative currents are then
written as a power series in such gradients up to a given order and the
individual terms involved in the decomposition of the shear-stress tensor
are also symmetric, trace-free, and orthogonal to \(\boldsymbol{u}\).

A resummation procedure is then applied to obtain relaxation-type
equations; for example, in the case of the shear-stress tensor, the
first-order relation $\pi^{\mu \nu}=-2\eta \sigma^{\mu \nu}$ is used to
derive
\begin{align} 
u^{\lambda}\nabla_{\lambda} \pi^{\left\langle \mu \nu \right\rangle} \propto 
-2 \eta u^{\lambda}\nabla_{\lambda} \sigma^{\left\langle \mu \nu\right\rangle} -
b_{15} \sigma^{\mu \nu} \Theta \, , \label{eq:BRSSSresumtrick} 
\end{align}
which then leads to hyperbolic equations of motion if the term
$u^{\lambda}\nabla_{\lambda} \sigma^{\left\langle \mu \nu\right\rangle}$
is substituted using (\ref{eq:BRSSSresumtrick}) in the corresponding
parabolic equations of motion. Note that $b_{15}$ is a scalar function of
$\rho$ and $p$ that can be absorbed in the definitions of the transport
coefficients, and that Eq. (\ref{eq:BRSSSresumtrick}) is accurate to
second order in the gradients, so that the acausal behaviour is cured and
causality is recovered.

%------------------------------------------------------------------------
\subsection{Comparing different GRDHD formulations}
\label{app:grdhd_comparison}
%------------------------------------------------------------------------

\begin{table}
  \centering
  \renewcommand{\arraystretch}{1.5}
  \setlength{\tabcolsep}{9pt} % Default value: 6pt
\begin{tabular}{|lccccccc|}
\hline
\hline
\multirow{2}{*}{\textbf{Formulation}}

& \multirow{2}{*}{\textbf{Order}}

& \multicolumn{2}{c|}{\textbf{bulk-pressure equation}} &     \multicolumn{2}{c|}{\textbf{heat-current equation}} & \multicolumn{2}{c}{\textbf{shear-tensor equation}}\\
&

& \textcolor{red}{\Large{$-$}} & \textcolor{blue}{\Large{$+$}} & \textcolor{red}{\Large{$-$}} & \textcolor{blue}{\Large{$+$}} & \textcolor{red}{\Large{$-$}} & \textcolor{blue}{\Large{$+$}}\\
\hline
\multirow{3}{*}{\scriptsize \textbf{IS79} \small}
&
\multirow{3}{*}{${ \scriptstyle \mathcal{O}_{_{\mathrm{RK}}} }$}
&
\textcolor{red}{\( \Pi \Theta \)}  
& 
&
\textcolor{red}{\( q^{\mu} \Theta \)}  
&
\textcolor{blue}{\( \omega^{\mu \nu}q_{\nu}\)}  
&
\textcolor{red}{
\( \pi^{\mu \nu}\Theta \)}  
&
\textcolor{blue}{\( \pi^{\lambda \langle \mu }{\omega^{\nu \rangle}}_{\lambda} \)}  
\\
&
&
\textcolor{red}{\( q_{\mu} {\mathcal{I}}^{\mu} \)}  
&
&
\textcolor{red}{\( \Pi {\mathcal{I}}^{\langle \mu \rangle} \)}  
&
&
\textcolor{red}{\( q^{\langle \mu }{\mathcal{I}}^{ \nu\rangle} \)}  
&
\\
&
&
&
&
\textcolor{red}{\( {\pi^{\mu}}_{\nu}{{\mathcal{I}}^{\nu}} \)}  
&
&
&
\\
\hline
\multirow{3}{*}{\scriptsize \textbf{DMNR12} \small}
&
\multirow{3}{*}{${ \scriptstyle \mathcal{O}_{_{\mathrm{RK}}} }$}
&
&
\textcolor{blue}{\( \pi^{\mu \nu}\sigma_{\mu \nu} \)}  
&
&
\textcolor{blue}{\( \omega^{\mu \nu}q_{\nu}\)}  
&
&
\textcolor{blue}{\( \pi^{\lambda \langle \mu }{\omega^{\nu \rangle}}_{\lambda} \)}  
\\
&
&
&
&
&
\textcolor{blue}{\( \sigma^{\mu \nu}q_{\nu} \)}  
&
&
\textcolor{blue}{\( \pi^{\lambda \langle \mu }{\sigma^{\nu \rangle}}_{\lambda} \)}  
\\
&
&
&
&
&
&
&
\textcolor{blue}{\( \Pi \sigma^{\mu \nu} \)}  
\\
\hline
\multirow{13}{*}{{\scriptsize \textbf{rBRSSS08} }}
&
\multirow{4}{*}{${ \scriptstyle \mathcal{O}_{_{\mathrm{RK}}} }$}
&
\textcolor{red}{
\(  \Pi \Theta \)}  
&
&
\textcolor{red}{N/A}
& 
\textcolor{blue}{N/A}
&
\textcolor{red}{
\( \nabla^{\langle \mu }q^{ \nu \rangle} \)}  
&
\textcolor{blue}{\( \pi^{\lambda \langle \mu }{\omega^{\nu \rangle}}_{\lambda} \)}  
\\
&
&
\textcolor{red}{
\( \nabla_{\mu}q^{\mu} \)}  
&
&
&
&
\textcolor{red}{\( q^{\langle \mu }{a}^{ \nu \rangle} \)}  
& 
\\
&
&
\textcolor{red}{\( {q}_{\mu}{a}^{\mu} \)}  
&
&
&
&
\textcolor{red}{\( q^{\langle \mu }{\mathcal{I}}^{ \nu \rangle} \)}  
&
\\
&
&
\textcolor{red}{\( q_{\mu}{\mathcal{I}}^{\mu} \)}  
&
&
&
&
& 
\\
\cline{2-8}
&
\multirow{5}{*}{${ \scriptstyle \mathcal{O}_{_{2\mathrm{K}}} }$}
&
&
\textcolor{blue}{\( \Theta^2 \)}
&
&
&
&
\textcolor{blue}{\( \omega^{\lambda \langle \mu }{\omega^{\nu \rangle}}_{\lambda} \)}  
\\
&
&
&
\textcolor{blue}{\( \omega^{\mu \nu}\omega_{\mu \nu}\)}  
&
&
&
&
\textcolor{blue}{\( {a}^{\langle \mu }{a}^{ \nu \rangle} \)}    
\\
&
&
&
\textcolor{blue}{\( {a}_{\mu}{a}^{\mu} \)} 
&
&
&
&
\textcolor{blue}{\( {a}^{\langle \mu }{\mathcal{I}}^{ \nu \rangle} \)}
\\
&
&
&
\textcolor{blue}{\( {a}_{\mu}{\mathcal{I}}^{\mu}  \)}
&
&
&
&
\textcolor{blue}{\( {\mathcal{I}}^{\langle \mu }{\mathcal{I}}^{ \nu \rangle} \)}
\\
&
&
&
\textcolor{blue}{\({\mathcal{I}}_{\mu}{\mathcal{I}}^{\mu}\)}
&
&
&
& 
\\
\cline{2-8}
&
\multirow{2}{*}{${ \scriptstyle \mathcal{O}_{_{2\mathrm{R}}} }$}
&
&
\textcolor{blue}{\( \pi^{\mu \nu}\pi_{\mu \nu} \)}   
&
&
&
&
\textcolor{blue}{\( \pi^{\lambda \langle \mu }{\pi^{\nu \rangle}}_{\lambda} \)}
\\
&
&
&
\textcolor{blue}{\( \Pi^2 \) }
&
&
&
&
\\
\cline{2-8}
&
\multirow{2}{*}{N/A}
&
&
\textcolor{blue}{
\( R\)}
&
&
&
&
\textcolor{blue}{
\( R^{\langle \mu \nu \rangle}\)}
\\
&
&
&
\textcolor{blue}{\( u^{\mu}u^{\nu}R_{\mu \nu}\)}
&
&
&
&
\textcolor{blue}{\( u_{\lambda}u_{\rho}R^{\lambda \langle \mu \nu \rangle
    \rho }\)}\\
\hline
\hline
\end{tabular}
\caption{Comparison between different formalisms. Terms that are included
  in the \textbf{HL83} formulation but are missing in the given
  formulation are coloured in red and listed under
  '\textcolor{red}{\(-\)}'. Similarly, those terms that are absent in the
  \textbf{HL83} formulation but are included in the other approaches are
  coloured in blue and listed under '\textcolor{blue}{\(+\)}'. Terms that
  differ only by a scalar function of \(\rho\) and \(p\) are not counted
  as missing, as potential differences can be absorbed in the definition
  of the transport coefficients. For each formulation we also use
  different rows to reflect the classification within the scheme with
  respect to the order of Knudsen or inverse Reynolds number
  \eqref{eq:classification1}. Finally, terms without any symbol are not
  classified because they do not occur in the corresponding approach.}
\label{tab:comparison}
\end{table}

After having introduced and reviewed four different second-order
formulations of the equations of GRDHD, namely, \textbf{IS79},
\textbf{HL83}, \textbf{DMNR12} and \textbf{rBRSSS08}, we next proceed
with a comparison of the sets of evolution equations for the dissipative
currents. In order to facilitate the identification of the transport
coefficients in our notation with the transport coefficients in the
original notations, we have distinguished between the different currents
\({I_{i}}^{\mu}\) and the kinematic acceleration \(a^{\mu}\). However, as
one can see from the first-order relations
(\ref{eq:firstorder1})--(\ref{eq:firstorder3}), a further reduction is
possible by using Eqs. (\ref{eq:redHL1})--(\ref{eq:redHL7}),
(\ref{eq:redDMNR1})--(\ref{eq:redDMNR4}) and (\ref{eq:redRR1}),
(\ref{eq:redRR2}), respectively. To this scope, we choose $p$ and $\rho$
as our thermodynamical variables and write every gradient as a linear
combination of $a^{\mu}$ and $\partial^{\left\langle \mu
  \right\rangle}\rho$. We apply this reduction in order to compare
\textbf{IS79}, \textbf{DMNR12} and \textbf{rBRSSS08} with \textbf{HL83}
in Table \ref{tab:comparison} after introducing the new definition
\begin{align}
\mathcal{I}^{\mu} := \nabla^{\mu}\rho \,.
\end{align}
This reduction procedure leads to differences in some transport
coefficients, which are captured by the previously introduced scalar
functions $\{b_{i}\}$ with $i\in [1,14]$.

Table \ref{tab:comparison} presents a quick overview of the various
formulations considered, whose denomination appears in the first
column. The table is written such that terms that are included in the
\textbf{HL83} formulation but are missing in the set we want to compare
with are coloured in red and listed under
'\textcolor{red}{\(-\)}'. Similarly, those terms that are absent in the
\textbf{HL83} formulation but are included in the other approaches are
coloured in blue and listed under '\textcolor{blue}{\(+\)}'.
Furthermore, terms that differ only by a scalar function of \(\rho\) and
\(p\) are not counted as missing, as potential differences can be
absorbed in the definition of the transport coefficients. Furthermore, to
facilitate the classification of the various terms within the scheme
\eqref{eq:classification1}, \ie with respect to the order of Knudsen or
inverse Reynolds number \(\mathcal{O}_1\), \(\mathcal{O}_{_{\mathrm{RK}}}\),
\(\mathcal{O}_{2\mathrm{R}}\), and \(\mathcal{O}_{2\mathrm{K}}\), they are
collected on different rows. Finally, terms without any symbol are not
classified because they do not occur in the corresponding approach.

Note that the \textbf{rBRSSS08} formulation includes second-order terms,
but does not distinguish between \(\eta \sigma^{\mu \nu}\Pi\) and \(\zeta
\pi^{\mu \nu}\Theta\), which are not the only second-order terms present
in the \textbf{rBRSSS08} formulation. Another difference is of course the
presence of terms associated with the curvature of spacetime, which are
absent in \textbf{DMNR12}. Clearly, these terms cannot be classified by a
certain order within the scheme \eqref{eq:classification1}. However, if
$g_{\mu \nu}$ is considered as an equilibrium fluid variable, then these
terms can be seen as of second order in Knudsen number. Finally, we
recall that the \textbf{rBRSSS08} formulation is derived in the Landau
frame, where the heat currents are zero.

\begin{table}
\centering
\begin{tabular}{p{0.22\textwidth}|p{0.15\textwidth}|p{0.13\textwidth}|c|c|c}

\textbf{IS79} & \textbf{HL83} & \textbf{DMNR12} & \textbf{rBRSSS08} & \textbf{local} & \textbf{source} \\

\hline
\hline
\(-1/3 ~ \zeta_V u^{\mu}_{E|\mu}\) & \(-\zeta \Theta\) & \(-\zeta^r \Theta\) & \(-\zeta \nabla_{\mu}^{\perp}u^{\mu}\) & \(\Pi_{\ns}\) & -\\

\hline
\(\displaystyle \frac{\kappa T}{\eta \beta} ~ \Delta^{\mu}_{\lambda}\alpha_{|\mu}\) & \(-\kappa T(\nabla^{\langle \mu \rangle}\ln T +a^{\mu})\) & \(\displaystyle \frac{\kappa_q^r \tau^r_W}{\psi^W_r\tau^r_V\beta_0^2h_0^2}~\nabla^{\langle \mu \rangle}\alpha_0\) & N/A & \(q^{\phantom{\ns}\mu}_{\ns}\) & -\\

\hline 
\(-2\zeta_S\Delta^{\alpha}_{\langle \lambda}(u_{E})\Delta^{\beta}_{\mu \rangle}(u_E)u^E_{\alpha | \beta}\) & \(-2\eta \sigma^{\mu\nu}\) &  \(2\eta^{r}\sigma^{\mu \nu}\) &  \(-\eta \sigma^{\mu \nu}\) &  \(\pi^{\phantom{\ns}\mu\nu}_{\ns}\) & -\\

\hline
\hline
\({1}/{3} ~ \zeta_{V} \beta_0\) & \(\zeta \beta_0\) & \(\tau^r_{\Pi}\) & \(\tau_{\Pi}\) & \(\tau_{\Pi}\) & \(\dot{\Pi}\)\\

\hline
\(\kappa T \beta_1\) & \(\kappa T \beta_1\) & \(\tau^r_W\) & N/A & \(\tau_q\) & \(\dot{q}^{\langle \mu\rangle}\)\\

\hline
\(2\zeta_S \beta_2\) & \(2 \eta \beta_2\) & \(\tau^r_{\pi}\) & \(\tau_{\pi}\) & \(\tau_{\pi}\) & \(\dot{\pi}^{\langle \mu \nu \rangle}\)\\

\hline
\hline
0 & \(- {1}/{2} ~ \tau_{\Pi}\) & \( - \tau^r_{\Pi}\delta^r_{\Pi \Pi}\) & 0 & \(\Delta^{^\Pi}_1\) & \(\Pi \Theta\)\\

\hline
\(1/3 ~ \zeta_V a_0'\) & 0 & \(\tau^r_{\Pi}\tau^r_{\Pi W}\) & 0 & \(\delta_1^{^\Pi}\) & \(q^{\mu}a_{\mu}\)\\
\hline
0 & 0 & \(\tau^r_{\Pi}\lambda^{r}_{\Pi \pi}\) & 0 & \(\delta^{^\Pi}_2\) & \(\pi^{\mu \nu}\sigma_{\mu \nu}\)\\

\hline
0 & 0 & 0 & \(\xi_3\) & \(\delta^{^\Pi}_3\) & \(\omega^{\mu \nu}\omega_{\mu \nu}\)\\

\hline
0 & \(- {1}/{2} ~ \tau_{q}\) & \(- \tau^r_W \delta^r_{WW}\) & N/A & \(\Delta^q_1\) & \(q^{\mu} \Theta\)\\

\hline
\(\kappa T a_0\) & 0 & \(-\tau^r_W \tau^r_{q\Pi}/\psi^W_r\) & N/A & \(\delta^q_1\) & \(\Pi a^{\mu}\)\\

\hline
\(\kappa T a_1\) & 0 & \(-\tau^r_W\tau^r_{q\pi}/\psi^W_r\) & N/A & \(\delta^q_2\) & \(\pi^{\mu \nu}a_{\nu}\)\\

\hline
\(\kappa T \beta_1=\tau_{q}\) & 0 & \(\tau^r_W\) & N/A & \(\delta^q_3\) & \(\omega^{\mu \nu}q_{\nu}\)\\

\hline
0 & 0 & \(- \tau^r_W\lambda^r_{WW}\) & N/A & \(\delta^q_4\) & \(\sigma^{\mu \nu}q_{\nu}\)\\

\hline
0 & \(-{1}/{2} ~ \tau_{\pi}\) & \(-2\tau^r_{\pi}\delta^r_{\pi \pi}\) & \(-(\tau_{\pi}d+\overline{\tau}_{\pi}^*)/(d-1)\) & \(\Delta^{\pi}_1\) & \(\pi^{\mu \nu}\Theta\)\\

\hline
\(2\zeta_s a_1'\) & 0 & \(2\tau^r_{\pi W}\tau^r_{\pi}\) & 0 & \(\delta^{\pi}_{1}\) & \(q^{\left\langle \mu \right.} a^{\left. \nu \right\rangle}\)\\

\hline
\(4\zeta_S\beta_2=2\tau_{\pi}\) & 0 & \(2\tau^r_{\pi}\) & \(-\lambda_2/\eta\) & \(\delta^{\pi}_{2}\) & \(\pi^{\lambda \left\langle \mu \right.}{\omega^{\left. \nu \right\rangle}}_{\lambda}\)\\

\hline
0 & 0 & \(-2\tau^r_{\pi}\lambda^{r}_{\pi \pi}\) & 0 & \(\delta^{\pi}_{3}\) & \(\pi^{\lambda \left\langle \mu \right.}{\sigma^{\left. \nu \right\rangle}}_{\lambda}\)\\

\hline
0 & 0 & \(2\tau^r_{\pi}\lambda^r_{\pi \Pi}\) & 0 & \(\delta^{\pi}_{4}\) & \(\Pi \sigma^{\mu \nu}\)\\

\hline
0 & 0 & \(2\tau^r_{\pi}\lambda^r_{\pi \Pi}\) & \(\lambda_3\) & \(\delta^{\pi}_{5}\) & \(\omega^{\lambda \left\langle \mu \right.}{\omega^{\left. \nu \right\rangle}}_{\lambda}\)\\

\hline
\hline
\(1/3 ~ \zeta_V \alpha_0\) &\(\zeta \alpha_0\) &  \(- \tau^r_{\Pi}l^r_{\Pi W}\) & 0 & \(L^{\Pi}_1\) & \(\nabla_{\mu}q^{\mu}\)\\

\hline
\(\kappa T \alpha_0\) & \(\kappa T \alpha_0\) & \(l^r_{q \Pi}\tau^{r}_{W}/\psi^W_r\) & N/A & \(L_1^q\) & \(\nabla^{\langle \mu \rangle}\Pi\)\\

\hline
\(\kappa T \alpha_1\) & \(\kappa T \alpha_1\) & \(- l^r_{q \pi}\tau^{r}_{W}/\psi^W_r\) & N/A & \(L_2^q\) & \(\nabla_{\nu}\pi^{\langle \mu \rangle \nu}\)\\

\hline
\(2\zeta_S a_1\) & \(2\eta \alpha_1\) & \(2l^r_{\pi W}\tau^r_{\pi}\) & 0 & \(L^{\pi}_1\) & \(\nabla^{ \langle \mu }q^{ \nu \rangle}\)\\
\hline
\hline
\end{tabular}
\caption{Table of transport coefficients I: we here match the transport
  coefficients used in this work with the corresponding transport
  coefficients in the original papers of the \textbf{IS79},
  \textbf{HL83}, \textbf{DMNR12}, and \textbf{rBRSSS08} formulations.}
\label{tab:transport1}
\end{table}

\begin{table}
\centering
\begin{tabular}{c|c|c|c|c|c}
\textbf{IS79} & \textbf{HL83} & \textbf{DMNR12} & \textbf{rBRSSS08} & \textbf{local} & \textbf{source}\\
\hline
\hline
0 & \(-1/2 ~ \zeta T\) & 0 & 0 & \(\Lambda^{^\Pi}_1\) & \(\Pi u_{\mu}{{I_{i}}}^{\mu}\)\\

\hline
0 & \(\zeta \gamma_0 T\) & \(\tau^r_{\Pi}\lambda^r_{\Pi W}\) & 0 & \(\Lambda^{^\Pi}_2\) & \(q^{\mu}{I_{i}}_{\mu}\)\\

\hline
0 & 0 & 0 & \(\overline{\xi}_4\) & \(\lambda^{^\Pi}_1\) & \({I_{i}}_{\mu}{I_{i}}^{\mu}\)\\

\hline
0 & \(-1/2 ~ \kappa T^2\) & 0 & N/A & \(\Lambda^q_1\) & \(q^{\mu}u^{\nu}{I_{i}}_{\nu}\)\\

\hline
0 & \(\kappa T^2 (1-\gamma_0)\) & \(\lambda^r_{q\Pi}\tau^r_W/\psi^W_r\) & N/A & \(\Lambda^q_2\) & \(\Pi {I_{i}}^{\left\langle \mu \right\rangle}\)\\

\hline
0 & \(\kappa T^2 (1-\gamma_1)\) & \(\lambda^r_{q\pi}\tau^r_W/\psi^W_r\) & N/A & \(\Lambda^q_3\) & \(\pi^{\mu \nu}{I_{i}}_{\nu}\)\\

\hline
0 & \( - 1/2 ~ \eta T\) & 0 & 0 & \(\Lambda^{\pi}_1\) & \(\pi^{\mu \nu}u^{\lambda}{I_{i}}_{\lambda}\)\\

\hline
0 & \(2\eta \gamma_1 T\) & \(-2\tau^r_{\pi}\lambda^r_{\pi W}\) & 0 & \(\Lambda^{\pi}_2\) & \(q^{\left\langle \mu \right.} {I_{i}}^{\left. \nu \right\rangle}\)\\

\hline
0 & 0 & 0 & \(\overline{\lambda}_4\) & \(\lambda^{\pi}_1\) & \({I_{i}}^{\left\langle \mu \right.} {I_{i}}^{\left. \nu \right\rangle}\)\\

\hline
\hline
0 & 0 & 0 & \(\xi_1/\eta^2\) & \(\varphi^{^\Pi}_1\) & \(\pi^{\mu \nu}\pi_{\mu \nu}\)\\

\hline
0 & 0 & 0 & \(\overline{\xi}_2/\zeta^2\) & \(\varphi^{^\Pi}_2\) & \(\Pi^2\)\\

\hline
0 & 0 & 0 & \(\lambda_1/\eta^2\) & \(\varphi^{\pi}_1\) & \(\pi^{\lambda \left\langle \mu \right.}{\pi^{\left. \nu \right\rangle}}_{\lambda}\)\\

\hline
\hline
0 & 0 & 0 & \(\xi_5\) & \(g^{^\Pi}_1\) & \(R\)\\

\hline
0 & 0 & 0 & \(\xi_6\) & \(g^{^\Pi}_2\) & \(u^{\mu}u^{\nu}R_{\mu \nu}\)\\

\hline
0 & 0 & 0 & \(\kappa\) & \(g^{\pi}_1\) & \(R^{\left\langle \mu \nu \right\rangle}\)\\

\hline
0 & 0 & 0 & \(2(\kappa^*-\kappa)\) & \(g^{\pi}_2\) & \(u_{\lambda}u_{\rho}R^{\lambda \left\langle \mu \nu \right\rangle \rho}\)\\
\hline
\hline
\end{tabular}
\caption{Table of transport coefficients II: the same as Table
  \ref{tab:transport1} but for those terms missing in the previous
  Table. Note that the definitions of the currents \({I_{i}}^{\mu}\)
  differ for different formalisms and a summary is given in Table
  \ref{tab:fluxes}.}
\label{tab:transport2}
\end{table}

\begin{table}
  \centering
  \begin{tabular}{c|c|c|c}
    \textbf{IS79} & \textbf{HL83} & \textbf{DMNR12} & \textbf{rBRSSS08}
    \\[1em] \hline N/A & \({I_{i}}^{\mu}=\nabla^{\mu}\left(c_i/T\right),
    \quad i \in \{1,\dots,5\}\) &
    \({I_6}^{\mu}=\partial^{\mu}\left(\frac{\mu}{T}\right)\) &
    \({I_7}^{\mu}=\nabla^{\mu}\ln e\)\\
  \end{tabular}
  \caption{Summary of the various definitions of the currents
    \({I_{i}}^{\mu}\) given for the four formulations considered
    here. Note that the \textbf{IS79} formulation neglects gradients of
    transport coefficients and that the coefficients $c_i$ appearing in
    the currents of the \textbf{HL83} formulation are given by
    \(c_{\{1,\dots,5\} } = \{\tau_{\Pi}/\zeta, \alpha_0, \tau_q/\kappa
    T, \alpha_1, \tau_{\pi}/\eta\}\). }\label{tab:fluxes}
\end{table}

As a corollary to this comparison, we report in Tables
\ref{tab:transport1} and \ref{tab:transport2} a list of all the transport
coefficients in our and in the original notation, while the definition of
all \({I_{i}}^{\mu}\) is given in Table \ref{tab:fluxes}.

%------------------------------------------------------------------------
%------------------------------------------------------------------------
\section{Details on Viscous Black-Hole Accretion}
\label{sec:vis_acc_details}
%------------------------------------------------------------------------
%------------------------------------------------------------------------

Given the very limited use and knowledge of the stationary solution of
the spherically symmetric equations of GRDHD in a Schwarzschild
spacetime, we here review the basic mathematical expressions and the
strategy employed to obtain a numerical solution in the presence of a
critical point.

%------------------------------------------------------------------------
\subsection{Equations of GRDHD}
%------------------------------------------------------------------------

We recall that the equations of GRDHD are given by
Eqs. (\ref{eq:continuity}), (\ref{eq:em_conservation}) where
$J^{\mu}=J_{_{\textrm{PF}}}^{\mu}$, and $T^{\mu \nu}$ is given by Eq.
(\ref{eq:energy_momentum}), together with Eq. (\ref{eq:bulk_implement})
in the case in which the heat current and the shear-stress tensor are set
to zero. By demanding the equations to be stationary and spherically
symmetric, \ie all variables are functions of the circumference radius
\(r\) only and the fluid four-velocity has the form
\(u^{\mu}=(u^t,u,0,0)^T\), it is possible to obtain the following
coupled, nonlinear system of ODEs in Schwarzschild coordinates
\begin{align}
\frac{d\rho}{dr} &= -\frac{\rho}{r} ~ \frac{M/(\mathcal{E}^2r)-\Pi r/\left[(\rho h +
  \Pi)\tau_{_{\Pi}} u\right]
  -2u^2/\mathcal{E}^2}{c^2_{s,t}-u^2/\mathcal{E}^2}\,,\label{eq:lin_sol_rho}\\ \frac{du}{dr}
&= \frac{u}{r} ~ \frac{M/(\mathcal{E}^2r)-\Pi r/\left[(\rho h + \Pi)\tau_{_{\Pi}} u \right]
  -2c^2_{s,t}}{c^2_{s,t}-u^2/\mathcal{E}^2}\,,\label{eq:lin_sol_u}\\ \frac{d\Pi}{dr}
&= -
\frac{\Pi\left(c^2_{s,t}-u^2/\mathcal{E}^2\right)/(u\tau_{_{\Pi}}) 
  +\zeta\left[M/(\mathcal{E}^2r)-\Pi r/\left[(\rho h +
    \Pi)\tau_{_{\Pi}} u\right]
    -2u^2/\mathcal{E}^2\right]/(\tau_{_{\Pi}} r)}{c^2_{s,t}-u^2/\mathcal{E}^2}\,,\\ \frac{dh}{dr}
&= - \frac{\rho h + \Pi}{\rho r} ~
\frac{\left[c^2_{s,t}-(\zeta/\tau_{_{\Pi}}-\Pi)/(\rho h +
    \Pi)\right] \left[M/(\mathcal{E}^2r)-\Pi r/\left[(\rho h + \Pi)
    \tau_{_{\Pi}} u\right]-2u^2/(\mathcal{E}^2)\right]}{c^2_{s,t}
  -u^2/\mathcal{E}^2}\,,\label{eq:lin_sol_h}
\end{align}
where \(\mathcal{E}:=u_{t}=-\sqrt{1-{2M}/{r}+u^2}\). Furthermore, the
system is characterised by two conserved quantities, namely, the
mass-accretion rate \(\dot{M}\) and the ``viscous'' Bernoulli constant
\(\mathcal{B}\)
\begin{align}
\dot{M} &:= 4\pi \rho u r^2\,,\\
\mathcal{B} &:= (\rho h +\Pi)\mathcal{E}/\rho\,.
\end{align}
Note that in the inviscid limit \(\lim_{_{\Pi \rightarrow 0}}\mathcal{B}
= \mathcal{B}_{_{\textrm{PF}}} := h \mathcal{E} = h u_t\), where
\(\mathcal{B}_{_{\textrm{PF}}}\) denotes the relativistic, inviscid
Bernoulli constant. Using \(\dot{M}\) and \(\mathcal{B}\), we can 
express \(\rho\) and \(\Pi\) in terms of \(u\), \(h\) and \(r\)
\begin{align}
\rho &= \dot{M}/(4\pi ur^2)\,,\label{eq:final_1}\\
\Pi &= \rho \mathcal{B}/\mathcal{E} - \dot{M}h/(4\pi ur^2)\,,
\end{align}
so that the viscous speed of sound becomes
\begin{align}
c^2_{s,t}=(\gamma_e-1)\frac{\mathcal{B}-\mathcal{E}}{\mathcal{B}} +
\frac{\zeta}{\tau_{_{\Pi}}}\frac{4\pi \mathcal{E} u
  r^2}{\mathcal{B}\dot{M}}\,,
\end{align}
and Eqs. (\ref{eq:lin_sol_u}) and (\ref{eq:lin_sol_h}) simplify to
\begin{align}
\frac{du}{dr}&=\frac{u}{r}~\frac{M/(\mathcal{E}^2r)-(\mathcal{B} - \mathcal{E}
  h)r/\left(\mathcal{B}\tau_{_{\Pi}} u \right)
  -2c^2_{s,t}}{c^2_{s,t}-u^2/\mathcal{E}^2}\,, \label{eq:final_4}\\ \frac{dh}{dr}
&= - \frac{1}{r}~\frac{\left[(\gamma_e-1)
    (\mathcal{B}-\mathcal{E})/\mathcal{E} +(\mathcal{B}-h\mathcal{E})/\mathcal{E}
    \right]\left[M/(\mathcal{E}^2r)-(\mathcal{B}-\mathcal{E} h)r/\left(\mathcal{B}
    \tau_{_{\Pi}} u\right) -2 u^2/\mathcal{E}^2
    \right]}{c^2_{s,t}-u^2/\mathcal{E}^2}\,.\label{eq:final_5}
\end{align}
Equations (\ref{eq:final_4})--(\ref{eq:final_5}) do not have an analytic
solution and their non-trivial numerical solution is discussed in more
detail in the next section.

%------------------------------------------------------------------------
\subsection{Singular point analysis and integration}
%------------------------------------------------------------------------

The first step in the solution of Eqs. (\ref{eq:final_4}),
(\ref{eq:final_5}) is to choose the location of the sonic point \(r_s\),
which is defined as the radial coordinate where
$u^2/\mathcal{E}^2=c_{s,t}^2$. This corresponds to the position where the
infalling fluid velocity equals that of the speed of sound, and where
Eqs. (\ref{eq:final_4}) and (\ref{eq:final_5}) become singular. Because
of that, we choose to start the numerical integration procedure directly
at $r_s$ and integrate inward (\ie supersonic portion) and outward (\ie
subsonic portion) from there. In this way, we can make use of exact
results for the fluid conditions and their first derivatives at the sonic
point in order to initialise numerical integration.

Using the inviscid solution with a polytropic constant (see discussion in
Sec. \ref{sec:accretion}), it is possible to obtain values for
\(\dot{M}\) and \(\mathcal{B}_{_{\textrm{PF}}}\). In the inviscid case,
all state variables at the sonic point are automatically determined by
choosing \(r_s\) \citep[see][for details]{Hawley84a}. Setting
\(\mathcal{B}=\mathcal{B}_{_{\textrm{PF}}}\) and using the value for
\(\dot{M}\) found for the inviscid case, we can compute our initial
values at \(r_s\) for the viscous case. As in the inviscid case, we
demand that the derivatives at the sonic point constitute a removable
singularity. Hence, from Eqs. (\ref{eq:final_4}) and (\ref{eq:final_5})
we find the conditions
\begin{align}
0 &= \frac{M}{\mathcal{E}^2r}-\frac{\mathcal{B}-\mathcal{E}
  h}{\mathcal{B}}\frac{r}{\tau_{_{\Pi}} u}-2 c^2_{s,t}\,,\\ 0 &=
c^2_{s,t}-\frac{u^2}{\mathcal{E}^2}\,.
\end{align}
Numerical root finding via Mathematica \citep{Mathematica} yields the
values of the radial four-velocity component and of the specific enthalpy
at the sonic radius, \(u_s\) and \(h_s\). Because multiple solutions
exist, we select the solution satisfying the conditions: \(u_s < 0\) and
\(h_s > 1 \), which was found to be unique for the choice of constant
$\zeta$ and \(\tau_{_{\Pi}}\). Next, accurate values for the derivatives
\(du/dr\) and \(dh/dr\) are needed at the sonic point because we would
otherwise rely on the evaluation of the right-hand side of
Eqs. (\ref{eq:final_4}) and (\ref{eq:final_5}) at \(r_s\) for the first
step of the integration procedure.

Even though the singularity is removable, numerical errors cause singular
behaviour and unreliable values for the derivatives. Following
\cite{Mandal2007} (but see also \citealp{Ray2002, Afshordi2003}) we can
treat Eqs. (\ref{eq:final_4})--(\ref{eq:final_5}) assuming that each
function $u$, $r$, and $h$ can be expressed in terms of a parameter
$\xi$, \ie $u=u(\xi), r=r(\xi), h=h(\xi)$, such that when writing
Eqs. (\ref{eq:final_4})--(\ref{eq:final_5}) symbolically as $du/dr=A/B$
and $dh/dr=C/B$, they can be effectively rewritten as $du/d\xi=A$,
$dr/d\xi=B$, and $dh/d\xi=C$. More specifically, we express
Eqs. (\ref{eq:final_4})--(\ref{eq:final_5}) symbolically as
\begin{align}
\frac{dr}{d\xi}&=r
\left(c^2_{s,t}-\frac{u^2}{\mathcal{E}^2}\right)\,,\label{eq:lin1}\\
\frac{du}{d\xi} &= u \left(
\frac{M}{\mathcal{E}^2r}-\frac{\mathcal{B}-\mathcal{E}
  h}{\mathcal{B}}\frac{r}{\tau_{_{\Pi}} u} -2
c^2_{s,t}\right)\,,\label{eq:lin2}\\
\frac{dh}{d\xi} &=
-\left[(\gamma_e-1)\frac{\mathcal{B}-\mathcal{E}}{\mathcal{E}} +
  \frac{\mathcal{B}-h\mathcal{E}}{\mathcal{E}}\right]
\left[\frac{M}{\mathcal{E}^2r}-\frac{\mathcal{B}-\mathcal{E}
    h}{\mathcal{B}}\frac{r}{\tau_{_{\Pi}} u}-2
  \frac{u^2}{\mathcal{E}^2}\right]\,.
\label{eq:lin3}
\end{align}
Note that $\xi$ does not have a physical interpretation and should be
seen simply as a mathematical parameter. However, with this
parameterisation, each of the solutions $r(\xi), u(\xi)$, and $h(\xi)$
can be thought as consisting of two branches on either side of the sonic
point: \ie one branch for $r<r_s$ and one for $r>r_s$, and the solution
at the sonic point can be obtained in the limit $\xi \rightarrow \pm
\infty$\footnote{This condition follows from the requirement that the
perturbation behaves as $\sim e^{\lambda \xi}$, where $\lambda$ is a
constant eigenvalue that can be either positive or negative, and that it
vanishes at the sonic point; see also below.}.

Equations (\ref{eq:lin1})--(\ref{eq:lin3}) effectively represent an
independent system whose paths in the \((u,h,r)\)-space, \ie the
\textit{phase space} of Eqs. (\ref{eq:lin1})--(\ref{eq:lin3}), correspond
to solutions of the original system, \ie Eqs. (\ref{eq:final_4}) and
(\ref{eq:final_5}). Thus, the original system can be viewed as
differential equations for the \textit{phase paths} of
Eqs. (\ref{eq:lin1})-- (\ref{eq:lin3}). The solution $(u_s,h_s)$ at the
sonic point \(r_s\) constitutes a solution $(u_s,h_s,r_s)$ of
Eqs. (\ref{eq:lin1})--(\ref{eq:lin3}) at the so-called equilibrium point,
where \(du/d\xi=dh/d\xi=dr/d\xi=0\). Notice the dependence on the
specific choice of \(\zeta=\zeta(u,h,r)\) and
\(\tau_{_{\Pi}}=\tau_{_{\Pi}}(u,h,r)\).

Equations (\ref{eq:lin1})--(\ref{eq:lin3}) can be linearised at the sonic
point, \ie \(u \approx u_s + \delta u\), \(h \approx h_s + \delta h\) and
\(r \approx r_s + \delta r\), to obtain a local solution
\citep{Jordan2007}. This linearisation procedure and the ansatz \(\delta
u = \delta_{u} \exp(\lambda \xi)\), \(\delta h = \delta_{h} \exp(\lambda
\xi)\) and \(\delta r = \delta_{r} \exp(\lambda \xi)\) leads to a an
eigenvalue problem, which can be solved numerically. The outcome is a
set of three eigenvalues \(\{\lambda_1,\lambda_2,\lambda_3\}\) and
eigenvectors with the properties \(\lambda_1 < 0\), \(\lambda_2 > 0\) and
\(\lambda_3 \approx 0\). The corresponding set of local solutions is then
\begin{align}
\delta u_1 (\xi) = (\delta_u)_1 \exp(\lambda_1 \xi)\,;~\delta h_1 (\xi)
= (\delta_h)_1 \exp(\lambda_1 \xi)\,; ~\delta r_1 (\xi) = (\delta_r)_1
\exp(\lambda_1 \xi)\,, \label{eq:set_1}\\ \delta u_2 (\xi) =
(\delta_u)_2 \exp(\lambda_2 \xi)\,;~\delta h_2 (\xi) = (\delta_h)_2
\exp(\lambda_2 \xi)\,; ~\delta r_2 (\xi) = (\delta_r)_2 \exp(\lambda_2
\xi)\,,\label{eq:set_2}
\end{align}
where $\left[(\delta_u)_{i},(\delta_h)_{i},(\delta_r)_i\right]$ denotes
the eigenvector which belongs to the $i$-th eigenvalue $\lambda_{i}$.
Because we require the linearised solution to pass through
$(u_s,h_s,r_s)$, we look for linearised solutions that fulfil $\delta
u_{i}(\xi) = \delta h_{i}(\xi) = \delta r_{i}(\xi) = 0$ for an arbitrary
$\xi$. Because $\lambda_3 \approx 0$, it is not possible to fulfil the
requirement $\delta u_3 = \delta h_3 = \delta r_3 = 0$ for any value of
$\xi$; hence, we neglect this eigenvalue. Similarly, because $\lambda_1$
and $\lambda_2$ are of opposite signs, superpositions of both sets are
not allowed for the same reason. Thus, we recover the desired solution in
the limit $\xi \rightarrow \infty $ for the set (\ref{eq:set_1}) and in
the limit $ \xi \rightarrow - \infty$ for the set (\ref{eq:set_2}). For
each eigenvalue $i = 1,2$, the derivatives are given by $(d u/d
r)_s=(\delta_u)_i/(\delta_r)_i$ and $d h/d r=(\delta_h)_i/(\delta_r)_i$,
respectively, and we select the eigenvalue whose eigenvector yields $(d
u/d r)_s > 0$.

Finally, we start our numerical integration by calculating the first
steps on either side of the sonic point, \ie at $r_l=r_s-\Delta r$ and
$r_r=r_s+\Delta r$, employing forward and backward finite-differences of
first order with radial stepsize $\Delta r$. The integration then
proceeds with a fourth-order $L$-stable singly diagonally implicit
Runge-Kutta (SDIRK) method (see Table 6.5. of \citealt{HairerBookODE_2}
and the corresponding Butcher tableau). It is worth noticing that this
method includes an embedded third-order formula and a continuous
solution, both of which can be found in \citet{HairerBookODE_2}. We use a
simple fixed-point iteration procedure to solve the implicit equation at
each stage, \ie given the solution at a specific radius $r_n$, where
$r_{n+1}=r + \Delta r $ marks the next grid point, we use the solution at
$r_n$ as an initial guess and iterate on the $k_{i}$'s, which are defined
by the following general formulas derived assuming the ODE has the form
$dy/dr=f(r,y)$ and an $s$-stage Runge-Kutta method
\begin{align}
\label{eq:sdirk}
y_{n+1}&=y_{n}+\Delta
r\sum_{i=1}^{s}b_{i}k_{i}\,,\\ k_{i}&=f\left(r_n+c_{i}\Delta
r,y_n+h\sum_{j}^{s}a_{ij}k_{j}\right)\,, \qquad i=1,\ldots,s\,,
\end{align}
where the coefficients \(\{c_{i},a_{ij},b_{i}\}\) are given in the form
of Butcher tableaus in Table \ref{tab:butcher}. The numerical integration
is obtained using 700,000 points on a grid that has a higher resolution
in the most delicate portions of the solution, \ie at the event horizon
and at the sonic point. After a cubic-spline interpolation, the numerical
solutions of the ODEs are then used as initial conditions for the
\texttt{BHAC} evolution.

\begin{table}
\centering
\begin{tabular}{c | c | c | c | c | c}

${1}/{4}$ & ${1}/{4}$ & & & & \\[5pt]

${3}/{4}$ & ${1}/{2}$ & ${1}/{4}$ & & & \\[5pt]

${11}/{20}$ & ${17}/{50}$ & $-{1}/{25}$ & ${1}/{4}$ & & \\[5pt]

${1}/{2}$ & ${371}/{1360}$ & $-{137}/{2720}$ & ${15}/{544}$ & ${1}/{4}$ & \\[5pt]

$1$ & ${25}/{24}$ & $-{49}/{48}$ & ${125}/{16}$ & $-{85}/{12}$ & ${1}/{4}$ \\[5pt]

\hline

 & ${25}/{24}$ & $-{49}/{48}$ & ${125}/{16}$ & $-{85}/{12}$ & ${1}/{4}$ \\

\end{tabular}
\caption{Butcher tableau for a \(L\)-stable SDIRK method of order 4 [\cf
    Eq. \eqref{eq:sdirk}].}
  \label{tab:butcher}
\end{table}

%%%%%%%%%%%%%%%%%%%%%%%%%%%%%%%%%%%%%%%%%%%%%%%%%%

% Don't change these lines
\bsp % typesetting comment
\label{lastpage}
\end{document}